\documentclass[useAMS,usenatbib,manuscript]{mn2e}
\usepackage{mn2e-breakabs}
\usepackage{rotating}       
\usepackage{graphicx}
\usepackage{amsmath} 
\usepackage{amssymb} 
\usepackage{natbib}
\usepackage{lscape}
\usepackage{color}

\usepackage{xspace}

\def\kms{\mbox{km~s$^{-1}$}}
\def\kpc{\mbox{kpc}}

\def\kpch{\mbox{$h^{-1}$\,kpc}}

\def\Mpc{\mbox{Mpc}}
\def\mpch{\mbox{$h^{-1}$\,Mpc}}
\def\Mpch{\mbox{$h^{-1}$\,Mpc}}

\def\M200{\mbox{$M_{\rm 200}$}}

\def\R200{\mbox{$R_{\rm 200}$}}

\def\Vmax{\mbox{$V_{\rm max}$}}
\def\Vpeak{\mbox{$V_{\rm peak}$}}
\def\V200{\mbox{$V_{\rm 200}$}}

\title[Clustering of BOSS-CMASS galaxies]{The clustering of galaxies
  at $z\approx0.5$ in the SDSS-III Data Release 9 BOSS-CMASS sample: a test for the $\Lambda$CDM
  cosmology}

\author[Nuza et al.]{Sebasti\'an\,E.\,Nuza$^{1}$\thanks{E-mail: snuza@aip.de,
    arielsan@mpe.mpg.de, fprada@iaa.es}, Ariel\,G.\,S\'anchez$^{2}$,
  Francisco\,Prada$^{3,4,5,6}$, Anatoly\,Klypin$^{7}$, \newauthor 
  David\,J.\,Schlegel$^{8}$, Stefan\,Gottl\"ober$^{1}$, Antonio\,D.\,Montero-Dorta$^{5}$, Marc\,Manera$^{9}$, \newauthor
  Cameron\,K.\,McBride$^{10}$, Ashley\,J.\,Ross$^{9}$, Raul\,Angulo$^{11}$, Michael\,Blanton$^{12}$, \newauthor 
  Adam\,Bolton$^{13}$, Ginevra\,Favole$^{5}$, Lado\,Samushia$^{9}$, Francesco\,Montesano$^{2}$, \newauthor 
  Will\,J.\,Percival$^{9}$, Nikhil\,Padmanabhan$^{14}$, Matthias\,Steinmetz$^{1}$, Jeremy\,Tinker$^{12}$, \newauthor
  Ramin\,Skibba$^{15}$, Donald\,P.\,Schneider$^{16,17}$, Hong\,Guo$^{18}$, Idit\,Zehavi$^{18}$, Zheng\,Zheng$^{13}$, \newauthor 
  Dmitry\,Bizyaev$^{19}$, Olena\,Malanushenko$^{19}$, Viktor\,Malanushenko$^{19}$, Audrey\,E.\,Oravetz$^{19}$, \newauthor
  Daniel\,J.\,Oravetz$^{19}$, Alaina\,C.\,Shelden$^{19,20}$\\
  \\
  $^1$ Leibniz-Institut f\"ur Astrophysik Potsdam (AIP), An der Sternwarte 16, 14482 Potsdam, Germany\\
  $^2$ Max-Planck-Insitut f\"ur Extraterrestrische Physik, Giessenbachstra\ss{}e, 85748 Garching, Germany\\
  $^3$ Campus of International Excellence UAM+CSIC, Cantoblanco, E-28049 Madrid, Spain  \\
  $^4$ Instituto de F\'{\i}sica Te\'orica, (UAM/CSIC), Universidad Aut\'onoma de Madrid, Cantoblanco, E-28049 Madrid, Spain  \\
  $^5$ Instituto de Astrof\'{\i}sica de Andaluc\'{\i}a (CSIC), Glorieta de la Astronom\'{\i}a, E-18080 Granada, Spain \\
  $^6$ Visiting scientist at the Max Planck Institut f\"ur Astrophysik (MPA), Karl-Schwarzschild-Str. 1, 85741 Garching, Germany \\
  $^7$ Astronomy Department, New Mexico State University, Las Cruces, NM, USA\\
  $^8$ Lawrence Berkeley National Laboratory, 1 Cyclotron Road, Berkeley, CA, USA\\
  $^{9}$ Institute of Cosmology \& Gravitation, Dennis Sciama Building, University of Portsmouth, Portsmouth, PO1 3FX, UK\\
  $^{10}$ Department of Physics, Vanderbilt University, Nashville, TN, USA \\
  $^{11}$ Max Planck Institut f\"ur Astrophysik (MPA), Karl-Schwarzschild-Str. 1, 85741 Garching, Germany\\
  $^{12}$ Center for Cosmology and Particle Physics, New York University, NY, USA\\
  $^{13}$ Department of Physics and Astronomy, University of Utah, 115 south 1400 East, Salt Lake City, UT 84112, USA\\
  $^{14}$ Yale Center for Astronomy and Astrophysics, Yale University, New Heaven, CT 06511, USA\\
  $^{15}$ Steward Observatory, University of Arizona, 933 N. Cherry Ave., Tucson, AZ 85721, USA\\
  $^{16}$ Department of Astronomy and Astrophysics, The Pennsylvania State University, University Park, PA 16802, USA\\
  $^{17}$ Institute for Gravitation and the Cosmos, The Pennsylvania State University, University Park, PA 16802, USA\\
  $^{18}$ Department of Astronomy, Case Western Reserve University, 10900 Euclid Avenue, Cleveland, OH 44106, USA\\ 
  $^{19}$ Apache Point Observatory, P.O. Box 59, Sunspot, NM 88349-0059, USA\\
  $^{20}$ Department of Astronomy, University of Florida, 211 Bryant Space Science Center, Gainesville, FL 326711-2055, USA\\
}

\begin{document}
\maketitle
\begin{abstract}
  We present results on the clustering of $282,068$ galaxies in
  the Baryon Oscillation Spectroscopic Survey (BOSS) 
  sample of massive galaxies with redshifts $0.4<z<0.7$ which is part of the Sloan Digital Sky 
  Survey III project. Our results cover a large range of scales from 
  $\sim500\;\kpch$ to $\sim90\;\Mpch$. We compare these estimates with the
  expectations of the flat $\Lambda$CDM standard cosmological model
  with parameters compatible with WMAP7 data.
  We use the MultiDark cosmological simulation, one of the largest $N$-body runs presently
  available, together with a simple halo abundance matching technique,
  to estimate galaxy correlation functions, power spectra,
  abundance of subhaloes and galaxy biases. We find that the $\Lambda$CDM
  model gives a reasonable description to the observed correlation functions
  at $z\approx0.5$, which is a remarkably good agreement considering
  that the model, once matched to the observed abundance of BOSS 
  galaxies, does not have any free parameters. However, we find a $\gtrsim10\%$ 
  deviation in the correlation functions for scales
  $\lesssim1\;h^{-1}\,\Mpc$ and $\sim10$--$40\;h^{-1}\,\Mpc$. 
  A more realistic abundance matching model and better statistics from upcoming 
  observations are needed to clarify the situation. We also estimate that 
  about $12\%$ of the ``galaxies'' in the abundance-matched sample are 
  satellites inhabiting central haloes with mass $M\gtrsim10^{14}$ $h^{-1}$
  M$_{\sun}$. Using the MultiDark simulation we also study 
  the real space halo bias $b$ of the matched catalogue 
  finding that $b=2.00\pm0.07$ at large scales, consistent with the one 
  obtained using the measured BOSS projected correlation function. 
  Furthermore, the linear large-scale bias, defined using the 
  extrapolated linear matter power spectrum, depends on the number density $n$ of 
  the abundance-matched sample
  as $b= -0.048 - \left(0.594\pm0.02\right)\log_{10}\left(n/\,h^{3}\,\Mpc^{-3}\right)$.
  Extrapolating these results to BAO scales we measure a scale-dependent damping of 
  the acoustic signal produced by non-linear evolution that leads to $\sim 2$--$4\%$ dips 
  at $\gtrsim 3\sigma$ level for wavenumbers $k\gtrsim0.1\,h\,{\rm Mpc}^{-1}$ 
  in the linear large-scale bias.
\end{abstract}  


\section{Introduction}
\label{sec:intro}

The clustering of galaxies is a fundamental measure of the statistical
properties of the cosmic density field through cosmic time. In the
last decade, it became possible to determine the clustering strength
of galaxy populations at spatial scales out to tens of Mpc and beyond
with reasonable accuracy by means of massive galaxy surveys such as
the Two-Degree Field Galaxy Redshift Survey
\citep[e.g.,][]{2001MNRAS.328.1039C} and Sloan Digital Sky Survey
\citep[SDSS-I/II; e.g.,][]{Gunn1998, 2000AJ....120.1579Y, Gunn2006}. These and previous
studies have shown that the correlation function is not a simple power-law
and that the correlation length of luminous and massive galaxies
is larger than that of less luminous ones \citep[see][ and references therein]{Zehavi2011}. 

The Baryon Oscillation Spectroscopic Survey \citep[BOSS; e.g.,][]{Bolton2012,Smee2012,Dawson2012}, 
a branch of the ongoing SDSS-III \citep[][]{Eisenstein2011}, is considerably increasing the size 
of available galaxy samples.
BOSS consists of galaxy and quasar spectroscopic surveys over
a sky area of 10,000\,deg$^2$ and its main goal is to measure the BAO
feature at high precision. Specifically, BOSS aims at measuring the redshifts of 
about 1.5 million galaxies out to $z=0.7$. It will also acquire 
about 150,000 Ly$\alpha$ forest spectra of quasars in the range $2.2<z<4$, to map 
the large-scale distribution of galaxies at these earlier epochs
\citep[see][]{2011arXiv1104.5244S}. The effective volume of the galaxy
survey is expected to be about 7 times higher than that of the 
SDSS-I/II Luminous Red Galaxy (LRG) sample which consisted of $\sim100,000$ LRGs out to
$z=0.45$. The selection criteria of the BOSS targets result in a sample 
of massive, and hence highly clustered systems, which are suitable candidates 
for a reliable detection of the Baryon Acoustic Oscillation (BAO) clustering signal 
that can be used to constrain the expansion history of the 
Universe \citep[e.g.,][]{Anderson2012,Reid2012,Sanchez2012}. 
Additionally, the project also provides a wealth of other information on clustering 
and physical properties of galaxies.  

Requirements for theoretical predictions of galaxy 
clustering in BOSS are extreme: one needs accurate predictions for very large volumes 
in order to compare with observations. Therefore, the combination of large-volume 
cosmological $N$-body simulations with prescriptions to associate galaxies 
with dark matter haloes turns out to be the most efficient way to generate 
the required model galaxy samples. Recently, \citet{White2011} 
presented clustering results for scales in the range $\sim0.5$--$20\,\mpch$ based on $\sim44,000$ 
galaxies in the redshift range $0.4<z<0.7$ obtained during the first semester of BOSS operation. 
To compare these observational results with theory, the authors combined 
large, albeit low-resolution, $N$-body simulations with the Halo Occupation Distribution (HOD) 
model \citep[e.g.,][]{Berlind02,KravtsovHOD04,Zentner2005,Skibba2009,RossBrunner2009,RPB2010}. 
Their results suggest that the majority of 
BOSS galaxies are central systems living in haloes with a mass of
$\sim10^{13}~h^{-1}$ M$_{\sun}$, while about 10\% of them are
satellites typically residing in haloes $\sim10$ times more massive.

The HOD approach is the most often used framework to make predictions for 
the large-scale distribution of galaxies. Alternatively, HODs can also be measured in observations
\citep{Zehavi05,Abazajian2005,Brown08,Zheng09}. The
main component of classical HOD models is the probability, $P(N|M)$, that a 
halo of virial mass $M$ hosts $N$ galaxies with some specified properties. 
In general, theoretical HODs require the fitting of a function with several parameters
\citep[e.g.,][]{KravtsovHOD04,Zheng2005}, which gives some freedom to
match the observed clustering of galaxies. These models also depend on
the theoretical approach adopted to predict the galaxy number $N$ inside
haloes of mass $M$. 
For example, \citet{Zheng2005} used SPH 
simulations and semi-analytical models to measure the number of
galaxies as a function of hosting halo mass, which is definitely a
challenging theoretical exercise. 
\citet{White2011} tuned five HOD free parameters 
to fit the observed clustering of galaxies. In this case a random
fraction of dark matter particles is selected from the simulations with
a fraction following the optimized HOD. This prescription will 
have the best match to observations hence producing good-quality
mock catalogues. 
However, different choices of the underlying cosmology might lead 
to different HOD parameters.
\citet{KravtsovHOD04} use a different approach: they identify subhaloes 
in high-resolution $N$-body simulations in order to associate 
them with satellite galaxies. This is an 
attractive path, which can be further perfected by more accurate
simulations and more elaborate prescriptions for ``galaxies'' in dark
matter-only simulations \citep[e.g.,][]{Trujillo-Gomez}.

Halo Abundance Matching (HAM) has recently emerged as an 
alternative to HOD in order to bridge the gap between dark matter haloes and
galaxies \citep{KravtsovHOD04,Tasitsiomi04,Vale04,conroy06,Kim08,
Guo10,Wetzel10,Trujillo-Gomez,Reddick2012}. Abundance-matching resolves the issue of
connecting observed galaxies to simulated dark matter 
haloes and subhaloes by setting a one-to-one correspondence between the red-band 
luminosity or stellar and dynamical masses: more luminous galaxies are assigned
to more massive (sub)haloes (in what follows we will 
write ``(sub)haloes'' to account for all haloes and subhaloes). 
By construction, the method reproduces the observed
luminosity function (or stellar mass function). It also reproduces the
scale dependence of galaxy clustering over a large range of epochs
\citep{conroy06,Guo10}. When abundance matching is used for the
observed stellar mass function \citep{li09}, it gives also a reasonable 
fit to lensing results \citep{Mandelbaum06} and to 
the relation between stellar and virial mass
\citep{Guo10}. \citet{Guo10} also attempted to reproduce the observed
relation between the stellar mass and the maximum circular velocity with
partial success finding deviations both in shape and 
amplitude between model and observations. For instance, at circular velocities 
in the range $100$--$150$~km~s$^{-1}$ the 
estimated circular velocity was $\sim 25$\% lower than the observed
one. They proposed that this disagreement is likely due to the fact that
they did not include the effect of baryons. Indeed,
\citet{Trujillo-Gomez} show that accounting for baryons drastically
improves the situation.

Just like as with HODs, there are different flavours of HAMs. Generally, one does 
not expect a pure monotonic relation between stellar and dynamical
masses. There should be some degree of stochasticity in this relation
due to deviations in the merger history, angular momentum, and
halo concentration. Even for haloes (or subhaloes) with the same mass, these 
properties should be different for different systems, which would lead to
deviations in stellar mass. Observational errors are also responsible in part 
for the non-monotonic relation between halo and stellar masses.
Most of modern HAM models already implement prescriptions to account
for the stochasticity
\citep{Tasitsiomi04,Behroozi10,Trujillo-Gomez,Leauthaud11}.  The
difference between monotonic and stochastic models depends on the
magnitude of the scatter and on the stellar mass. The typical value of
the scatter in the $r$-band is expected to be $\Delta M_r =0.3$--$0.5$ mag
\citep[e.g.,][]{Trujillo-Gomez}. For the Milky-Way-size galaxies the
differences are practically negligible \citep{Behroozi10}, but they
increase for very massive galaxies such as those targeted with BOSS due 
to the strong dependence of the bias with mass.

\begin{figure}
      \includegraphics[width=91mm]{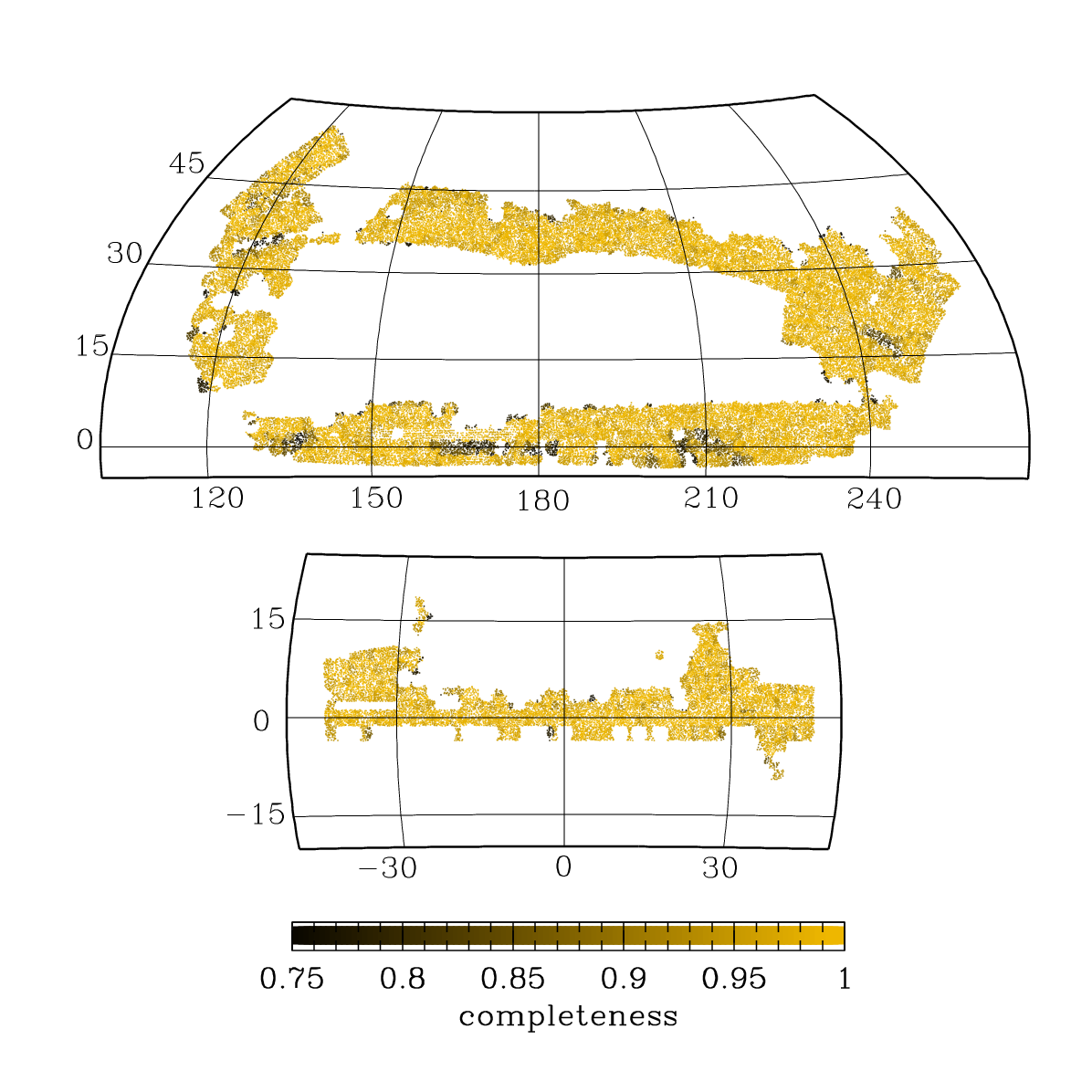}
      \caption{Sky area covered by the DR9 BOSS-CMASS sample
        shown in Aitoff projection colour-coded by completeness (see text). 
        The upper and lower maps display the northern and southern galactic 
        caps respectively. 
      }
\label{fig:map}
\end{figure}

Almost two years after the start of the project, BOSS has obtained the
spectra of about 536,000 galaxies and 102,000 quasars \citep[][]{Ahn2012}. 
Using the SDSS-III Data Release 9 (DR9) BOSS data 
we present results on the two-dimensional, projected and redshift-space correlation 
functions on scales from $\sim500\,\kpch$ to $\sim90\,\Mpch$ including fibre collision 
corrections. In order to account for the $\Lambda$CDM cosmological model
we use a large high-resolution $N$-body simulation with a resolution 
high enough to resolve subhaloes, which is very important for the 
HAM prescription. When connecting haloes with galaxies we use a
stochastic HAM model.

This paper is organized as follows. In Section~\ref{sec:cmass} we
present the BOSS galaxy sample studied here, dubbed ``CMASS'', and the
measurements of the two-dimensional, projected and redshift-space
galaxy clustering in observations. In Section~\ref{MultiDark_clustering} we
present the details of the MultiDark simulation, the halo catalogues 
and the HAM technique adopted here. 
In Sections~\ref{results} and~\ref{sec:satellites} we compare the 
clustering measures with observations and study the occupation distribution 
given by our halo catalogue. We also discuss the comparison between our halo 
occupation distribution with that obtained by \cite{White2011} using an HOD model. 
In Section~\ref{bias} we study the scale-dependent bias of galaxy
clustering of the CMASS sample as inferred from our HAM model and 
MultiDark simulation both in real and Fourier space. 
Finally, in Section~\ref{conclusions} we close the paper with the summary 
and conclusions. Throughout this paper we assume MultiDark cosmological parameters 
(see Section~\ref{MultiDark}).

In Appendix~\ref{app:a} we discuss several effects that 
can affect the clustering power, whereas in Appendix~\ref{app:b} we 
present tables and correlation matrices for the observed correlation 
functions. We also include our HAM results in the correlation function tables.

\section{Observations}
\label{sec:cmass}

\subsection{The CMASS sample}
\label{sec:cmass_sample}

\begin{figure}
  \includegraphics[width=0.45\textwidth]{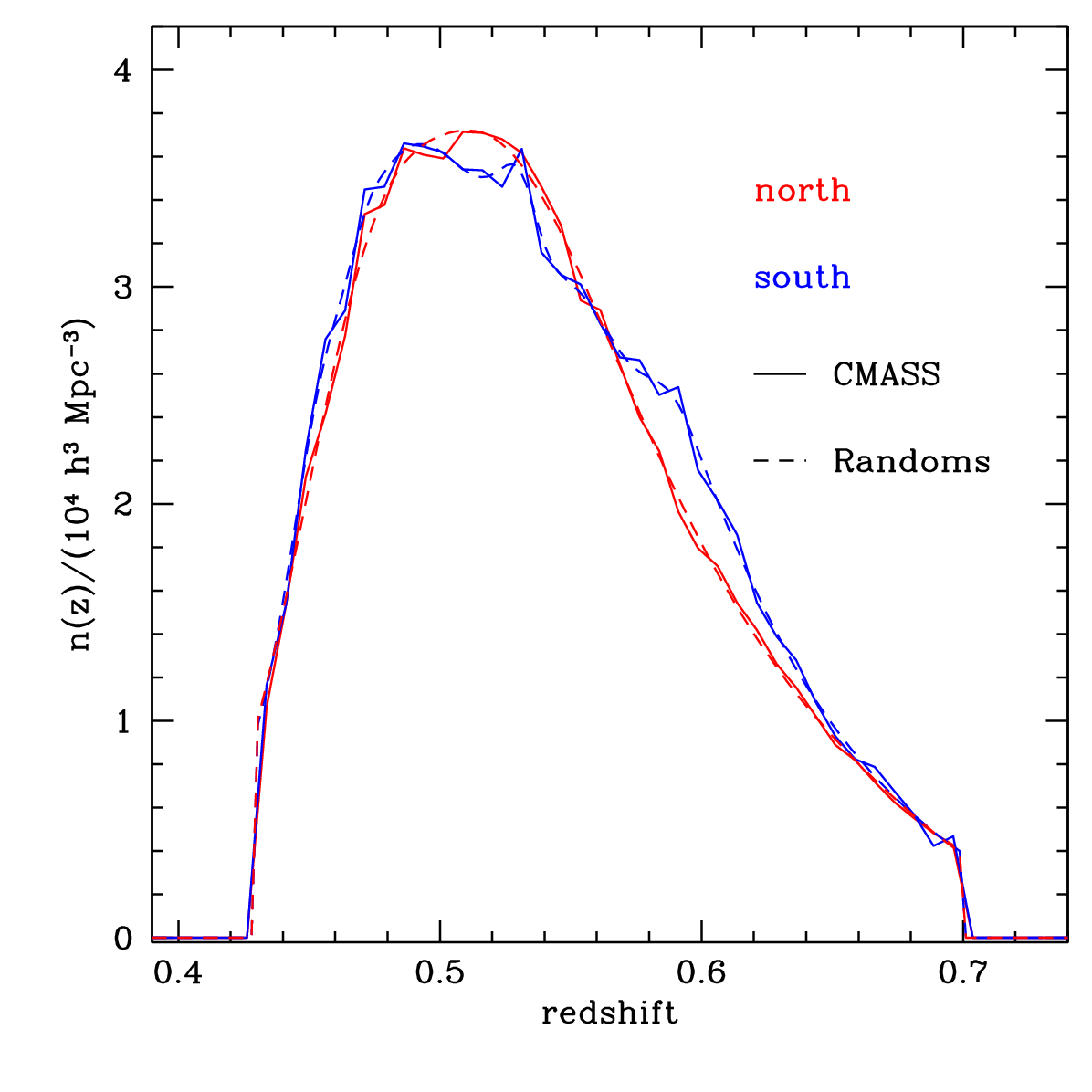}
  \caption{The comoving number density of galaxies in the DR9
    BOSS-CMASS sample both for the north and south subsamples in the
    redshift range $0.4<z<0.7$. Dashed lines show the smoothed
    distributions used to create the Poisson distribution of particles
    when computing the correlation functions (see text).  }
\label{fig:CMASS}
\end{figure}

In this section we introduce the BOSS sample of massive galaxies
analyzed in this work. The target galaxies are selected in such a way
that the stellar mass of the systems is approximately constant over
the entire redshift range of interest. As a consequence, the resulting
galaxy sample is usually dubbed `constant mass' (CMASS) sample. These
galaxies are characterized by high-luminosities which translate in a
rather low comoving space density of about $3\times10^{-4}\,\Mpch$.
The sample can be obtained by applying the following colour 
cuts to the observations \cite[see e.g.][]{Eisenstein2011,Anderson2012}:

\begin{eqnarray}
 17.5 \,\,\,\, < \,\,\,\, i_{\rm cmod} &<& 19.9{\rm,}                    \nonumber \\
 r_{\rm mod}-i_{\rm mod}            &<& 2{\rm,}                                   \nonumber \\
 i_{\rm fiber2} &<& 21.5{\rm,}                                \nonumber \\
 d_{\perp}      &>& 0.55{\rm,}                                \nonumber \\
 i_{\rm cmod}              &<& 19.86+1.6\times\left(d_{\perp}-0.8\right) 
\end{eqnarray}

\noindent where $d_{\perp} = \left(r_{\rm mod}-i_{\rm mod}\right) -
\left(g_{\rm mod}-r_{\rm mod}\right)/8$ and $i_{\rm fiber2}$ is the
$i$ magnitude measured with the 2$''$ BOSS fiber within the SDSS
$ugriz$ photometric system \citep[][]{1996AJ....111.1748F}. The
subscripts $_{\rm cmod}$ and $_{\rm mod}$ denote ``cmodel'' and ``model''
magnitudes respectively. These cuts are chosen to pick out massive red
galaxies at $z\gtrsim0.4$. In particular, the condition
$d_{\perp}>0.55$ selects systems with observed red $r-i$ colours,
whereas the conditions imposed on the $i$-magnitude is designed to
identify an approximately complete galaxy sample down to a limiting
stellar mass. Most of these galaxies ($\sim75\%$) show an early-type morphology
with a characteristic stellar mass of $M_*\sim10^{11}$ $h^{-1}$
M$_{\sun}$ and an absolute $r$-band magnitude 
of $M_{r}-5\log h\lesssim-20.7$ \citep[][]{Masters2011}.

\citet{Schlafly2010} and \citet{Schlafly2011} found systematic offsets
between the colours of SDSS objects in the southern and northern
Galactic hemispheres which might reflect a combination of percent
calibration errors in the SDSS photometry and errors in the
corrections for Galactic extinction. The \citet{Schlafly2011} results
suggest a systematic offset in the value of $d_{\perp}$ of 0.0064
between the north and south. As the CMASS selection criteria depends
on $d_{\perp}$, this offset leads, in principle, to a difference in the galaxy
samples selected for spectroscopic observations in the two
hemispheres. \citet{Ross2011} found a 2\% difference in the number density of 
CMASS targets between the northern and southern hemispheres, which 
reduces to 0.3\% when this offset is applied to the 
galaxies in the south before applying the CMASS selection criteria. 
However, \citet{Ho2012} found no appreciable north/south colour offsets in their sample. 
In this work we do not apply a colour offset to the selection 
of CMASS galaxies in the south. Although we present results obtained from
the combined (north$+$south) CMASS sample, we also analyse the data from the northern and
southern hemispheres separately in order to avoid potential systematics
that could be associated with the use of slightly different selection
criteria.

For a number of reasons it is not possible to obtain reliable redshifts
for all the galaxies satisfying the CMASS selection criteria (see
Section \ref{sec:clustering}).  We estimate the completeness
$c=n_z/n_{\rm t}$, where $n_{t}$ is the number of galaxy targets and
$n_z$ the number of these with reliable redshift estimates (weighted
as described in Section \ref{sec:clustering}) for each sector of the
survey mask, that is, the areas of the sky covered by a unique set of
spectroscopic tiles \citep[][]{Blanton2003,Tegmark2004} which
we characterize using the {\sc Mangle} software 
\citep[][]{HamiltonTegmark2004, Swanson2008}. The average completeness of 
the combined CMASS sample is 98.2\%. We trim the final area of our sample to all
sectors with completeness $c \geq 0.75$, producing our final sample of 
282,068 galaxies, of which 219,773 and 62,295 are located in the northern
and southern galactic caps respectively. Fig.~\ref{fig:map} shows an
Aitoff projection of the resulting survey mask in the northern (upper
panel) and southern (lower panel) regions, with effective areas
$A_{\rm eff}=\sum_i c_i \Omega_i$, where the sum extends over all
sectors contained in the mask and $\Omega_i$ corresponds to their
solid angles, 2502 deg$^2$ and 688 deg$^2$ respectively. The redshift
distribution of the CMASS sample can be seen in Fig.~\ref{fig:CMASS}
both for the north and south subsamples. The dashed lines show
the smoothed distributions used to create the random samples of points
for our clustering analysis (see Section~\ref{sec:clustering}). As
shown in this figure the galaxy number density peaks at $z\simeq0.52$
having a value of $\bar{n}_{\rm g}\simeq3.6\times10^{-4}\,h^3\,\Mpc^{-3}$ 
and a mean redshift of $\bar{z}=0.55$.

\subsection{Clustering measures}
\label{sec:clustering}

We characterize the clustering of the CMASS galaxy sample by means of
two-point statistics in configuration space. We measure the
angle-averaged redshift-space correlation function $\xi(s)$ and the
full two-dimensional $\xi(\sigma,\pi)$, where $\sigma$ and $\pi$ are
the components in the direction perpendicular and parallel to the line
of sight of the total separation vector ${\bf s}$. These measurements
are affected by redshift-space distortions. 
In order to obtain a clustering measure that is less sensitive to these effects 
we also compute the projected correlation function \citep[][]{DavisPeebles83}
\begin{equation}
 \Xi(\sigma)=2\int_0^{\infty}\xi(\sigma,\pi)\,{\rm d}\pi.
 \label{int_xi}
\end{equation}
In practice, we sum all pairs with $\pi_{\rm max}<200\,h^{-1}\,{\rm Mpc}$ 
using linearly-spaced bins. We have checked that the projected correlation has already converged 
for $\pi_{\rm max}\approx100\,h^{-1}\,{\rm Mpc}$.

We compute the full correlation functions $\xi(\sigma,\pi)$
using the \citet{LandySzalay93} estimator
\begin{equation}
\xi(\sigma,\pi) = \frac{DD- 2 DR + RR}{RR}
\end{equation}
where $DD$, $DR$, and $RR$ are the suitably normalized numbers of
data-data, data-random, and random-random pair counts in each bin of
$(\sigma,\pi)$.
In order to measure these quantities without introducing systematic effects,
a few important corrections must be taken into account. Here we give a brief 
description of the main issues that should be considered while a more detailed 
discussion will be presented in \citet{Ross2012}.

As described in the previous section, the spectroscopic CMASS sample
is constructed from a target list drawn from the SDSS photometric
observations. Even though the overall completeness of the CMASS sample
is high, it is not possible to obtain reliable redshifts for all
galaxies satisfying the selection criteria specified in
Section~\ref{sec:cmass_sample}. Which galaxies are observed
spectroscopically is determined by an adaptive tiling algorithm, based
on that of \citet{Blanton2003}, which attempts to maximize the number
of measured spectra over the survey area. As a result of this
algorithm, not all galaxies satisfying the CMASS criteria are selected
as targets for spectroscopy. Even when a fibre is assigned to a galaxy
and a spectrum is observed, it might not be possible to obtain a
reliable estimation of the redshift of the object, leading to what is
called a {\it redshift failure}. These tend to occur for fibres
located near the edges of the observed plates. This implies that it 
is not possible to simply consider these redshift failures as an extra
component affecting the overall completeness of the sector since their
probability is not uniform across the field.
In order to correct for this effect we define a set of weights, $w_{\rm zf}$, 
whose default value is one for all galaxies in the sample. 
For every galaxy with a redshift failure, we increase by one the value of
$w_{\rm zf}$ of the nearest galaxy with a good redshift measurement. 
The application of these weights effectively corrects for
the non-uniformity effects produced by redshift failures.

The main cause for the loss of objects is, however, fibre collisions
\citep[][]{Zehavi2002, Masjedi2006}. The BOSS spectrographs are fed by
optical fibres plugged on plates, which must be separated by at least
$62''$ (in the concordance cosmology this corresponds to a distance of $\sim0.27\,\Mpch$ 
at $z\sim0.5$). It is then impossible, in any given observation, to obtain
spectra of all galaxies with neighbours closer than this angular
distance. The problem is alleviated in regions covered by multiple
exposures, but it is in general not possible to observe all objects in
crowded regions. 

In this work we correct for the impact of fibre collisions on our clustering measurements
by applying the correction presented in \citet{Guo2011}. Using this method the total galaxy sample $D$
is divided into two subsamples, dubbed as $D_1$ and $D_2$. These are constructed following the targeting
algorithm of the catalogue in a way that guarantees that group $D_1$ is not affected by fibre collisions,
while $D_2$ contains all collided galaxies.
Any clustering measurement of the combined sample can be obtained as a combination of the contributions
from these two groups. Based on tests on mock galaxy catalogues, \citet{Guo2011} showed that the
application of this method can accurately recover the projected and redshift-space correlation functions on
scales both below and above the fibre collision scale, providing a substantial improvement over the commonly
used nearest neighbour and angular correction methods.

We constructed random catalogues for subsamples $D_1$ and $D_2$ for the northern and southern
hemispheres with 40 times more objects than the real data 
following their respective angular completenesses. The redshifts of
these random points were generated in order to follow the 
distributions of the real samples, which were obtained by a smoothing 
spline interpolation of the observed redshift distributions 
(see dashed lines in Fig~\ref{fig:CMASS}).

With the increasing size of current galaxy surveys, and the corresponding
improvement on the statistical uncertainties, the contribution of systematic errors 
to the total error budget of any clustering statistic becomes increasingly important.
Due to its large volume and high number density BOSS is perhaps one of the best 
examples of this. \citet{Ross2012} 
present a detailed analysis of the systematic effects that could potentially
affect any clustering measurement based on the CMASS sample and show that, 
besides redshift failures and fibre collisions, other important
systematics must be considered in order to obtain accurate clustering measurements.
The main result from this analysis is that these systematics can be corrected for
by applying a set of weights, $w_{\rm sys}$, which depend on both, the galaxy properties and their
positions in the sky. We consider these weights in all our clustering measurements.

Finally, we also include a set of weights to reduce the variance of the
estimator that are given by
\begin{equation}
w = (1 + n(z)J_{w})^{-1}, 
\end{equation}
where $n(z)$ is the mean galaxy density at redshift $z$ and $J_w$ is a free parameter.
\citet{Hamilton93} showed that setting $J_{w}=4\pi J_3(s)$,
where $J_3(s)=\int_0^s\xi(s')s'^2{\rm d}s'$, minimizes the
variance on the measured correlation function for the given scale $s$. 
Here we follow the standard practice and use a scale-independent value of $J_w=2\times 10^4$.

\begin{figure}
  \includegraphics[width=0.50\textwidth]{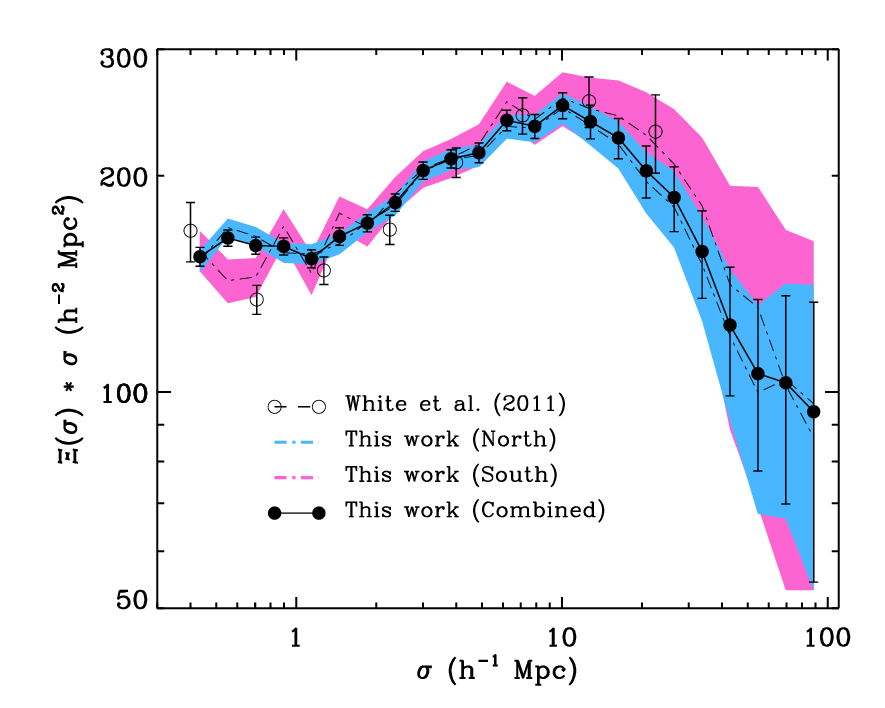}
  \caption{Projected correlation function times the projected distance
    for the DR9 BOSS-CMASS galaxy sample in the redshift range
    $0.43<z<0.7$. The blue and red shaded areas correspond to 
    the north and south subsamples and give an estimate of their standard 
    deviation. The dot-dashed lines display their mean value. 
    The result of combining both subsamples is shown as filled circles. 
    Standard deviation for the projected correlations of all samples are 
    estimated using an ensemble of 600 mock catalogues (see Section~\ref{sec:clustering}).
    For comparison the projected correlation inferred from the first 
    semester of the BOSS-CMASS data is also shown \citep[open circles;][]{White2011}.
  }
\label{fig:proj_CMASS_comp}
\end{figure}

Fig.~\ref{fig:proj_CMASS_comp} shows the projected correlation functions $\Xi(\sigma)$ 
times the projected distance of the north, south and combined CMASS samples. 
The combined sample gives a similar outcome to that of the north as a result of the higher 
statistics in the latter.
For comparison the projected correlation inferred from a CMASS sample 
corresponding to the first semester of the BOSS observations is also 
shown \citep[open circles;][]{White2011}. Besides the increase in the sample
size and the volume probed, there are differences at small and large
scales which are probably due to the different corrections for fibre
collisions and the use of the weights to correct for the systematics
affecting the galaxy density field.

Although the projected correlation functions of the north and
south subsamples agree within their respective uncertainties, they
show some intriguing differences. At scales in the range $\sim20$--$50\,\Mpch$ 
the amplitude of $\Xi(\sigma)$ in the south is higher than that of the
north. Similarly, the measurements of $\xi(s)$ show the same behaviour. 
However, in this case, the agreement of the mean values is somewhat 
closer. In Section~\ref{results} we present further results on these 
clustering measures for the north, south and combined BOSS-CMASS galaxy 
samples separately.

\subsubsection{Estimation of covariances in the data}
\label{cov_data}

To estimate covariance matrices for these clustering measures, we use 
a set of 600 mock catalogues designed to follow the same geometry and 
redshift distribution of the CMASS sample while mimicking their 
clustering properties at large scales \citep{Manera2013}. 
These mocks are inspired by the {\it PTHalos} method of \citet{Scoccimarro2002} which 
is aimed at computing the evolution of structure using Lagrangian perturbation theory 
including several prescriptions to account for haloes at smaller scales. However, we note 
that some important differences exist between both treatments.  
The resulting covariances are compatible with the results of $N$-body 
simulations (see Section~\ref{variance_MD}) while displaying 
convergence at a few percent-level for the range of scales studied here. 
For a detailed description about these mocks and their comparison with $N$-body results 
see \citet{Manera2013}\footnote{Mocks will be available in http://www.marcmanera.net/mocks/}.
In Appendix~\ref{app:b} we provide tables for the estimated correlation 
matrices of the projected and redshift-space correlation functions for the 
north, south and combined BOSS-CMASS samples.

\section{Clustering in the $\Lambda$CDM model}
\label{MultiDark_clustering}

\subsection{The MultiDark simulation}
\label{MultiDark}

The MultiDark run (MDR1) is an $N$-body cosmological simulation of the
$\Lambda$CDM model that was done using the Adaptive-Refinement-Tree
(ART) code \citep{ART1997,ART2008}. The simulation has
2048$^{3}\approx 8.6\times 10^9$ dark matter particles in a box of 
1\,$h^{-1}$\,Gpc on a side. The mass of the dark matter particle
is $8.72\times10^9$ $h^{-1}$ M$_{\sun}$. The cosmological parameters
adopted in the simulation are consistent with the latest WMAP7 results
\citep{2011ApJS..192...14J} and with other cosmological probes
\citep[see Table 1 of][]{Bolshoi}. Hence, we adopt a matter 
density parameter $\Omega_{\rm M}=0.27$ and a dimensionless Hubble 
parameter $h=0.7$. Initial conditions were set at the
redshift $z_{\rm init}=65$ using a power spectrum characterized by a scalar
spectral index $n_s=0.95$ and normalized to $\sigma_8=0.82$ in the same way as 
done for the Bolshoi simulation \citep[see][ for a detailed description of this
simulation]{Bolshoi}. Since the adopted cosmological parameters 
are very close to the latest observational estimations any departure from the 
true cosmology will not affect our main results.
The ART code is designed in such a way that the
physical resolution is nearly preserved over time with a value of
$\sim7\,h^{-1}$\,kpc for the redshift range between $z=0$--$8$.  For
further details on the ART code and MultiDark simulation 
see \citet{Prada2012} and references therein.

\subsubsection{Halo finding}

Dark matter haloes are identified in the simulation with a parallel
version of the Bound-Density-Maxima (BDM) algorithm
\citep{1997astro.ph.12217K,Riebe11}. The BDM is a Spherical
Overdensity (SO) code.  It finds all density maxima in the
distribution of particles using a top-hat filter with 20
particles. For each maximum the code estimates the radius within which
the overdensity has a specified value. Among all overlapping density
maxima the code finds the one having the deepest gravitational
potential. The position of this maximum is the centre of a
``distinct'' halo, which is a halo whose centre is not inside the virial radius
of a bigger one. Distinct haloes are also tracers of central galaxies.
Self-bound haloes with more than 20 particles lying inside the virial
radius of a distinct halo are classified as subhaloes. Subhalo
identification is more subtle since it requires the removal of unbound
particles and identification of fake satellites 
\citep[see][ for a detailed description of this method]{Riebe11}. 
The BDM algorithm was extensively tested and compared with other halo finders
\citep[][]{Knebe11,RockStar}. In Appendix~\ref{app:aa} we show a comparison 
between the real-space correlation function for halo catalogues selected 
both with BDM and RockStar halo finders (see Fig~\ref{app:A1}). 
The BDM halo catalogues for the MDR1 simulation 
are publicly available at the MultiDark Database: http://www.multidark.org.

The size of a distinct halo can be defined by means of the spherical 
radius within which the average density is $\Delta$ times higher than the
critical density of the Universe, $\rho_{\rm cr}(z)$. As a
consequence, the corresponding enclosed mass is given by

\begin{equation}
  M_{\Delta} =\frac{4\pi}{3}\Delta\rho_{\rm cr}(z)R_{\Delta}^3. 
  \label{eq:Delta} 
\end{equation}

\begin{figure}
      \includegraphics[width=85mm]{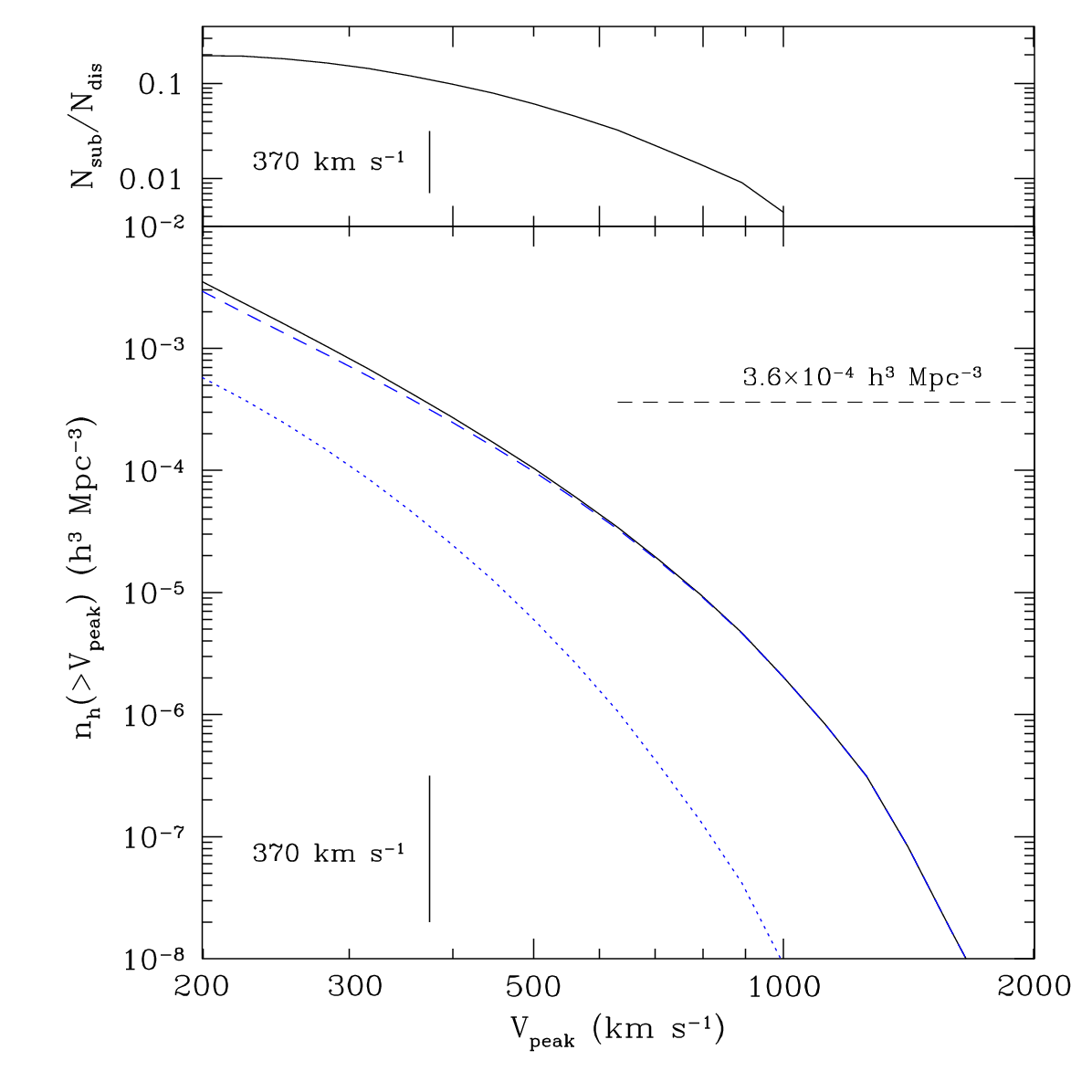}
      \caption{{\it Bottom panel:} The cumulative number density of
        distinct haloes (dashed line) and subhaloes (dotted line) in
        the MultiDark simulation at $z=0.53$ as a function of maximum
        circular velocity. The cumulative number for all haloes is
        also shown as a solid line. {\it Top panel:} The cumulative
        subhalo fraction as a function of halo maximum circular
        velocity. As a reference we indicate in both panels the mean number density
        of the BOSS-CMASS galaxy sample and as vertical lines the
        corresponding maximum circular velocity threshold ($V_{\rm cut}$) 
        used in the HAM procedure.}
\label{fig:cum_sat}
\end{figure}

\noindent We  use a threshold overdensity of
$\Delta=200$ that results in values for halo mass and radius of
$\M200$ and $\R200$ respectively. In addition, BDM catalogues also provide virial 
masses and radius ($M_{\rm vir}$ and $R_{\rm vir}$) defined using the standard
overdensity $360 \, \rho_{\rm back}(z)$, where $\rho_{\rm back}(z)$ is 
the background mean density of the Universe.

One of the most important characteristics of a (sub)halo is its
maximum circular velocity at redshift $z$:
  \begin{equation}
    V^2_{\rm max}(z)=\max\left[\frac{GM(<r,z)}{r}\right].
\end{equation}
There are several advantages of using $\Vmax$ at a given time to 
characterize the dynamical mass of a halo as opposed to the ``virial mass''. 
First, $\Vmax$ does not have the ambiguity related with the definition of mass. 
Virial mass and radius vary depending on the overdensity threshold used. 
For the oftenly employed overdensity 200 and ``virial'' overdensity thresholds, 
the differences in definitions result in changes in the halo radius from one
definition to another and, thus, in concentration, by a factor of
$1.2$--$1.3$, where the exact value depends on the halo concentration. 
Second, and more important, the maximum circular velocity $\Vmax$ is a 
better quantity to characterize haloes whenever is needed to relate 
them to their associated galaxies. For instance, for galaxy-size haloes the 
maximum circular velocity is defined at a typical radius of $\sim40$~kpc, i.e., 
much closer to the sizes of luminous parts of galaxies than the 
virial radius, which for the Milky-Way halo 
is of the order of $\sim 250~\kpc$ \citep{Klypin2002}.

\subsection{Bridging the gap between galaxies and haloes}
\label{HAM}

To select a simulated halo sample representing BOSS-CMASS galaxies 
we apply the HAM technique.
Once we have the maximum circular velocities for distinct haloes and
subhaloes the implementation of the HAM prescription is simple: we start with 
a monotonic assignment. We count all haloes and subhaloes, which have
maximum circular velocity $V_{\rm max}$ larger than the threshold 
$V_{\rm cut}$, and gradually decrease the value of $V_{\rm cut}$ 
until the number density of (sub)haloes is equal to that of BOSS 
galaxies at $z\approx 0.5$. 

A usual choice for $\Vmax$ is to 
take the halo maximum circular velocity at $z=0$ as a measure of  
the dynamical mass of the system since this is a quantity that can 
be easily obtained from simulations. 
However, it is generally accepted that, for subhaloes, a better characteristic would 
be the peak value of the maximum circular velocity, $V_{\rm peak}$, during the 
entire subhalo evolution \citep[e.g.,][]{conroy06,Trujillo-Gomez}. 
The latter is related to the tidal stripping effect: once a halo falls into the 
potential well of a larger one some of its material can be stripped away, thus 
lowering the value of $V_{\rm max}$. Since in real galaxies stars occupy the inner 
regions of subhaloes, where tidal forces are much weaker, their circular velocities 
should be, in general, less influenced by this effect. For instance, \citet{Watson2012}, 
based on a subhalo evolution model applied to clustering measurements in the SDSS, suggest 
that tidal stripping of stars in luminous galaxies (as those presumably expected in the 
BOSS-CMASS sample) is much less efficient than in less luminous systems. 
Nevertheless, when it comes to matching (sub)haloes to the galaxy abundance 
one should use the best theoretically motivated parameter, which in this case is $V_{\rm peak}$. 
Hence, in what follows, we will adopt this quantity as a measure 
of the dynamical mass of (sub)haloes in the simulation.

The bottom panel of Fig.~\ref{fig:cum_sat} shows the number
density of (sub)haloes at $z=0.53$ in the MultiDark simulation 
as a function of $\Vpeak$. A number density close to that
of the BOSS-CMASS sample corresponds to (sub)haloes with a 
peak maximum circular velocity above 370\,\kms, which is sufficiently 
larger than the completeness limit of the  MultiDark simulation, i.e., $\sim180\,\kms$. 
This means that (sub)haloes hosting BOSS-CMASS galaxies are well resolved. 
The top panel of Fig.~\ref{fig:cum_sat} shows the cumulative
subhalo fraction as a function of maximum circular velocity. 
For values of $V_{\rm max}>400\,\kms$ the subhalo fractions are typically less than $10\%$. 
We will return to this point again in Section \ref{sec:satellites}.

\subsubsection{Halo stochasticity}

There are a number of arguments why there should be some degree of
stochasticity in the stellar mass -- circular velocity relation
\citep[e.g.,][]{Tasitsiomi04,Behroozi10,Trujillo-Gomez}.  In our case
the stochasticity means that some haloes above the velocity cut host
galaxies with stellar masses smaller than the corresponding stellar mass
cut of the BOSS sample and should not be included into the sample. Simultaneously, 
some smaller haloes may host galaxies with a larger stellar
mass, and should be considered. Because the number density of galaxies is
fixed by observations, the numbers of included and excluded haloes must be
equal.  Following \citet{Trujillo-Gomez} we implement this process
using a Gaussian spread with an offset. If $V_{\rm cut}$ is the
velocity cut in the monotonic assignment, then a (sub)halo is taken if
its peak maximum circular velocity $\Vpeak$ satisfies the condition

\begin{equation}
\Vpeak\left[1+ \mathcal{N}(0,\sigma)\right]-\Delta V > V_{\rm cut},
\end{equation}

\noindent where $\mathcal{N}(0,\sigma)$ is a Gaussian random number with mean zero and 
{\it rms} $\sigma$. The offset $\Delta V$ is needed to compensate the larger
influx of smaller haloes. We use $\sigma =0.3$ and $\Delta V
=40\,\kms$, which are consistent with the values adopted by \citet{Trujillo-Gomez}.
Note that the offset $\Delta V$ and the spread $\sigma$ are not free
parameters. The offset is just a normalization. The value of $\sigma$ is
defined by the spread of the observational Baryonic Tully-Fisher
relation (or its equivalent for early-type galaxies), which has
 uncertainties \citep[e.g.,][]{Trujillo-Gomez}. The stochastic
assignment has a very small effect on clustering for scales larger than
$0.5\;\Mpch$ decreasing the correlation functions no more than $\sim 8\%$.

\begin{figure}
      \includegraphics[width=85mm]{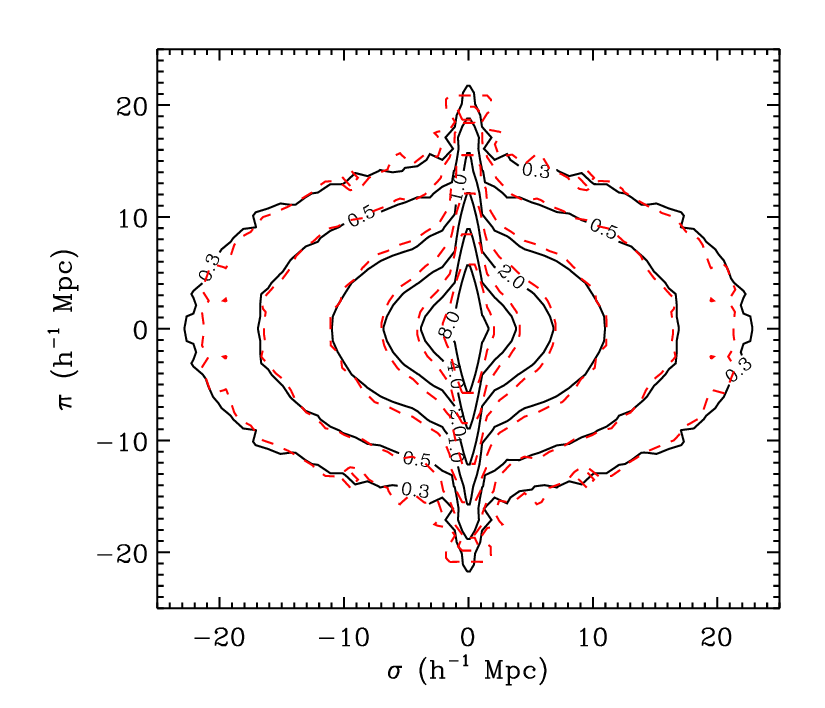}
      \caption{
      Contours of the two-dimensional correlation function
      $\xi(\sigma,\pi)$ estimated from the DR9 BOSS-CMASS north
      galaxy sample (dashed contours) at $0.4<z<0.7$ and for our MultiDark
      halo catalogue constructed using the HAM technique at $z=0.53$
      (solid contours).
             }
\label{fig:sig_pi_MD}
\end{figure}

\subsection{Modelling BOSS-CMASS clustering}
\label{Modelling}

We use the MultiDark BDM catalogues constructed for the overdensity $360
\, \rho_{\rm back}(z)$ to facilitate the comparison with the HOD
modelling presented in \citet{White2011}. However, as stated before, our results do not
depend on halo mass definition since halo matching is done using the peak maximum 
circular velocity of either distinct haloes or subhaloes. 
For the HAM to the BOSS-CMASS sample we choose a redshift of $z=0.53$ 
and an effective number density of $n_* \equiv 3.6\times10^{-4}\,h^{3}\,\Mpc^{-3}$ 
(see Fig.~\ref{fig:CMASS}). 
However, our clustering results are mostly insensitive 
to small deviations around these fiducial values as it is shown in 
appendices~\ref{app:ab} and~\ref{app:ac} respectively.

To model the effect of galaxy peculiar velocities in the redshift
measurements, we transform the coordinates of our simulated (sub)haloes to
redshift-space using ${\bf s}={\bf x}$ + ${\bf v\cdot\hat{r}}/(aH)$,
where ${\bf x}$ and ${\bf v}$ are their position and peculiar velocity
vectors respectively, $a$ is the scale factor and $H$ is the Hubble constant. 
We compute the two-dimensional correlation function $\xi(\sigma,\pi)$ of 
our catalogue counting the number of ``galaxy'' tracers in bins parallel
($\pi$) and perpendicular ($\sigma$) to the line-of-sight. 
When estimating the projected correlation function, we count all pairs
along the parallel direction out to $\pi_{\rm max}=200\,\Mpch$ using 
linearly-spaced bins. 
We have checked that our model projected correlation function has already 
converged for $\pi_{\rm max}\approx100\,h^{-1}\,{\rm Mpc}$.

\begin{figure*}
      \includegraphics[width=86mm]{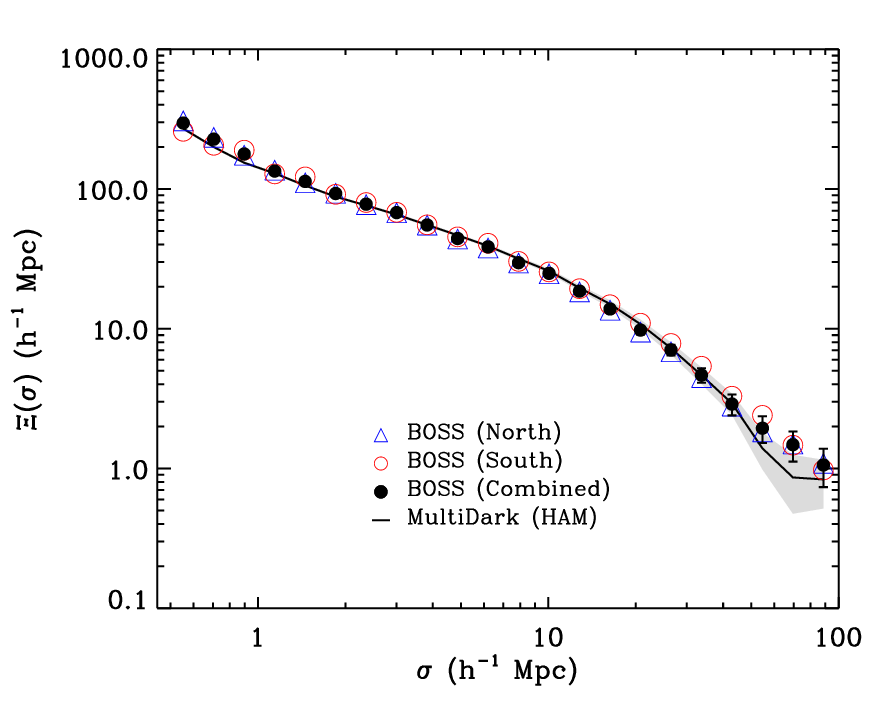}
      \includegraphics[width=86mm]{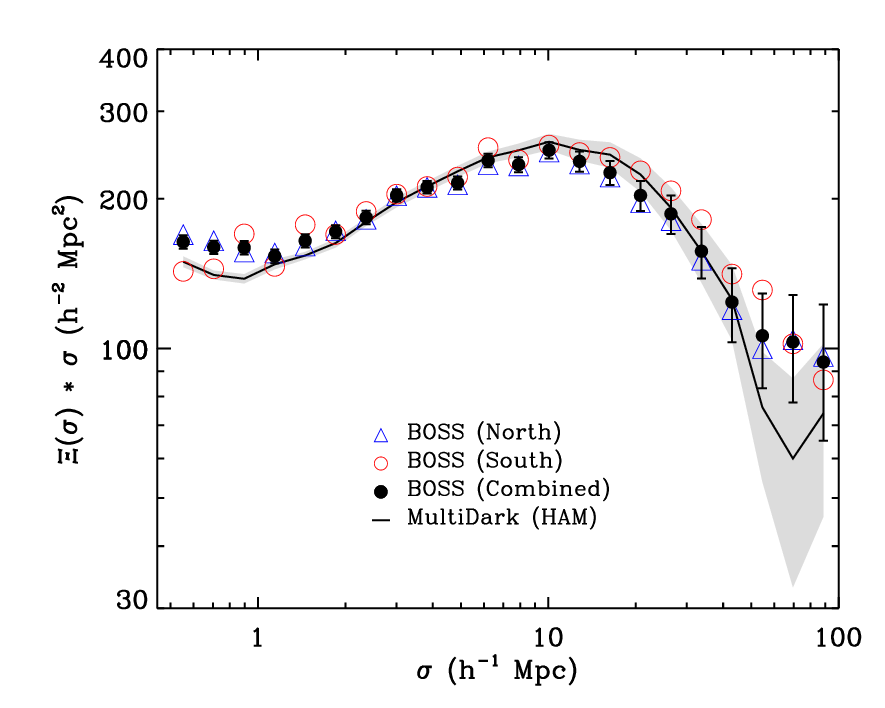}
      \caption{{\it Left panel:}
      Projected correlation function for the $0.4<z<0.7$ DR9
      BOSS-CMASS north, south and combined galaxy samples (open blue triangles, 
      open red circles and filled black circles respectively) and the MultiDark catalogue selected
      with the HAM procedure at $z=0.53$ (solid line). The shaded area for MultiDark gives an 
      estimate of the cosmic variance. BOSS-CMASS error bars were estimated using an
      ensemble of 600 mock galaxies (see Section~\ref{sec:clustering}). For clarity, only error bars for the 
      combined sample are shown. The corresponding ones for the north and south are a factor of about 1.13 and 2.15 
      times larger respectively. The transition between the one-halo and two-halo terms can be 
      seen at $\sim1\,\Mpch$. Flattening of the signal at intermediate scales and bending
      at large scales are also evident features. {\it Right panel:}
      Detailed differences between the $\Lambda$CDM model and BOSS
      clustering are better seen when plotting the quantity
      $\Xi(\sigma)\,\sigma$.}
\label{fig:wp_MD}
\end{figure*}

\begin{figure*}
      \includegraphics[width=88mm]{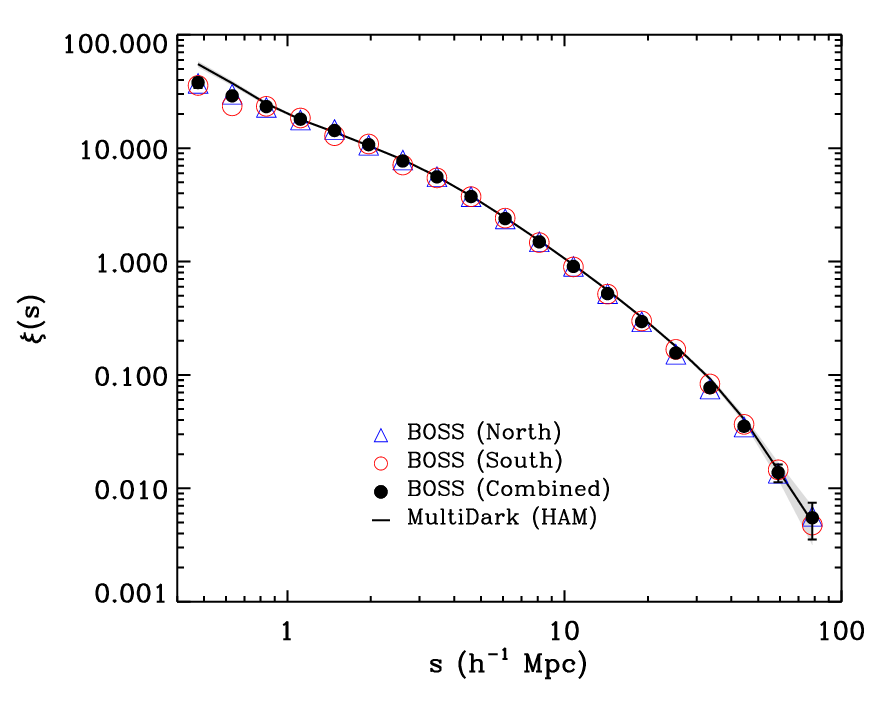}
      \includegraphics[width=88mm]{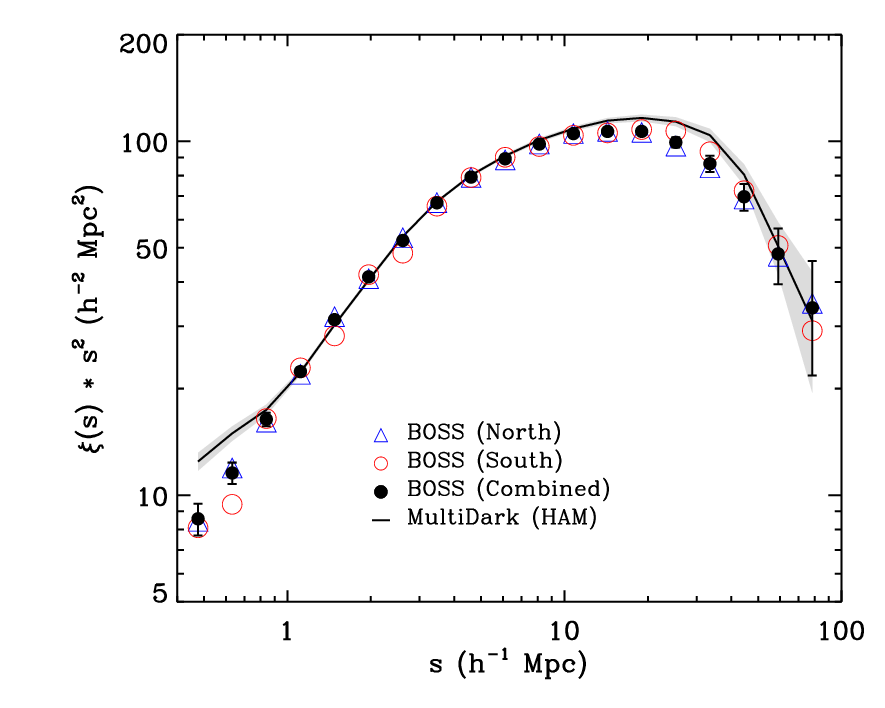}
      \caption{{\it Left panel:}
      Redshift-space correlation function for the $0.4<z<0.7$ DR9
      BOSS-CMASS north, south and combined galaxy samples (open blue triangles,
      open red circles and filled black circles respectively) and the MultiDark catalogue selected
      with the HAM procedure at $z=0.53$ (solid line). Standard deviation for model and observations 
      are shown in the same way as in Fig.~\ref{fig:wp_MD}. {\it Right panel:} Shown is the quantity
      $\xi(s)\,s^2$ which better reflects the differences between our $\Lambda$CDM model and 
      BOSS clustering measures.}
\label{fig:z_MD}
\end{figure*}

\subsubsection{Estimation of cosmic variance in MultiDark}
\label{variance_MD}

Our estimation of the (sub)halo clustering corresponding to the BOSS number density 
in a $\Lambda$CDM cosmology is limited by the finite volume of the MultiDark simulation. 
To estimate the expected level of cosmic variance we use a set of low-resolution 
simulations from the Large Suite of Dark Matter Simulations 
({\it LasDamas}; see http://lss.phy.vanderbilt.edu/lasdamas/). In particular, we use 
the {\it Carmen} boxes, which are 40 dark matter-only runs done with $1120^3$ particles 
in a periodic cube with $1\,h^{-1}$\,Gpc on a side. The dark matter density and 
scalar spectral index of the {\it Carmen} simulations display a difference of 
about $8\%$ in comparison to the corresponding values of MultiDark. 
However, since here we only want to obtain an estimate for the magnitude 
of the cosmic variance, we consider this approach as good enough for our purpose.

For all {\it Carmen} runs we used halo catalogues matched to the abundance 
of the BOSS-CMASS sample at $z\approx0.5$ as explained in Section~\ref{HAM}. This allowed us 
to get the magnitude of cosmic variance in the clustering signal.
In this way, we can get a simple estimate of the expected {\it rms} deviations of our 
MultiDark results due to random fluctuations in the 
intial conditions of the universe.

As mentioned in Section~\ref{cov_data}, the estimation of cosmic variance in 
the observed correlation functions is done using the covariance matrices of a set of 600 
galaxy mocks designed to follow the same geometry and redshift distribution of 
the CMASS sample, while mimicking its clustering properties at large scales. 
\cite{Manera2013} show that the covariances for the correlation functions of $N$-body 
simulations are consistent with those resulting from the mocks. However, it is 
important to keep in mind that the clustering measures of the mocks are not a good 
representation of the clustering in the BOSS-CMASS sample at the smallest scales. 
This is due to the fact that the mocks are constructed using Lagrangian perturbation 
theory including approximations which break down at small scales.
However, we checked that the magnitude of the variance obtained both from the {\it PTHalos} mock 
catalogues and {\it LasDamas} set of $N$-body simulations displays good consistency, after rescaling 
them to take into account for the difference in their effective volumes.
We conclude then that it is safe to compare the cosmic variance of MultiDark 
(estimated from the {\it Carmen} set of simulations) with that resulting from the 
mock galaxy catalogues. In Section~\ref{results} we present the MultiDark HAM clustering results in comparison with 
our observational clustering estimates.

\section{Clustering of galaxies in the BOSS-CMASS sample: results from model and observations}
\label{results}

The two-dimensional correlation function $\xi(\sigma,\pi)$ for
the north subsample of BOSS-CMASS is presented in Fig.~\ref{fig:sig_pi_MD} for
distances up to $\sim20\,\Mpch$ (dashed contours).  The Finger-Of-God elongation along
the line-of-sight direction at small perpendicular separations, which
is due to galaxy small-scale random velocities, is clearly seen. The
flattening of contours at larger projected scales is due to the Kaiser
effect caused by large-scale infall velocities
\citep{Kaiser1987}. The clustering of ``galaxies'' obtained 
with the MultiDark cosmological simulation (solid contours)
produce a fair representation of the measured clustering in the
CMASS sample. Nevertheless, there are some deviations. At small
separations, $\sigma\lesssim 1\,\Mpch$, observations show more clustering
as compared with results from the simulation. The situation reverses at large 
scales ($\sigma\approx20\,\Mpch$), where our cosmological simulation results in a
slightly stronger clustering.

The projected and redshift-space correlation functions for the 
observed and halo samples considered in this work are presented 
in Figs.~\ref{fig:wp_MD} and~\ref{fig:z_MD}. The north, south and combined 
CMASS samples (symbols) are shown together with the result of our simple 
HAM model (solid lines).
The shaded area for MultiDark gives an estimate of the cosmic variance 
as computed from {\it LasDamas} suite of simulations.

As noted before, there are some noticeable discrepancies at 
small and intermediate scales. The detailed differences between the projected
correlation function and MultiDark 
can be better seen in the right panel of Fig.~\ref{fig:wp_MD}, where differences in the
correlations are amplified after multiplying by the corresponding projected distance. 
The disagreement at scales $\lesssim1\,\Mpch$ is of the order 
of $10\%$. At larger scales ($\sim 10$--$30\,\Mpch$) the 
theoretical estimates lie slightly above those of the 
north galaxy subsample (which has about 
four times larger statistics than the corresponding southern sample) 
but they are still consistent with each other within $\sim1\sigma$ level.

The redshift-space clustering results, both for the CMASS sample and
the $\Lambda$CDM model given by the MultiDark simulation, 
are shown in Fig.~\ref{fig:z_MD}. As before, 
the shaded area represents cosmic variance estimates and differences 
between model and observations are better seen in the right panel. 
Peculiar velocities of galaxies inside virialized systems reduce the 
clustering signal thus lowering the slope of the correlation function 
at scales of $1$--$2\,\Mpch$.

\begin{figure}
      \includegraphics[width=90mm]{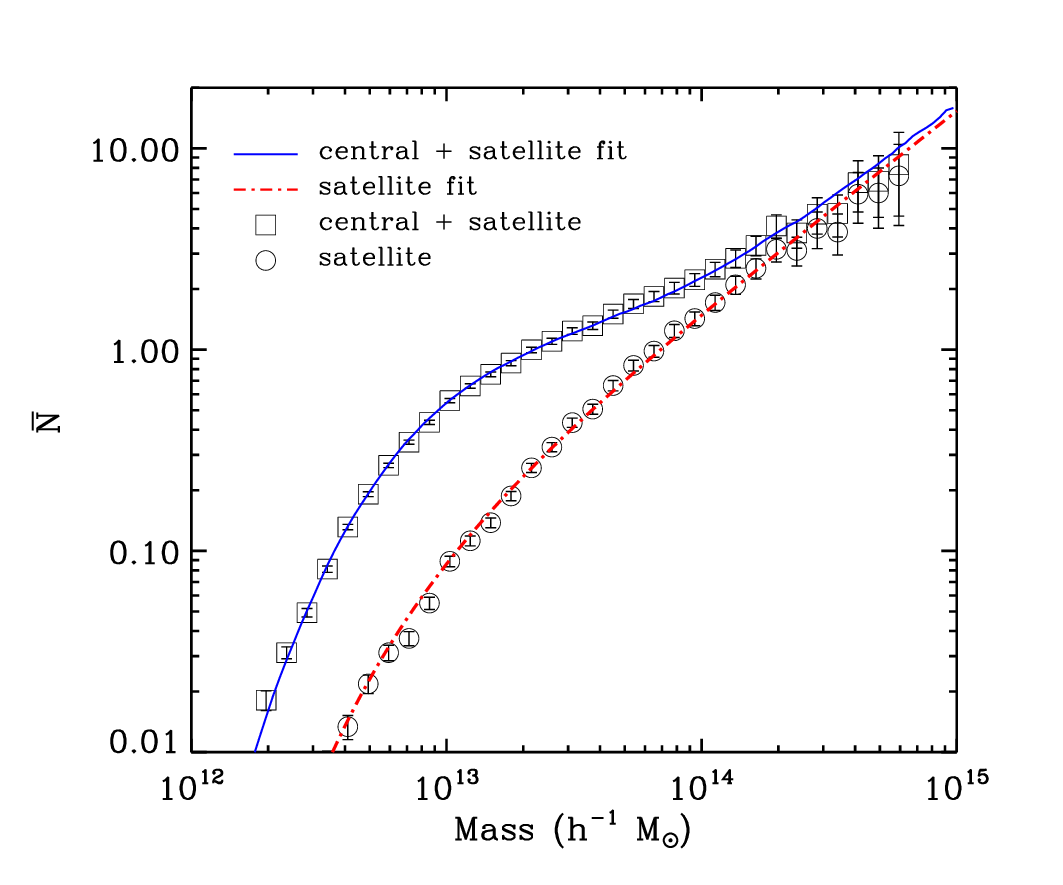}
      \caption{
      The mean occupancy of all haloes in our MultiDark sample used to
      match the BOSS-CMASS observations as a function of halo mass
      (open squares). Open circles and dashed line correspond to
      satellites and central haloes respectively. Error bars 
      are calculated assuming Poisson statistics in the counting. 
      The fit given by Eq.~(\ref{eq:fsat}) is shown as a 
      dot-dashed line.}
\label{fig:sat2}
\end{figure}

For scales in the range $0.5$--$0.8\,\Mpch$ our simple HAM overpredicts the observed values 
by an amount of the order of $30\%$. At larger scales ($\gtrsim0.8\,\Mpch$) the matching 
between the model and observations improves significantly. Differences are less than $3\%$ for a 
wide range of distances ranging from $0.8\,\Mpch$ to about $20\,\Mpch$. At $20$--$40\,\Mpch$ the 
MultiDark result overpredicts the observed clustering by $\gtrsim 10\%$. Statistically, the
differences are significant: the effect is about $3\sigma$ at $\sim30\,\Mpch$ 
(e.g., at $s=33.49\,\Mpch$ the redshift-space correlation function for the combined CMASS sample 
and MultiDark gives $\xi_{\rm N+S}(s)=0.077\pm0.003$ and $\xi_{\rm MD}(s)=0.093\pm0.003$, respectively). 
At scales $\gtrsim 40\,\Mpch$ our HAM model and observations are consistent 
within $\sim1\sigma$. 
In tables~\ref{tab:proy_cf} and~\ref{tab:reds_cf} (see Appendix~\ref{app:b}) 
we present measurements and standard deviations of the correlations shown 
in Figs.~\ref{fig:wp_MD} and~\ref{fig:z_MD}.    

Despite of differences the overall match is 
quite good considering the simplicity of our method: the only free parameter used in our HAM is the 
abundance of (sub)haloes present in the simulation. The disagreement between the $N$-body 
results and observations could be related to the simple stochastic HAM adopted here and may be 
alleviated using a more sophisticated procedure including, for instance, light-cone effects 
and a match to the stellar mass distribution of the sample at these redshifts. On the other hand, 
the mismatch could also be due to some difference between the true cosmology and the one adopted 
for our simulation. 
In the following sections we will exploit the information encoded in 
our BOSS-CMASS abundance-matched halo sample as a way to shed some light on the actual 
trends of the real galaxy population.

\section{The mean halo occupancy of the abundance-matched halo sample}
\label{sec:satellites}

After fixing the abundance of (sub)haloes to that given by observations 
the simulated subhalo distribution is completely determined by the cosmological 
model adopted and the resolution of the simulation.
The main advantage of the MultiDark simulation is that it has sufficient 
resolution to resolve massive satellites around our central distinct haloes. 
Therefore, the satellite distribution around haloes can be directly 
studied from the matched halo catalogues. As shown previously in the top panel of
Fig.~\ref{fig:cum_sat}, the fraction of subhaloes around central haloes 
with a number density similar to that of the CMASS sample is close to 
$10\%$. In particular, for haloes having $V_{\rm max} \geq 370$ km
s$^{-1}$, which corresponds to a number density of $n_* \equiv 3.6\times10^{-4}\,h^3\,\Mpc^{-3}$, 
the resulting subhalo fraction is $12\%$ with negligible statistical uncertainties.
The HOD modelling by \citet{White2011}, using the first semester of
BOSS data, reported a satellite fraction $(10 \pm 2)\%$ which is
reduced to $(7 \pm 2)\%$ when they ignore in their fit to the
correlation function at small scales affected by fibre collisions. 
Note that our HAM procedure is non-parametric and provides 
subhalo fractions consistent with our $\Lambda$CDM cosmological
simulation. In that respect, our approach is different from HOD modelling: 
as mentioned in the introduction in the latter case the halo-occupancy distribution and 
satellite fractions are obtained from a fit to the empirical correlation function.

\begin{figure}
      \includegraphics[width=90mm]{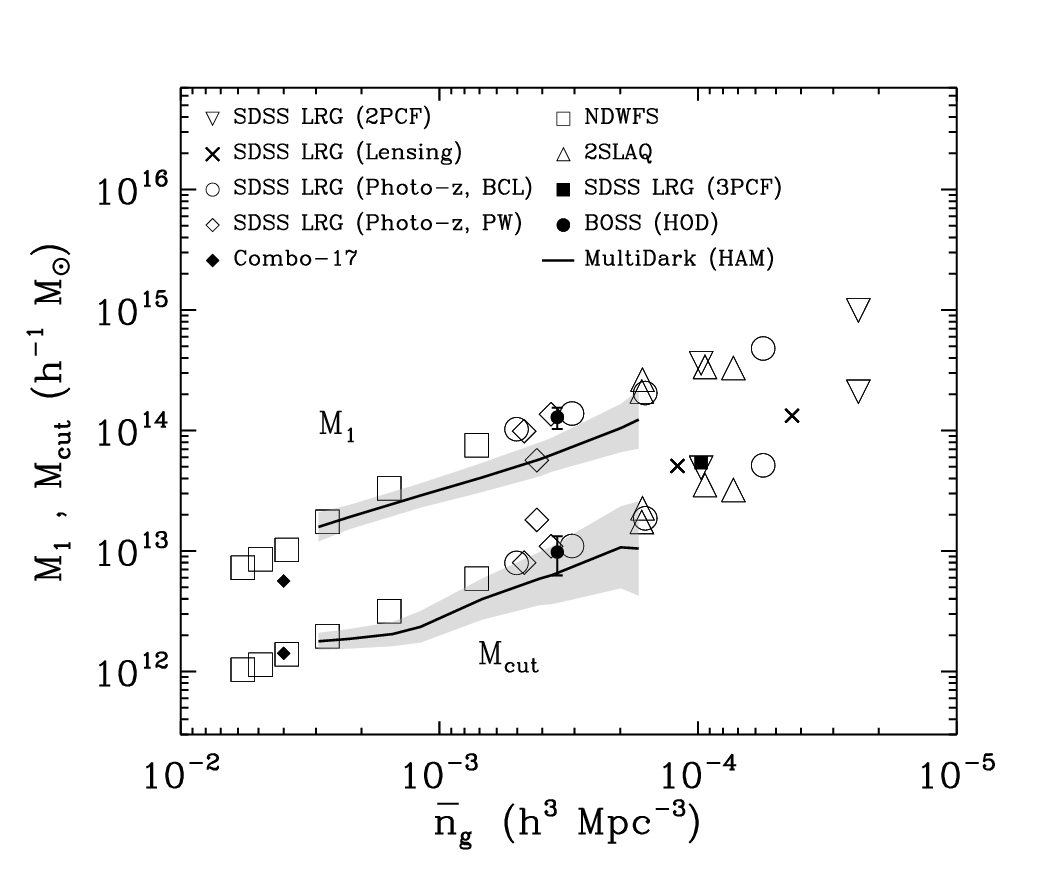}
      \caption{
      MultiDark HOD parameters, $M_{\rm cut}$ and $M_{1}$, as a
      function of number density (solid line) using the simple HAM
      prescription at $z=0.53$.  We compare our results with a variety 
      of intermediate redshift massive galaxy samples \citep[see][]{Phleps06, Mandelbaum06, 
      Kulkarni07, RossN07, Blake08, Brown08, Padmanabhan09, Wake08, Zheng09}. 
      Filled circles with error bars show results from \citet{White2011} 
      HOD's analysis of early BOSS data (see text).
      }
      \label{fig:M1_Mcut}
\end{figure}

Fig.~\ref{fig:sat2} shows the mean occupancy of haloes 
as obtained from the MultiDark halo abundance-matching
scheme. The open circles represent the contribution of subhaloes while 
open squares correspond to the total occupancy 
of haloes, including both central and satellite ``galaxies'' from our halo 
catalogue. Distinct haloes display a clear transition around 
$M_{\rm vir}\gtrsim10^{13}$ $h^{-1}$ M$_{\sun}$ (see solid line).
The mean number of subhaloes as a function of halo mass 
can be accurately described by a function of the 
form \citep[e.g.,][]{Wetzel10}

\begin{equation}
   \bar{N}_{\rm sat}(M)=\left(\frac{M}{M_1}\right)^{\alpha} e^{-M_{\rm cut}/M_1},
   \label{eq:fsat} 
\end{equation}

\noindent where $\log M_{\rm cut}=12.80\pm0.24$, 
$\log M_{1}=13.80\pm0.14$ and $\alpha=1.00\pm0.18$ are the 
best fit values (dot-dashed line). 
Here, $M_{1}$ is the halo mass which hosts, approximately, one 
satellite and $M_{\rm cut}$ governs the
strength of the transition between systems with and without satellites. 
For high halo masses, fluctuations in the determination of
the satellite occupancy arise because we are dealing with small number
statistics as a result of the fixed volume of the simulation. 
The solid line in Fig.~\ref{fig:sat2} shows the total mean halo occupancy but using in this 
case the best fit model for the satellite distribution in order to 
extrapolate the result towards higher masses.

In Fig.~\ref{fig:M1_Mcut} we compare the HOD parameters, $M_{\rm cut}$
and $M_{1}$, obtained from MultiDark at $z=0.53$ as a function of 
(sub)halo number density (solid lines) following our HAM scheme. 
This figure also shows estimates for a variety of intermediate redshift 
massive galaxy samples from the literature, including the HOD results from White et
al. (2011) for the early BOSS data sample. This compilation of 
different datasets has been kindly provided by M. White. 
Error bars on the individual points are typically $\gtrsim0.1$ dex.
The agreement between the MultiDark HAM estimations 
of $M_{1}$ and $M_{\rm cut}$ and those 
from different surveys is remarkable if one considers the differences in sample selection, redshift 
range and HOD methods. In particular, our estimates for the HOD parameters tend to be smaller 
than those of White et al.'s., which is consistent with the larger subhalo fraction found in our case. 
Nevertheless, both estimations marginally agree at the 1$\sigma$ level.

Finally, note that \citet{White2011} did not include the weights $w_{\rm star}$ considered by 
\citet{Ross2012} to correct for systematic effects in the CMASS galaxy density field. In principle, 
this  could have an impact in the estimation of their correlation functions and, as a consequence, 
on the derived parameters. On the other hand, it is important to keep in mind that our results rely 
completely on our halo catalogue.

\begin{figure}
      \includegraphics[width=85mm]{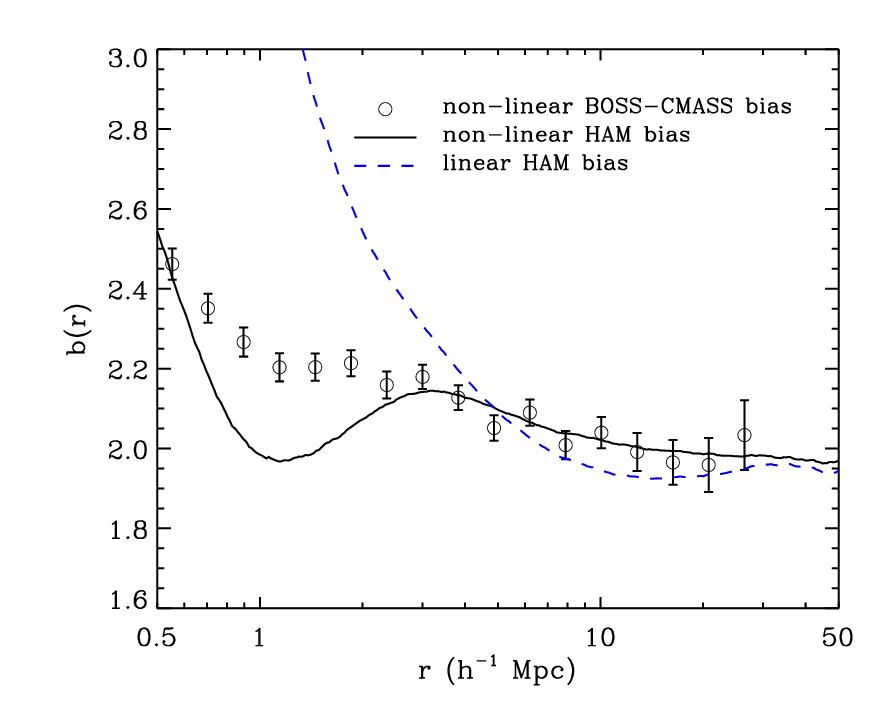}
      \caption{Scale-dependent bias. Shown are the abundance-matched halo 
        and BOSS-CMASS galaxy biases relative to the dark matter distribution of 
        MultiDark at $z=0.53$ (solid curve and empty circles respectively). The halo bias 
        (noted HAM) relative to the linear-theory estimation is shown as a dashed line (see text). 
        }
      \label{fig:bias_from_xi}
\end{figure}

\begin{figure*}
\centering
\includegraphics[width=0.45\textwidth]{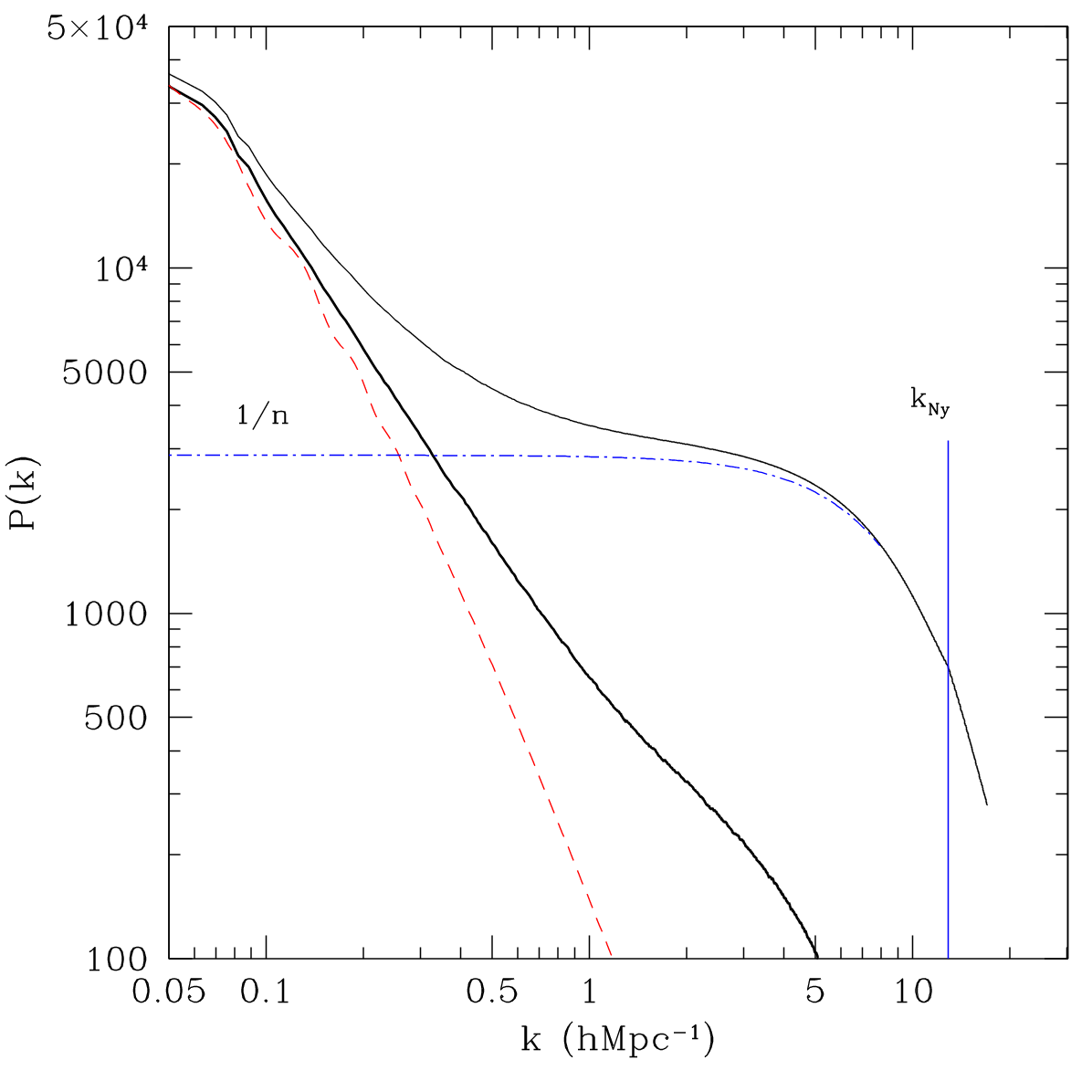}
\includegraphics[width=0.45\textwidth]{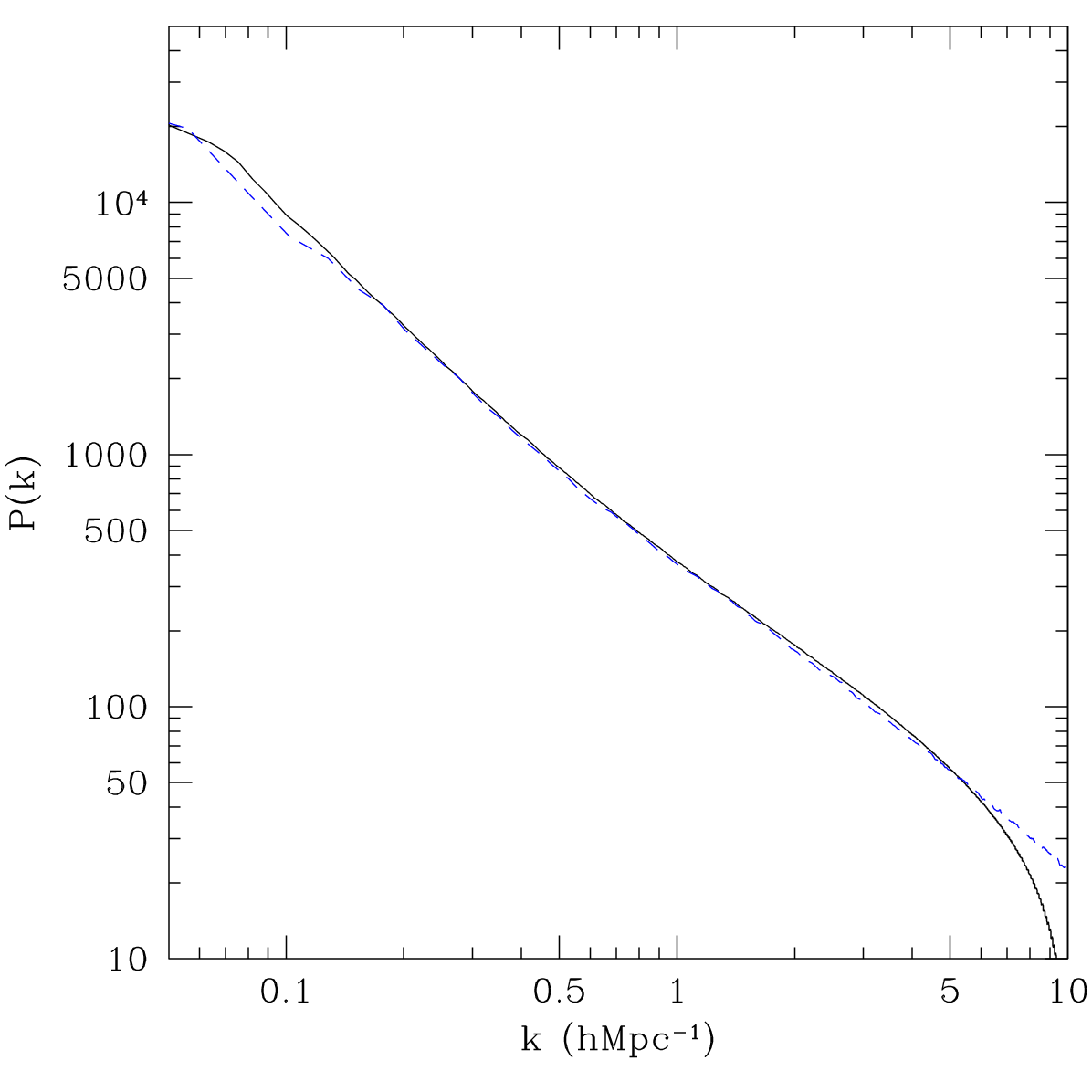}
\caption{{\it Left panel:} Recovering the power spectrum: shot-noise and
  density assignment corrections. The top solid thin curve shows the
  ``raw'' estimate of the power spectrum at $z=0.53$ for haloes and
  subhaloes with circular velocities larger than $\Vpeak>370~\kms$
  corresponding to a number density close to that of galaxies in the
  BOSS-CMASS sample $n_* \equiv 3.6\times10^{-4}\,h^3\,\Mpc^{-3}$.  The dot-dashed line
  is the combined correction in Eq.~(\ref{eq:noise}) due to the
  shot-noise and the density assignment. The vertical line shows the
  Nyquist frequency.  The thick solid line is the recovered power
  spectrum. The dashed line shows the linear power spectrum of dark
  matter density perturbations scaled up to match the amplitude of the
  recovered power spectrum at long waves. {\it Right panel:} Comparison 
  between the recovered power spectra for haloes+subhaloes
  with $\Vpeak>200~\kms$ in the MultiDark (solid line) and the
  Bolshoi (dashed line) simulations at $z=0$. Deviations at $k<0.1\,h\,{\rm Mpc}^{-1}$ are due
  to cosmic variance. The deviations at $k>5\,h\,{\rm Mpc}^{-1}$ are due to
  density assignment effects in the MultiDark simulation. However, for wave-numbers
  in the range $0.2\,h\,{\rm Mpc}^{-1}<k<5\,h\,{\rm Mpc}^{-1}$ the resulting 
  power spectra are not affected by cosmic variance and resolution and the agreement 
  between simulations is excellent, with deviations less than just few percent. }
\label{fig:powerStart}
\end{figure*}

\section{Biases of the abundance-matched halo and BOSS-CMASS samples}
\label{bias}

In Section~\ref{bias_from_xi} we focus our work on the estimation of the abundance-matched halo 
and BOSS-CMASS galaxy biases, and their comparison, using the real-space and projected correlation 
functions respectively; whereas Section~\ref{bias_from_pk} focuses on the abundance-matched 
halo bias from power spectra.

\subsection{Abundance-matched halo and galaxy biases from correlation functions}
\label{bias_from_xi}

Using the resulting halo sample and dark matter particles from the simulation 
we can estimate the real-space bias, $b(r)$, of the halo population with respect to the underlying 
mass distribution by the following relation

\begin{equation}
b^2(r) \equiv \frac{\xi_{\rm HAM}(r)}{\xi_{\rm m}(r)}{\rm ,}
\label{eq:bias_xi} 
\end{equation}

\noindent where $\xi_{\rm HAM}(r)$ and $\xi_{\rm m}(r)$ are the real-space correlation 
functions (i.e., no redshift-space distortions) for the MultiDark haloes and dark matter 
in the volume at the considered redshift. 
This is shown in Fig.~\ref{fig:bias_from_xi} as a function of spatial scale (solid line). 
At the transition scale of $\sim1\,\Mpch$ the bias reaches a local minimum, 
increasing strongly towards smaller scales where galaxies are more strongly clustered 
with respect to the dark matter. The bump-like feature between $\sim 1$--$10\,\Mpch$ is 
related to the transition between the one- and two-halo terms in the correlation function,
while for larger scales the bias factor tends to decrease. 
For scales $\gtrsim1\,\Mpch$ we can constrain the abundance-matched halo bias 
to be in the range $b\approx2$--$2.2$, approaching $b\approx2$ for the largest radii 
(see also section \ref{bias_large_scales_pk}).

\begin{figure}
\centering
\includegraphics[width=0.45\textwidth]{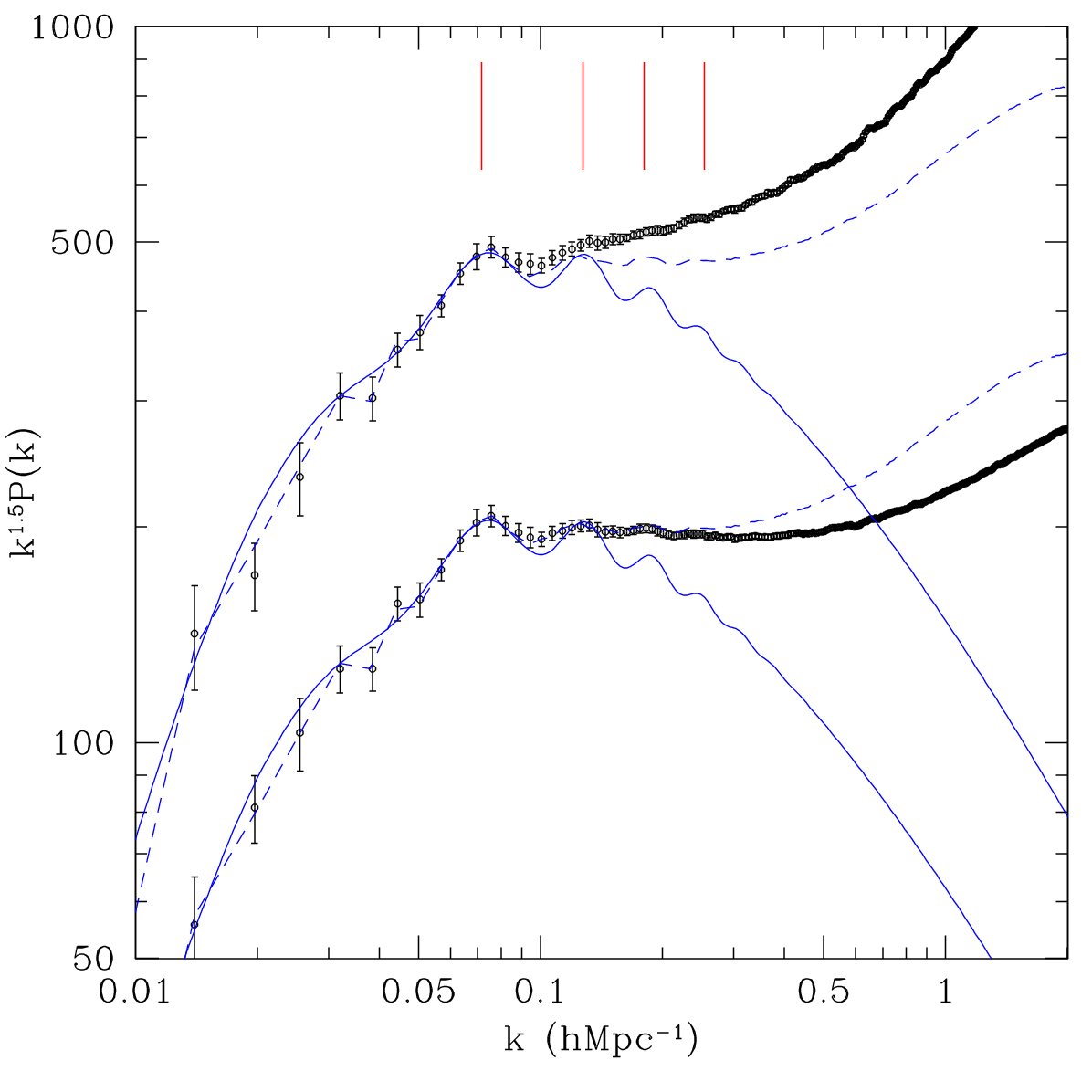}
\caption{Power spectra (multiplied by $k^{1.5}$) of dark matter haloes in real space (open circles with
  error bars) for haloes with $V_{\rm peak}>370~\kms$ (top) and
  $V_{\rm peak}>180~\kms$ (bottom). Solid curves show the linear power spectra
  scaled to match the amplitude of fluctuations at long waves. For reference 
  we included non-linear dark matter power spectra scaled to fit the long-wave side of the 
  first BAO peak (dashed lines). The
  four vertical lines indicate the positions of maxima due to BAO. The BAO
  peaks in the linear spectrum give rise to peaks in the power
  spectrum of haloes.}
\label{fig:powerRaw}
\end{figure}

The linear bias estimation is shown as a dashed line, where the linear matter 
correlation function is used instead. The latter is computed from 
the initial matter power spectrum of the simulation scaled to the redshift of 
interest according to linear theory. 
As expected, the linear bias at small scales differs strongly from the non-linear 
result while approaching more similar values at larger scales. 

We also estimated the scale-dependent bias of the BOSS-CMASS sample 
up to $\sim 30\,\Mpch$ (circles in Fig. 10) given by the square root ratio of the 
observed projected correlation function (as presented in Fig.~\ref{fig:wp_MD}; solid circles) and the 
non-linear projected correlation function of matter given by MultiDark at $z=0.53$. 
This approach is, to first order, very close to that inferred from the real-space correlation 
functions by means of Eq.~(\ref{eq:bias_xi}). 
The error bars were estimated propagating the errors adopted in observations. 
As shown in Fig.~\ref{fig:bias_from_xi}, for scales larger than $\sim2\,\Mpch$ the abundance-matched 
halo and galaxy biases display a good agreement within the error bars; whereas for smaller 
scales the BOSS-CMASS galaxy bias lies above that of the abundance-matched halo sample in 
accordance with the mismatch found between the HAM and BOSS-CMASS projected correlation 
functions (see Fig.~\ref{fig:wp_MD}).

Interestingly, these results are in very good agreement with the findings of \citet{Ho2012}. 
These authors found a linear galaxy bias of $b=1.98\pm0.05$ in the redshift bin of $z=0.50$--$0.55$ 
by studying the angular clustering of the photometric CMASS sample. 
In what follows we will extend this analysis to Fourier space to better characterize 
the scale-dependence of the abundance-matched halo bias.

\subsection{Abundance-matched halo bias from power spectra}
\label{bias_from_pk}

Here we present the clustering bias of the 
abundance-matched halo sample in Fourier space by means of its power spectrum 
since it is well known that this statistics is better suited to 
separate effects on different scales.

We want to present an approximation of our numerical results for the 
comological model adopted in the simulation since it is usually more convenient to use 
analytical approximations instead of dealing with raw simulations. Additionally, the 
derived bias dependence can motivate further comparisons with observational results. 
Interestingly, the high quality of our results allows us to study effects which are 
difficult to measure with low-resolution simulations.

One should clearly understand the role of the standard $\Lambda$CDM model with
the particular set of cosmological parameters used for our simulations.
Our results show that, once we match the abundance of haloes, the
model reasonably reproduces a wide range of scales of the observed 
projected and redshift-space correlation 
functions despite of some discrepancies at small and medium scales as 
presented in Section~\ref{results}. 
In principle, one can invert the correlation function to obtain the power spectrum. 
However, in practice, a model-independent inversion is a technically complicated
process. This is why we chose a different approach: we use
the power spectrum of haloes in the model as an approximation 
of the actual power spectrum of galaxies in BOSS-CMASS for scales larger than 
$\sim1\,\Mpch$ and up to those close to BAOs.

Additionally, we use two other sets of simulations besides MultiDark. 
The first one is the already mentioned {\it Carmen} series of 40 simulations 
of the {\it LasDamas} set that allow us to estimate 
the effect of cosmic variance. Note that, as before, 
we use only relative model-to-model deviations in the {\it Carmen}
simulations: error bars in our results are obtained in this way. Secondly, we 
also use results of the Bolshoi simulation \citep{Bolshoi}. This simulation has a
factor of $\sim 5$ better mass and force resolution, but it was performed 
for a smaller simulation box ($250~\Mpch$ on a side). There is an overlap 
between the MultiDatk and Bolshoi simulations: the simulation volume of 
Bolshoi is large enough to study (sub)haloes with circular 
velocities of $\sim 200~\kms$. At the same time, these (sub)haloes are 
reasonably well resolved in the MultiDark simulation having more than 100 
particles. Comparison of MultiDark and Bolshoi power spectra for these 
haloes allows us to look for discrepancies between the simulations 
at scales $k>0.1\,h\,{\rm Mpc}^{-1}$.

\begin{figure}
\centering
\includegraphics[width=0.45\textwidth]{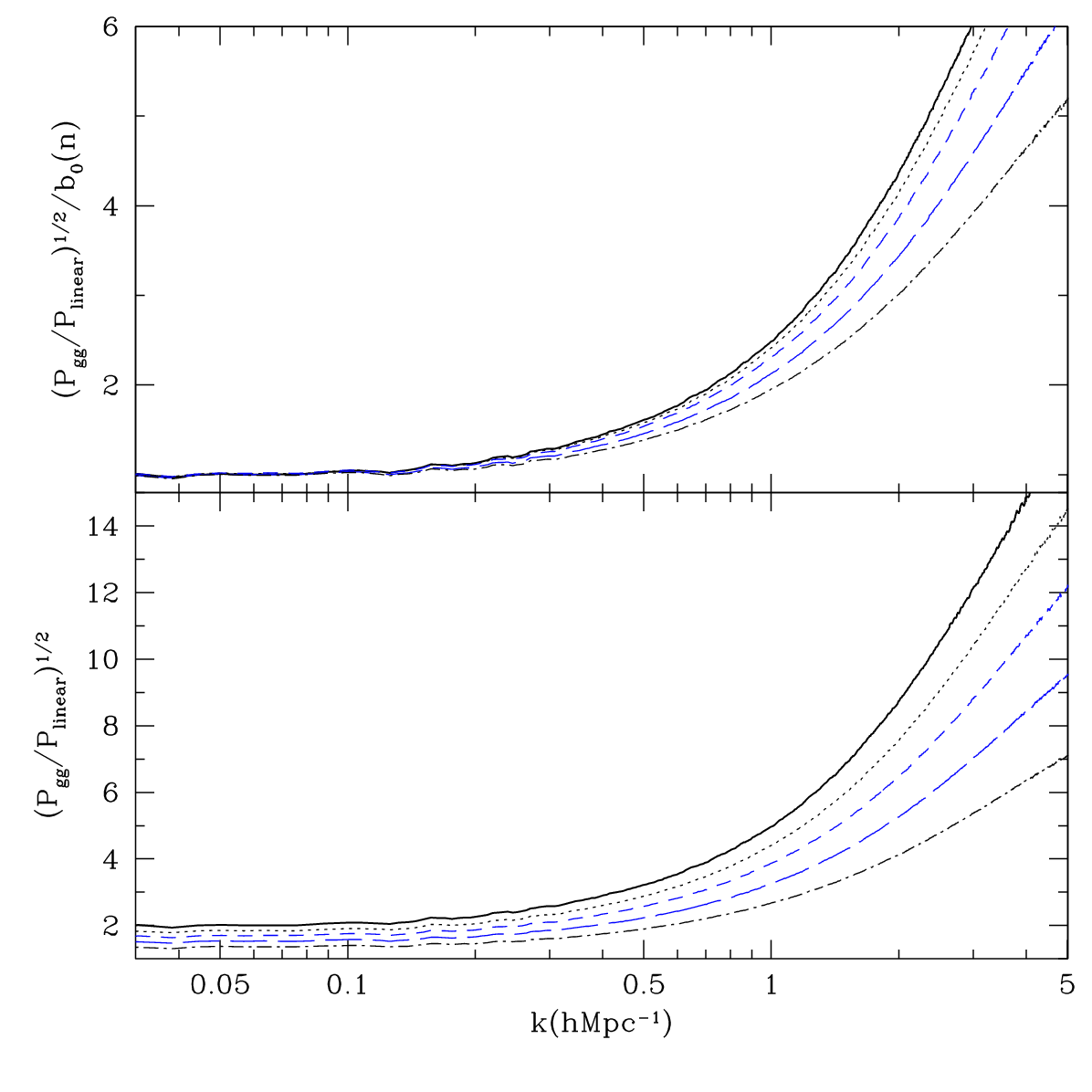}
\caption{{\it Bottom panel}: Real-space bias factor
  $b(k)=(P_{\rm HAM}/P_{\rm linear})^{1/2}$ for haloes with circular velocities
  $V_{\rm peak}=190,220,270,340$ and $370~\kms$ (from  bottom to top). 
  {\it Top panel}: Bias factor for different haloes normalized to unity at
  long-waves. There are small depressions in the bias factor at peaks of BAOs.
  When normalized to the long-wave value $b_0$, the bias factor is slightly smaller 
  for less massive haloes. However, the main effect is the overall shift $b_0$.}
\label{fig:biasVc}
\end{figure}

\subsubsection{Power spectra estimation}

To estimate power spectra, we use a large density mesh
of $4096^3$ cells and then we apply the standard FFT method. 
The Cloud-In-Cell density assignment scheme is used to 
calculate the density fields from the coordinates of haloes in the 
simulations. However, before the power spectra can be reliably used 
two corrections should be applied: a correction due to the density 
assignment \citep[][]{2005ApJ...620..559J} 
and the usual shot-noise correction. If the number density of objects 
is $n=N/L^3$ and the Nyquist wave-number 
is $k_{\rm Ny}=\pi N_{\rm grid}/L$, then
the corrected power spectrum is given by

\begin{equation}
    P(k) = P_{\rm raw}(k) -\frac{1}{n}\left[1-\frac{2}{3}
        \sin^2\left(\frac{\pi k}{2k_{\rm Ny}}\right) \right],
\label{eq:noise}
\end{equation}

\noindent where $L$ is the length of the computational box and 
$N_{\rm grid}=4096$. This approximation is known to work well for 
$k<0.7k_{\rm Ny}$ \citep[][]{2005ApJ...620..559J,2008ApJ...687..738C}.  
However, to remain on safe grounds we decided to limit our 
analysis to $k < 0.4k_{\rm Ny}=5\,h\,{\rm Mpc}^{-1}$. 
The left panel of Fig.~\ref{fig:powerStart} illustrates the procedure 
of shot-noise and density corrections using a halo 
sample with $V_{\rm max}>370$ km s$^{-1}$ extracted 
from the MultiDark simulation at $z=0.53$.

\begin{figure}
\centering
\includegraphics[width=0.47\textwidth]{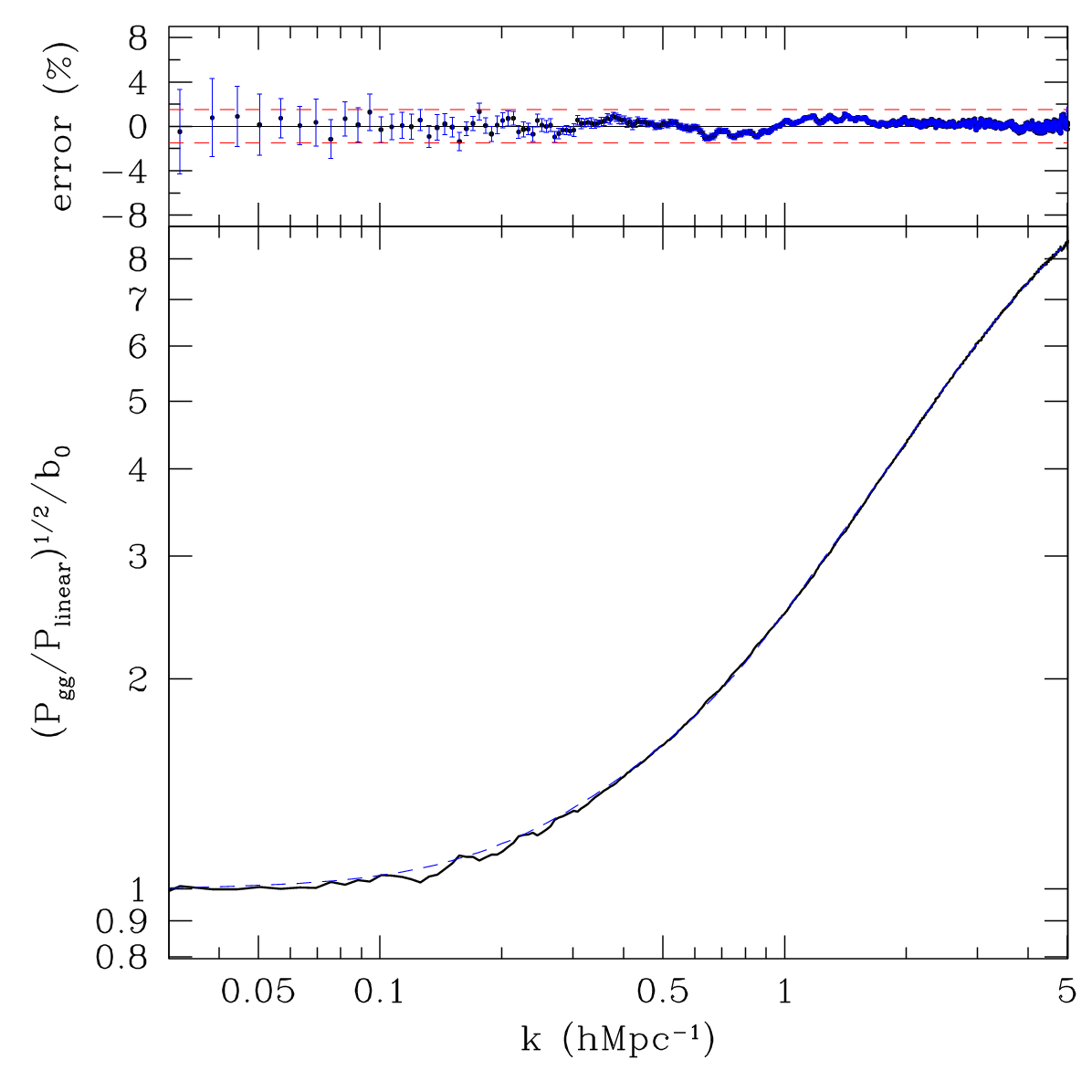}
\includegraphics[width=0.47\textwidth]{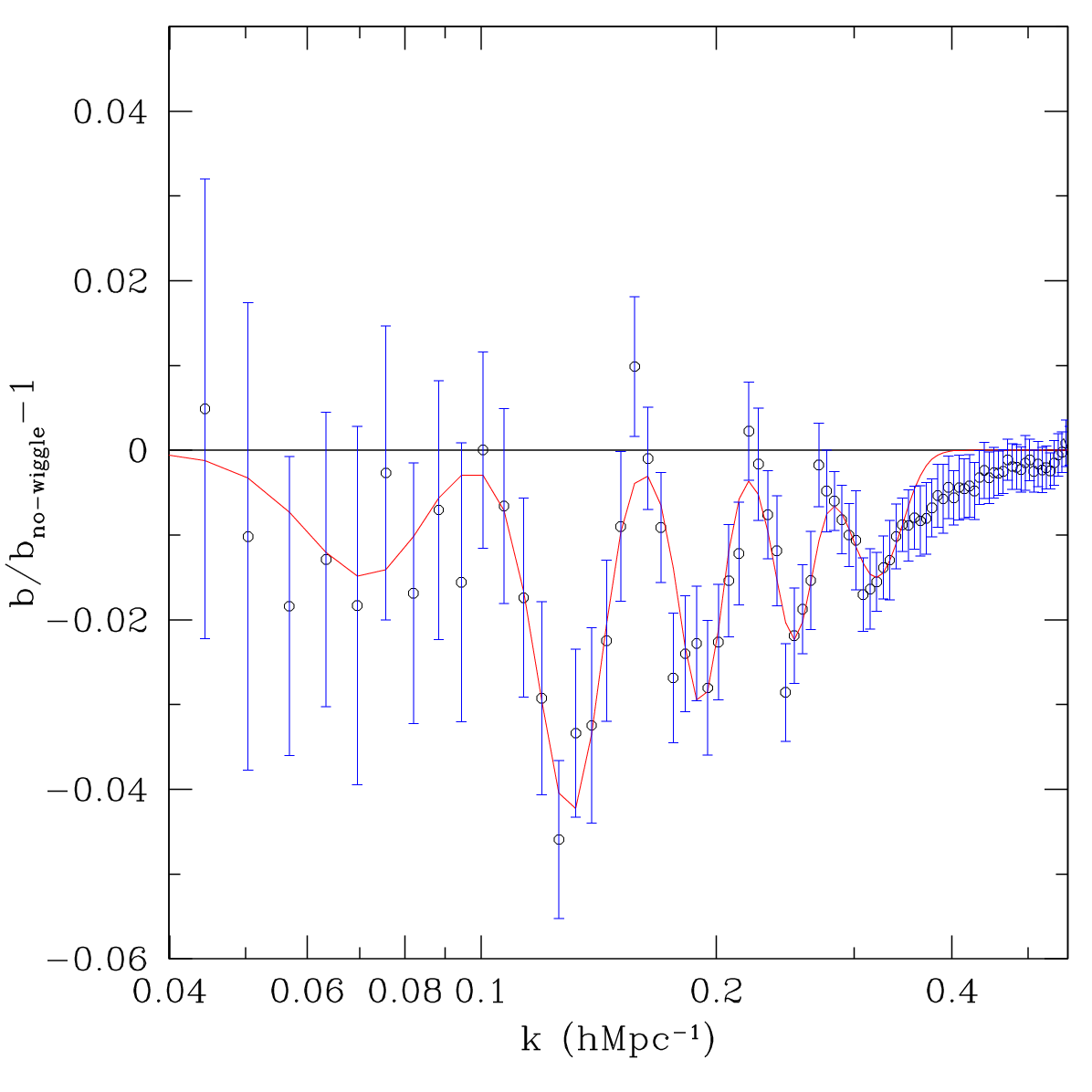}
\caption{Real-space bias factor for haloes with circular velocities
  larger than $V_{\rm max}=370~\kms$. {\it Top panel:} The linear bias factor
  normalized to the long-wave value $b_0$ (bottom plot; solid line) is compared
  with the analytical approximation given by
  Eq.~(\ref{eq:bk2}) (dashed line). Wiggles below the smooth dashed curve are due to BAOs. 
  The top plot displays the relative error in percentages of the analytical approximation (filled circles). 
  {\it Bottom panel:} Deviations of the linear bias from the ``de-wiggled'' component of the
  bias factor given by Eq.~(\ref{eq:bk}). Open circles show the
  relative deviations $b(k)/b_{\rm no-wiggle}-1$ for each wave-number. The
  solid line is an analytical model for the residuals which displays five peaks and results 
  from the sum of exponential terms in Eq.~(\ref{eq:bk2}). In both panels, error bars show
  the {\it rms} fluctuations due to cosmic variance.}
\label{fig:bias370}
\end{figure}

In the right panel of Fig.~\ref{fig:powerStart} we compare results
of the MultiDark and Bolshoi simulations. Just as one may expect,
there are some deviations at long waves due to the cosmic variance:
the Bolshoi box of $250~\Mpch$ is too small to accurately probe these
waves. There are also deviations at short waves that correspond 
to $k>7\,h\,{\rm Mpc}^{-1}$ that are mainly due to the difference in 
density assignment between both simulations. For the Bolshoi simulation, 
the adopted mesh sets a minimum physical scale four times higher in frequency 
in comparison to MultiDark. However, for wave-numbers in the 
range $0.2\,h\,{\rm Mpc}^{-1}<k<5\,h\,{\rm Mpc}^{-1}$ 
the agreement between the simulations is remarkably good. 
This agreement is especially important for short waves, where 
both resolution and shot-noise could have corrupted the results. 
However, since this has not happened, it indicates that 
our power spectrum estimates for MultiDark can be trusted 
up to, at least, $k=5\,h\,{\rm Mpc}^{-1}$.

Fig.~\ref{fig:powerRaw} shows power spectra of haloes with circular
velocity cuts $\Vpeak>370~\kms$ (top curves) and $\Vpeak>180~\kms$ 
(bottom curves). To highlight BAO features, we actually plot the power
spectra of the halo distribution multiplied by $k^{1.5}$. As a result, the first five 
peaks in the spectra are clearly seen in the plot. However, they are somewhat 
smeared out by the non-linear evolution. As expected, the smearing increases for 
larger wave-numbers where the non-linearity is more important.

\subsubsection{Abundance-matched halo bias}
\label{bias_large_scales_pk}

In what follows, we define the bias factor by

\begin{equation}
b_P(k,\Vpeak) \equiv \left[ \frac{P_{\rm HAM}(k,\Vpeak)}{P_{\rm linear}(k)}\right]^{1/2},
\label{eq:biasdef}
\end{equation}

\noindent where $P_{\rm linear}(k)$ is the linear power spectrum of the dark
matter and $P_{\rm HAM}(k,\Vpeak)$ is the power spectrum of haloes and
subhaloes with circular velocities larger than $\Vpeak$. In order to
distinguish the latter from the oftenly used non-linear 
dark matter power spectrum or from the power spectrum of distinct haloes only, 
we use subscript HAM to indicate that the results correspond to our HAM technique. 
By definition, the scale-dependent bias factor of Eq.~(\ref{eq:biasdef}) 
encodes the information of the non-linear power spectrum of the ``galaxy'' sample.

We start our analysis with the long-wave normalization of the bias parameter 
for different peak velocity cuts and, thus, for different number densities 
of our ``galaxies''. The bottom panel 
in Fig.~\ref{fig:biasVc} shows $b_P(k,\Vpeak)$ for different
velocities; at all wave-numbers it increases
with increasing $\Vpeak$. The top panel shows that when normalized to
the long-wave value, $b_0(\Vpeak)$, the bias factor is nearly the same.
However, there is some residual dependence on \Vpeak, i.e., the deviations
of the bias from one velocity cut to another can be as large as 15\% and this 
should be taken into account if an accurate fit is needed. An approximation 
for the real-space long-wave bias factor as a function of the average 
number density of dark matter haloes, $n(>\Vpeak)$, is presented below:

\begin{equation}
b_0(n)=-0.048-\left(0.594\pm0.02\right)\log_{10} n, \\
\label{eq:b0}
\end{equation}

\noindent where $n$ is in units of $h^3\,\Mpc^{-3}$.
We now focus our analysis on the bias factor of haloes with $\Vpeak>370~\kms$, whose abundance 
$n_* \equiv 3.6\times10^{-4}\,h^{3}\,\Mpc^{-3}$ is close that of BOSS-CMASS galaxies at $z=0.53$. 
The top panel of Fig.~\ref{fig:bias370} shows the bias factor of these haloes
normalized to the value found at long waves, i.e. $b_0(n_*)=2.00\pm0.07$.
Overall the bias factor is nearly flat at long waves and monotonically
increases to short waves. The following approximation for the smooth (i.e., ``de-wiggled'')
component of $b_P(k,n_*)$ gives percent-level accuracy (see dashed line 
in the top panel (bottom plot) of Fig.~\ref{fig:bias370}):

\begin{equation}
b_{\rm dw}(k) = 1+\log_{10}\left(1+11.2k^2 + 18.3k^6 + 0.59k^{11}\right)
\label{eq:bk}
\end{equation}

\noindent where the wave-number $k$ is in units of $h\,\Mpc^{-1}$ and the subscript `dw' stands for `de-wiggled'. 
However, this approximation misses an important effect of non-linearities: the damping of the BAO 
signal in Fourier-space.
The coupling between different Fourier modes washes out the acoustic oscillations, erasing the higher 
harmonic peaks \citep{meiksin1999,eisenstein07,angulo05,angulo08,sanchez08,montesano10}.
In recent years, there has been substantial progress in the theoretical understanding of non-linear distortions
in the BAO signal, which can now be accurately modelled 
\citep[see e.g.,][]{crocce06,Crocce08,matsubara08b,matsubara08a,taruya09},
and even partially corrected for \citep{eisenstein07b, seo10}.
As the bias factor in Eq.~(\ref{eq:biasdef}) is defined with respect to the extrapolated linear theory power spectrum, 
this damping leads to small wiggles in the bias at the 2--4\% level detected at high significance for 
$k\gtrsim0.1\,h\,{\rm Mpc}^{-1}$ (see bottom panel of Fig.~\ref{fig:bias370}). 
Therefore, in order to improve the fitting of the bias we also included the five 
main BAO peaks on top of the smooth component as follows:

\begin{equation}
b(k,n_*) = b_{\rm dw}(k)\times\prod_{i=1}^{5}\left[1-\alpha_i\exp\left(-\frac{(k-k_i)^2}{\sigma_i^2}\right)\right]
\label{eq:bk2}.
\end{equation}

\noindent Here each BAO peak is approximated as a small suppression of
the bias factor given by the last term of the equation, $k_i$ is the wave-number of
the peak and $\alpha_i\approx 0.01$--$0.05$ and $\sigma_i\approx
0.01$--$0.02$ are free parameters.  The typical errors given by this
approximation are smaller than 2\% (see the top panel (top plot) of
Fig.~\ref{fig:bias370}). The values of the parameters used in the
equation can be seen in Table~\ref{tab:baos}.

\begin{table}
\centering
  \caption{Parameters for the approximation of the real-space bias factor given by Eq.~(\ref{eq:bk2}).}
  \begin{tabular}{@{}lccc@{}}
  \hline
  BAO peak& $k\;(h\;{\rm Mpc}^{-1})$ &$\alpha_i$& $\sigma_i$ \\
  \hline
  1 & 0.071 & 0.015 & 0.017 \\
  2 & 0.130 & 0.043 & 0.017 \\
  3 & 0.191 & 0.030 & 0.017 \\
  4 & 0.251 & 0.022 & 0.020 \\
  5 & 0.310 & 0.015 & 0.035 \\
  \hline
  \end{tabular}
  \label{tab:baos}
\end{table}


\section{Conclusions}
\label{conclusions}
We presented an analysis of the clustering of $282,068$ galaxies in 
the DR9 sample of BOSS data for a wide range of scales ranging from 
$\sim500\,\kpch$ to $\sim90\,\Mpch$. We separately studied the clustering 
in the northern and southern hemispheres, as well as for the combined sky 
sample. We measured the two-dimensional, projected and redshift-space correlation 
functions and compare the results with those obtained from a large cosmological 
simulation with $1\,h^{-1}\,{\rm Gpc}$ on a side at a redshift of $z=0.53$. 
We also provide tables of the measured correlations together with estimates 
of the correlation matrices of the observed correlations (see Appendix~\ref{app:b}). 
The cosmological parameters adopted in the simulation are consistent with the latest 
WMAP7 results and several other probes. To bridge the gap between galaxies 
and dark matter haloes we use an HAM technique applied to the BOSS-CMASS sample.  
Our simulation, also known as MultiDark, is able to resolve the relevant subhalo 
masses needed to analyse the resulting satellite distribution. 
It is worth noting that the subhalo population in the simulation 
is completely determined by the cosmological model adopted not relying on a pure 
fit to the empirical correlation functions.

Our main results can be summarized as follows:
  
\begin{itemize}
\item  
There is a 10--20\% asymmetry in the projected and redshift-space 
correlation functions between the north and south subsamples at $\gtrsim20\,\Mpch$ scales, 
which is better seen in the case of the projected correlation function. 
However, for both subsamples, the mean values agree with each other 
within a $\sim1\sigma$ level of uncertainty.\\

\item As compared with the first-semester of BOSS results presented by
\citet{White2011}, we find a small increase in power in the projected correlation
function at scales smaller than $\sim1\,\mpch$ due to the improved treatment of fibre collisions
and new corrections for systematics. However, the correlation functions 
(projected and redshift-space) decline by 10--20\% at $10$--$30\,\Mpch$
scales in comparison with our HAM model. This is most noticeable for the north 
subsample which has about four times larger statistics than its southern 
counterpart. The comparison with the south subsample yields more consistent 
results with MultiDark at all scales, both in the projected and redshift-space correlations.\\

\item Our $N$-body results for the clustering of ``galaxies'' give 
a reasonable representation of the measured clustering in the CMASS sample 
given the simplicity of the HAM model used. 
The more consistent results between the north
and south subsamples for the redshift-space correlation function show
a remarkable agreement with theory on scales ranging from $0.8\,\Mpch$ up to $20\,\Mpch$: 
the differences are of the order of $\sim 3\%$. This result is more impressive when 
considering the fact that our simple HAM scheme does not include any free parameter. 
However, for $s<0.8\,\Mpch$ our matching tends to overpredict the clustering amplitude 
given by observations. Additionally, at distances in the range $20$--$40\,\Mpch$ we find some 
deviations ($\gtrsim 10\%$) when comparing the model to the combined galaxy sample. 
Statistically, this difference is important -- e.g., it
represents a $\sim3\,\sigma$ deviation at $\sim30\,\Mpch$. Future 
data and a more sophisticated theoretical modelling may help to
clarify the situation.\\

\item The distribution of (sub)haloes as a function of 
halo mass, as measured from our abundance-matched halo catalogue, 
points towards a galaxy population inhabiting 
haloes of mass $M\gtrsim10^{13}\,h^{-1}$ M$_{\sun}$, with about $12\%$ of 
them being satellites orbiting centrals with 
$M\gtrsim10^{14}\,h^{-1}$ M$_{\sun}$. We also derived values 
for the HOD parameters of the sample using our 
simulation, i.e. $\log M_{\rm cut}=12.80 \pm 0.24$ and $\log M_{1}=13.80 \pm 0.14$. \\

\item The scale-dependent real-space galaxy bias of BOSS-CMASS galaxies is likely to 
be $b\simeq2$ at scales $\gtrsim10\,h^{-1}$ Mpc (see Eq.~(\ref{eq:bias_xi})) 
as inferred both from the HAM and BOSS-CMASS observed correlations (see Fig. 10).   
Furthermore, using our simulation, we also computed a large-scale bias 
(defined as the ratio between the abundance-matched galaxy catalogue and 
the extrapolated linear matter power spectra; see Eq.~(\ref{eq:biasdef})) 
and found that it depends on the ``galaxy'' number density as 
$b_{0}(n_{\rm g})=-0.048-\left(0.594\pm0.02\right)\log_{10}\left(n_{\rm g}/h^{3}\;{\rm Mpc}^{-3}\right)$ 
for the cosmological model adopted in our simulation. 
Specifically, for a (sub)halo number density of 
$3.6\times10^{-4}\,h^{3}\,\Mpc^{-3}$, we get $b_0=2.00\pm0.07$.
\\

\item The large-scale galaxy bias, defined using Eq.~(\ref{eq:biasdef}), has $\sim2$--4\% dips 
at the positions of BAO peaks in the spectrum of fluctuations that are due to shifts caused 
by non-linear effects. In this case, we also provide a formula of the bias as a 
function of number density for the cosmology adopted here that can also be used to recover the 
non-linear ``galaxy'' power spectrum in terms of the extrapolated linear density field of matter.
\end{itemize}

\section*{Acknowledgments}

We thank the anonymous referee for several comments that helped to improve this work.

Funding for SDSS-III has been provided by the Alfred P. Sloan Foundation, the Participating Institutions, 
the National Science Foundation, and the U.S. Department of Energy. 
SDSS-III is managed by the Astrophysical Research Consortium for the Participating Institutions of the 
SDSS-III Collaboration including the University of Arizona, the Brazilian Participation Group, Brookhaven 
National Laboratory, University of Cambridge, University of Florida, the French Participation Group, the 
German Participation Group, the Instituto de Astrofisica de Canarias, the Michigan State/Notre Dame/JINA 
Participation Group, Johns Hopkins University, Lawrence Berkeley National Laboratory, Max Planck Institute 
for Astrophysics, New Mexico State University, New York University, Ohio State University, 
Pennsylvania State University, University of Portsmouth, Princeton University, the Spanish Participation Group, 
University of Tokyo, University of Utah, Vanderbilt University, University of Virginia, University of Washington, 
and Yale University. 

The MultiDark Database used in this paper and the web application providing online access to it were constructed 
as part of the activities of the German Astrophysical Virtual Observatory as a result of the collaboration between 
the Leibniz-Institute for Astrophysics Potsdam (AIP) and the Spanish MultiDark Consolider Project CSD2009-00064. 
The Bolshoi and MultiDark simulations were run on the NASA's Pleiades supercomputer at the NASA Ames Research Center.

S.E.N. and F.P. acknowledges support from the Spanish MICINN's Consolider grant MultiDark CSD2009-00064.
S.E.N. also acknowledges support by the Deutsche Forschungsgemeinschaft under the grant MU1020 16-1. 
F.P. also thanks the support of the MICINN Spanish grant AYA2010-21231-C02-01 and the Campus of 
International Excellence UAM$+$CSIC. A.K. acknowledges support from the NSF under a grant to NMSU. 

\bibliography{Nuza_et_al_2013}

\begin{thebibliography}{90}
\expandafter\ifx\csname natexlab\endcsname\relax\def\natexlab#1{#1}\fi

\bibitem[{{Abazajian} {et~al.}(2005)}]{Abazajian2005}
{Abazajian}, K., {et~al.} 2005, \apj, 625, 613

\bibitem[{{Ahn} {et~al.}(2012)}]{Ahn2012}
{Ahn}, C.~P., {et~al.} 2012, \apjs, 203, 21

\bibitem[{{Anderson} {et~al.}(2012)}]{Anderson2012}
{Anderson}, L., {et~al.} 2012, \mnras, 427, 3435

\bibitem[{{Angulo} {et~al.}(2005){Angulo}, {Baugh}, {Frenk}, {Bower},
  {Jenkins}, \& {Morris}}]{angulo05}
{Angulo}, R., {Baugh}, C.~M., {Frenk}, C.~S., {Bower}, R.~G., {Jenkins}, A., \&
  {Morris}, S.~L. 2005, \mnras, 362, L25

\bibitem[{{Angulo} {et~al.}(2008){Angulo}, {Baugh}, {Frenk}, \&
  {Lacey}}]{angulo08}
{Angulo}, R.~E., {Baugh}, C.~M., {Frenk}, C.~S., \& {Lacey}, C.~G. 2008,
  \mnras, 383, 755

\bibitem[{{Behroozi} {et~al.}(2010){Behroozi}, {Conroy}, \&
  {Wechsler}}]{Behroozi10}
{Behroozi}, P.~S., {Conroy}, C., \& {Wechsler}, R.~H. 2010, \apj, 717, 379

\bibitem[{{Behroozi} {et~al.}(2013){Behroozi}, {Wechsler}, {Wu}, {Busha},
  {Klypin}, \& {Primack}}]{RockStar}
{Behroozi}, P.~S., {Wechsler}, R.~H., {Wu}, H.-Y., {Busha}, M.~T., {Klypin},
  A.~A., \& {Primack}, J.~R. 2013, \apj, 763, 18

\bibitem[{{Berlind} \& {Weinberg}(2002)}]{Berlind02}
{Berlind}, A.~A., \& {Weinberg}, D.~H. 2002, \apj, 575, 587

\bibitem[{{Blake} {et~al.}(2008){Blake}, {Collister}, \& {Lahav}}]{Blake08}
{Blake}, C., {Collister}, A., \& {Lahav}, O. 2008, \mnras, 385, 1257

\bibitem[{{Blanton} {et~al.}(2003){Blanton}, {Lin}, {Lupton}, {Maley}, {Young},
  {Zehavi}, \& {Loveday}}]{Blanton2003}
{Blanton}, M.~R., {Lin}, H., {Lupton}, R.~H., {Maley}, F.~M., {Young}, N.,
  {Zehavi}, I., \& {Loveday}, J. 2003, \aj, 125, 2276

\bibitem[{{Bolton} {et~al.}(2012)}]{Bolton2012}
{Bolton}, A.~S., {et~al.} 2012, \aj, 144, 144

\bibitem[{{Brown} {et~al.}(2008)}]{Brown08}
{Brown}, M.~J.~I., {et~al.} 2008, \apj, 682, 937

\bibitem[{{Colless} {et~al.}(2001)}]{2001MNRAS.328.1039C}
{Colless}, M., {et~al.} 2001, \mnras, 328, 1039

\bibitem[{{Conroy} {et~al.}(2006){Conroy}, {Wechsler}, \&
  {Kravtsov}}]{conroy06}
{Conroy}, C., {Wechsler}, R.~H., \& {Kravtsov}, A.~V. 2006, \apj, 647, 201

\bibitem[{{Crocce} \& {Scoccimarro}(2006)}]{crocce06}
{Crocce}, M., \& {Scoccimarro}, R. 2006, PRD, 73, 063519

\bibitem[{{Crocce} \& {Scoccimarro}(2008)}]{Crocce08}
---. 2008, PRD, 77, 023533

\bibitem[{{Cui} {et~al.}(2008){Cui}, {Liu}, {Yang}, {Wang}, {Feng}, \&
  {Springel}}]{2008ApJ...687..738C}
{Cui}, W., {Liu}, L., {Yang}, X., {Wang}, Y., {Feng}, L., \& {Springel}, V.
  2008, \apj, 687, 738

\bibitem[{{Davis} \& {Peebles}(1983)}]{DavisPeebles83}
{Davis}, M., \& {Peebles}, P.~J.~E. 1983, \apj, 267, 465

\bibitem[{{Dawson} {et~al.}(2013)}]{Dawson2012}
{Dawson}, K.~S., {et~al.} 2013, \aj, 145, 10

\bibitem[{{Eisenstein} {et~al.}(2007{\natexlab{a}}){Eisenstein}, {Seo},
  {Sirko}, \& {Spergel}}]{eisenstein07b}
{Eisenstein}, D.~J., {Seo}, H.-J., {Sirko}, E., \& {Spergel}, D.~N.
  2007{\natexlab{a}}, \apj, 664, 675

\bibitem[{{Eisenstein} {et~al.}(2007{\natexlab{b}}){Eisenstein}, {Seo}, \&
  {White}}]{eisenstein07}
{Eisenstein}, D.~J., {Seo}, H.-J., \& {White}, M. 2007{\natexlab{b}}, \apj,
  664, 660

\bibitem[{{Eisenstein} {et~al.}(2011)}]{Eisenstein2011}
{Eisenstein}, D.~J., {et~al.} 2011, \aj, 142, 72

\bibitem[{{Fukugita} {et~al.}(1996){Fukugita}, {Ichikawa}, {Gunn}, {Doi},
  {Shimasaku}, \& {Schneider}}]{1996AJ....111.1748F}
{Fukugita}, M., {Ichikawa}, T., {Gunn}, J.~E., {Doi}, M., {Shimasaku}, K., \&
  {Schneider}, D.~P. 1996, \aj, 111, 1748

\bibitem[{{Gottl\"ober} \& {Klypin}(2008)}]{ART2008}
{Gottl\"ober}, S., \& {Klypin}, A. 2008, ArXiv e-prints

\bibitem[{{Gunn} {et~al.}(1998)}]{Gunn1998}
{Gunn}, J.~E., {et~al.} 1998, \aj, 116, 3040

\bibitem[{{Gunn} {et~al.}(2006)}]{Gunn2006}
---. 2006, \aj, 131, 2332

\bibitem[{{Guo} {et~al.}(2012){Guo}, {Zehavi}, \& {Zheng}}]{Guo2011}
{Guo}, H., {Zehavi}, I., \& {Zheng}, Z. 2012, \apj, 756, 127

\bibitem[{{Guo} {et~al.}(2010){Guo}, {White}, {Li}, \&
  {Boylan-Kolchin}}]{Guo10}
{Guo}, Q., {White}, S., {Li}, C., \& {Boylan-Kolchin}, M. 2010, \mnras, 404,
  1111

\bibitem[{{Hamilton}(1993)}]{Hamilton93}
{Hamilton}, A.~J.~S. 1993, \apj, 417, 19

\bibitem[{{Hamilton} \& {Tegmark}(2004)}]{HamiltonTegmark2004}
{Hamilton}, A.~J.~S., \& {Tegmark}, M. 2004, \mnras, 349, 115

\bibitem[{{Ho} {et~al.}(2012)}]{Ho2012}
{Ho}, S., {et~al.} 2012, \apj, 761, 14

\bibitem[{{Jarosik} {et~al.}(2011)}]{2011ApJS..192...14J}
{Jarosik}, N., {et~al.} 2011, \apjs, 192, 14

\bibitem[{{Jing}(2005)}]{2005ApJ...620..559J}
{Jing}, Y.~P. 2005, \apj, 620, 559

\bibitem[{{Kaiser}(1987)}]{Kaiser1987}
{Kaiser}, N. 1987, \mnras, 227, 1

\bibitem[{{Kim} {et~al.}(2008){Kim}, {Park}, \& {Choi}}]{Kim08}
{Kim}, J., {Park}, C., \& {Choi}, Y.-Y. 2008, \apj, 683, 123

\bibitem[{{Klypin} \& {Holtzman}(1997)}]{1997astro.ph.12217K}
{Klypin}, A., \& {Holtzman}, J. 1997, ArXiv Astrophysics e-prints

\bibitem[{{Klypin} {et~al.}(2002){Klypin}, {Zhao}, \&
  {Somerville}}]{Klypin2002}
{Klypin}, A., {Zhao}, H., \& {Somerville}, R.~S. 2002, \apj, 573, 597

\bibitem[{{Klypin} {et~al.}(2011){Klypin}, {Trujillo-Gomez}, \&
  {Primack}}]{Bolshoi}
{Klypin}, A.~A., {Trujillo-Gomez}, S., \& {Primack}, J. 2011, \apj, 740, 102

\bibitem[{{Knebe} {et~al.}(2011)}]{Knebe11}
{Knebe}, A., {et~al.} 2011, \mnras, 415, 2293

\bibitem[{{Kravtsov} {et~al.}(2004){Kravtsov}, {Berlind}, {Wechsler}, {Klypin},
  {Gottl{\"o}ber}, {Allgood}, \& {Primack}}]{KravtsovHOD04}
{Kravtsov}, A.~V., {Berlind}, A.~A., {Wechsler}, R.~H., {Klypin}, A.~A.,
  {Gottl{\"o}ber}, S., {Allgood}, B., \& {Primack}, J.~R. 2004, \apj, 609, 35

\bibitem[{{Kravtsov} {et~al.}(1997){Kravtsov}, {Klypin}, \&
  {Khokhlov}}]{ART1997}
{Kravtsov}, A.~V., {Klypin}, A.~A., \& {Khokhlov}, A.~M. 1997, \apjs, 111, 73

\bibitem[{{Kulkarni} {et~al.}(2007){Kulkarni}, {Nichol}, {Sheth}, {Seo},
  {Eisenstein}, \& {Gray}}]{Kulkarni07}
{Kulkarni}, G.~V., {Nichol}, R.~C., {Sheth}, R.~K., {Seo}, H.-J., {Eisenstein},
  D.~J., \& {Gray}, A. 2007, \mnras, 378, 1196

\bibitem[{{Landy} \& {Szalay}(1993)}]{LandySzalay93}
{Landy}, S.~D., \& {Szalay}, A.~S. 1993, \apj, 412, 64

\bibitem[{{Leauthaud} {et~al.}(2011){Leauthaud}, {Tinker}, {Behroozi}, {Busha},
  \& {Wechsler}}]{Leauthaud11}
{Leauthaud}, A., {Tinker}, J., {Behroozi}, P.~S., {Busha}, M.~T., \&
  {Wechsler}, R.~H. 2011, \apj, 738, 45

\bibitem[{{Li} \& {White}(2009)}]{li09}
{Li}, C., \& {White}, S.~D.~M. 2009, \mnras, 398, 2177

\bibitem[{{Mandelbaum} {et~al.}(2006){Mandelbaum}, {Seljak}, {Kauffmann},
  {Hirata}, \& {Brinkmann}}]{Mandelbaum06}
{Mandelbaum}, R., {Seljak}, U., {Kauffmann}, G., {Hirata}, C.~M., \&
  {Brinkmann}, J. 2006, \mnras, 368, 715

\bibitem[{{Manera} {et~al.}(2013)}]{Manera2013}
{Manera}, M., {et~al.} 2013, \mnras, 428, 1036

\bibitem[{{Masjedi} {et~al.}(2006)}]{Masjedi2006}
{Masjedi}, M., {et~al.} 2006, \apj, 644, 54

\bibitem[{{Masters} {et~al.}(2011)}]{Masters2011}
{Masters}, K.~L., {et~al.} 2011, \mnras, 418, 1055

\bibitem[{{Matsubara}(2008{\natexlab{a}})}]{matsubara08a}
{Matsubara}, T. 2008{\natexlab{a}}, PRD, 78, 083519

\bibitem[{{Matsubara}(2008{\natexlab{b}})}]{matsubara08b}
---. 2008{\natexlab{b}}, PRD, 77, 063530

\bibitem[{{Meiksin} {et~al.}(1999){Meiksin}, {White}, \&
  {Peacock}}]{meiksin1999}
{Meiksin}, A., {White}, M., \& {Peacock}, J.~A. 1999, \mnras, 304, 851

\bibitem[{{Montesano} {et~al.}(2010){Montesano}, {S{\'a}nchez}, \&
  {Phleps}}]{montesano10}
{Montesano}, F., {S{\'a}nchez}, A.~G., \& {Phleps}, S. 2010, \mnras, 408, 2397

\bibitem[{{Padmanabhan} {et~al.}(2009){Padmanabhan}, {White}, {Norberg}, \&
  {Porciani}}]{Padmanabhan09}
{Padmanabhan}, N., {White}, M., {Norberg}, P., \& {Porciani}, C. 2009, \mnras,
  397, 1862

\bibitem[{{Phleps} {et~al.}(2006){Phleps}, {Peacock}, {Meisenheimer}, \&
  {Wolf}}]{Phleps06}
{Phleps}, S., {Peacock}, J.~A., {Meisenheimer}, K., \& {Wolf}, C. 2006, \aap,
  457, 145

\bibitem[{{Prada} {et~al.}(2012){Prada}, {Klypin}, {Cuesta}, {Betancort-Rijo},
  \& {Primack}}]{Prada2012}
{Prada}, F., {Klypin}, A.~A., {Cuesta}, A.~J., {Betancort-Rijo}, J.~E., \&
  {Primack}, J. 2012, \mnras, 423, 3018

\bibitem[{{Reddick} {et~al.}(2012){Reddick}, {Wechsler}, {Tinker}, \&
  {Behroozi}}]{Reddick2012}
{Reddick}, R.~M., {Wechsler}, R.~H., {Tinker}, J.~L., \& {Behroozi}, P.~S.
  2012, ArXiv e-prints

\bibitem[{{Reid} {et~al.}(2012)}]{Reid2012}
{Reid}, B.~A., {et~al.} 2012, \mnras, 426, 2719

\bibitem[{{Riebe} {et~al.}(2011)}]{Riebe11}
{Riebe}, K., {et~al.} 2011, ArXiv e-prints

\bibitem[{{Ross} \& {Brunner}(2009)}]{RossBrunner2009}
{Ross}, A.~J., \& {Brunner}, R.~J. 2009, \mnras, 399, 878

\bibitem[{{Ross} {et~al.}(2010){Ross}, {Percival}, \& {Brunner}}]{RPB2010}
{Ross}, A.~J., {Percival}, W.~J., \& {Brunner}, R.~J. 2010, \mnras, 407, 420

\bibitem[{{Ross} {et~al.}(2011)}]{Ross2011}
{Ross}, A.~J., {et~al.} 2011, \mnras, 417, 1350

\bibitem[{{Ross} {et~al.}(2012)}]{Ross2012}
---. 2012, \mnras, 424, 564

\bibitem[{{Ross} {et~al.}(2007)}]{RossN07}
{Ross}, N.~P., {et~al.} 2007, \mnras, 381, 573

\bibitem[{{S{\'a}nchez} {et~al.}(2008){S{\'a}nchez}, {Baugh}, \&
  {Angulo}}]{sanchez08}
{S{\'a}nchez}, A.~G., {Baugh}, C.~M., \& {Angulo}, R. 2008, \mnras, 390, 1470

\bibitem[{{S{\'a}nchez} {et~al.}(2012)}]{Sanchez2012}
{S{\'a}nchez}, A.~G., {et~al.} 2012, \mnras, 425, 415

\bibitem[{{Schlafly} \& {Finkbeiner}(2011)}]{Schlafly2011}
{Schlafly}, E.~F., \& {Finkbeiner}, D.~P. 2011, \apj, 737, 103

\bibitem[{{Schlafly} {et~al.}(2010){Schlafly}, {Finkbeiner}, {Schlegel},
  {Juri{\'c}}, {Ivezi{\'c}}, {Gibson}, {Knapp}, \& {Weaver}}]{Schlafly2010}
{Schlafly}, E.~F., {Finkbeiner}, D.~P., {Schlegel}, D.~J., {Juri{\'c}}, M.,
  {Ivezi{\'c}}, {\v Z}., {Gibson}, R.~R., {Knapp}, G.~R., \& {Weaver}, B.~A.
  2010, \apj, 725, 1175

\bibitem[{{Scoccimarro} \& {Sheth}(2002)}]{Scoccimarro2002}
{Scoccimarro}, R., \& {Sheth}, R.~K. 2002, \mnras, 329, 629

\bibitem[{{Seo} {et~al.}(2010){Seo}, {Eckel}, {Eisenstein}, {Mehta},
  {Metchnik}, {Padmanabhan}, {Pinto}, {Takahashi}, {et~al.}}]{seo10}
{Seo}, H.-J., {Eckel}, J., {Eisenstein}, D.~J., {Mehta}, K., {Metchnik}, M.,
  {Padmanabhan}, N., {Pinto}, P., {Takahashi}, R., {et~al.} 2010, \apj, 720,
  1650

\bibitem[{{Skibba} \& {Sheth}(2009)}]{Skibba2009}
{Skibba}, R.~A., \& {Sheth}, R.~K. 2009, \mnras, 392, 1080

\bibitem[{{Slosar} {et~al.}(2011)}]{2011arXiv1104.5244S}
{Slosar}, A., {et~al.} 2011, \jcap, 9, 1

\bibitem[{{Smee} {et~al.}(2012)}]{Smee2012}
{Smee}, S., {et~al.} 2012, ArXiv e-prints

\bibitem[{{Swanson} {et~al.}(2008){Swanson}, {Tegmark}, {Hamilton}, \&
  {Hill}}]{Swanson2008}
{Swanson}, M.~E.~C., {Tegmark}, M., {Hamilton}, A.~J.~S., \& {Hill}, J.~C.
  2008, \mnras, 387, 1391

\bibitem[{{Taruya} {et~al.}(2009){Taruya}, {Nishimichi}, {Saito}, \&
  {Hiramatsu}}]{taruya09}
{Taruya}, A., {Nishimichi}, T., {Saito}, S., \& {Hiramatsu}, T. 2009, PRD, 80,
  123503

\bibitem[{{Tasitsiomi} {et~al.}(2004){Tasitsiomi}, {Kravtsov}, {Gottl{\"o}ber},
  \& {Klypin}}]{Tasitsiomi04}
{Tasitsiomi}, A., {Kravtsov}, A.~V., {Gottl{\"o}ber}, S., \& {Klypin}, A.~A.
  2004, \apj, 607, 125

\bibitem[{{Tegmark et al.}(2004)}]{Tegmark2004}
{Tegmark et al.}, M. 2004, \apj, 606, 702

\bibitem[{{Trujillo-Gomez} {et~al.}(2011){Trujillo-Gomez}, {Klypin}, {Primack},
  \& {Romanowsky}}]{Trujillo-Gomez}
{Trujillo-Gomez}, S., {Klypin}, A., {Primack}, J., \& {Romanowsky}, A.~J. 2011,
  \apj, 742, 16

\bibitem[{{Vale} \& {Ostriker}(2004)}]{Vale04}
{Vale}, A., \& {Ostriker}, J.~P. 2004, \mnras, 353, 189

\bibitem[{{Wake} {et~al.}(2008){Wake}, {Croom}, {Sadler}, \&
  {Johnston}}]{Wake08}
{Wake}, D.~A., {Croom}, S.~M., {Sadler}, E.~M., \& {Johnston}, H.~M. 2008,
  \mnras, 391, 1674

\bibitem[{{Watson} {et~al.}(2012){Watson}, {Berlind}, \&
  {Zentner}}]{Watson2012}
{Watson}, D.~F., {Berlind}, A.~A., \& {Zentner}, A.~R. 2012, \apj, 754, 90

\bibitem[{{Wetzel} \& {White}(2010)}]{Wetzel10}
{Wetzel}, A.~R., \& {White}, M. 2010, \mnras, 403, 1072

\bibitem[{{White} {et~al.}(2011)}]{White2011}
{White}, M., {et~al.} 2011, \apj, 728, 126

\bibitem[{{York} {et~al.}(2000)}]{2000AJ....120.1579Y}
{York}, D.~G., {et~al.} 2000, \aj, 120, 1579

\bibitem[{{Zehavi} {et~al.}(2002)}]{Zehavi2002}
{Zehavi}, I., {et~al.} 2002, \apj, 571, 172

\bibitem[{{Zehavi} {et~al.}(2005)}]{Zehavi05}
---. 2005, \apj, 630, 1

\bibitem[{{Zehavi} {et~al.}(2011)}]{Zehavi2011}
---. 2011, \apj, 736, 59

\bibitem[{{Zentner} {et~al.}(2005){Zentner}, {Berlind}, {Bullock}, {Kravtsov},
  \& {Wechsler}}]{Zentner2005}
{Zentner}, A.~R., {Berlind}, A.~A., {Bullock}, J.~S., {Kravtsov}, A.~V., \&
  {Wechsler}, R.~H. 2005, \apj, 624, 505

\bibitem[{{Zheng} {et~al.}(2009){Zheng}, {Zehavi}, {Eisenstein}, {Weinberg}, \&
  {Jing}}]{Zheng09}
{Zheng}, Z., {Zehavi}, I., {Eisenstein}, D.~J., {Weinberg}, D.~H., \& {Jing},
  Y.~P. 2009, \apj, 707, 554

\bibitem[{{Zheng} {et~al.}(2005)}]{Zheng2005}
{Zheng}, Z., {et~al.} 2005, \apj, 633, 791

\end{thebibliography}
\bibliographystyle{apj}

\appendix

\section{Dependence of clustering on different effects}
\label{app:a}

\begin{figure}
      \includegraphics[width=87mm]{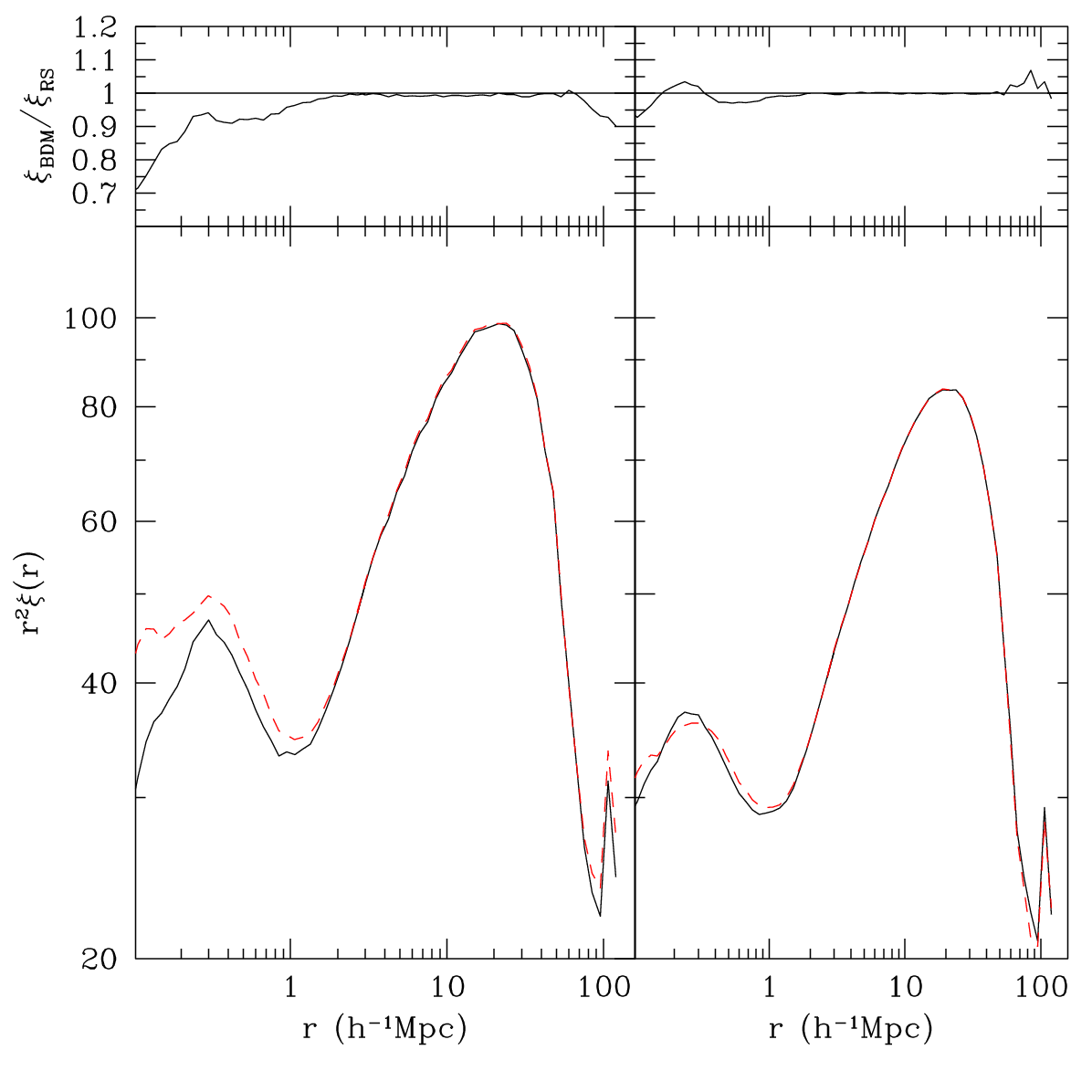}
      \caption{Comparison of the real-space correlation functions of (sub)haloes identified with the BDM 
               and RockStar halo finders at $z=0$ in the MultiDark simulation. Left panels 
               are for (sub)haloes with maximum circular velocity $\Vpeak>350\,\kms$ while right 
               panels are for $\Vpeak>300\,\kms$. Top panels present ratios of the correlation 
               functions. Solid (dashed) lines in the bottom panels show the BDM (RockStar) correlation 
               function multiplied by the square of radius.
              }
\label{app:A1}
\end{figure}

\subsection{Halo finding}
\label{app:aa}

The dependence of clustering with the halo finder used to identify 
virialized systems in the simulation is shown in Fig.~\ref{app:A1} for the 
BDM \citep{1997astro.ph.12217K,Riebe11} and RockStar \citep{RockStar} 
codes. As an example we select all (sub)haloes present in the Multidark simulation 
at $z=0$ with V$_{\rm max}>300$ km s$^{-1}$ and V$_{\rm max}>350$ km s$^{-1}$ 
in order to compute the real-space correlation function of the resulting halo catalogues. 
As can be seen in the figure the convergence between both halo finders is 
remarkable; for small (i.e., $\sim0.5$--$1\,h^{-1}\,\Mpc$) and large 
(i.e., $\gtrsim 70\,h^{-1}\,\Mpc$) scales the differences in power are typically of the 
order of $10\%$.

\subsection{Redshift evolution of BOSS-CMASS galaxies}
\label{app:ab}

\begin{figure}
      \includegraphics[width=87mm]{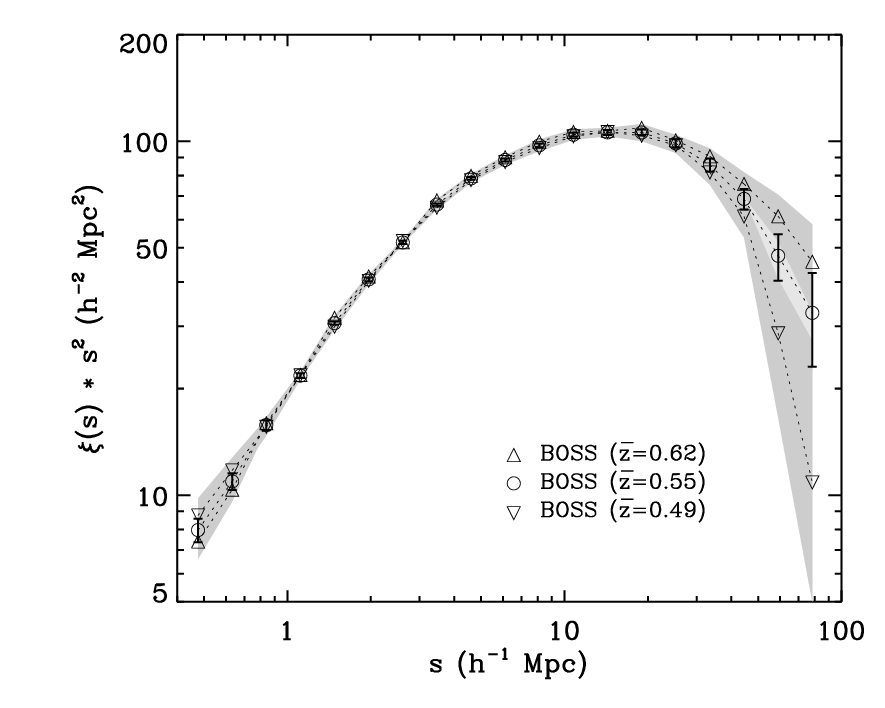}
      \caption{Redshift-space correlation function for three subsamples 
      drawn from the combined DR9 BOSS-CMASS galaxy sample at three different mean 
      redshifts. The error bars and the grey shaded 
      areas indicate estimates of the cosmic variance in observations for the 
      subsamples with $\bar{z}=0.55$ and $\bar{z}=0.49,0.62$
      respectively (see Section~\ref{Modelling}).      
      }
\label{app:A2}
\end{figure}

To assess the evolution of clustering with redshift within the BOSS-CMASS sample 
we splitted the combined (north+south) sample into three different subsamples with mean 
redshifts of $\bar{z}=0.49,0.55,0.62$ chosen to be around our fiducial value of $z=0.53$. 
Fig.~\ref{app:A2} shows that the clustering power is essentially independent of the corresponding 
mean redshift of the BOSS-CMASS sample for a large range of scales: the differences are 
negligible within the range $\sim0.7$--$40\,h^{-1}\;{\rm Mpc}$ but tend to increase 
for smaller and larger scales. At the largest scales the measurements of the subsample 
at $\bar{z}=0.49$ show the largest deviations in amplitude, as well as in the errors due 
to the smaller volume probed. However, the different correlation functions are still compatible with 
each other since deviations are within the $\sim1\sigma$ level given 
by cosmic variance, as indicated by the error bars and shaded areas (see Section~\ref{Modelling}).

\begin{figure}
      \includegraphics[width=87mm]{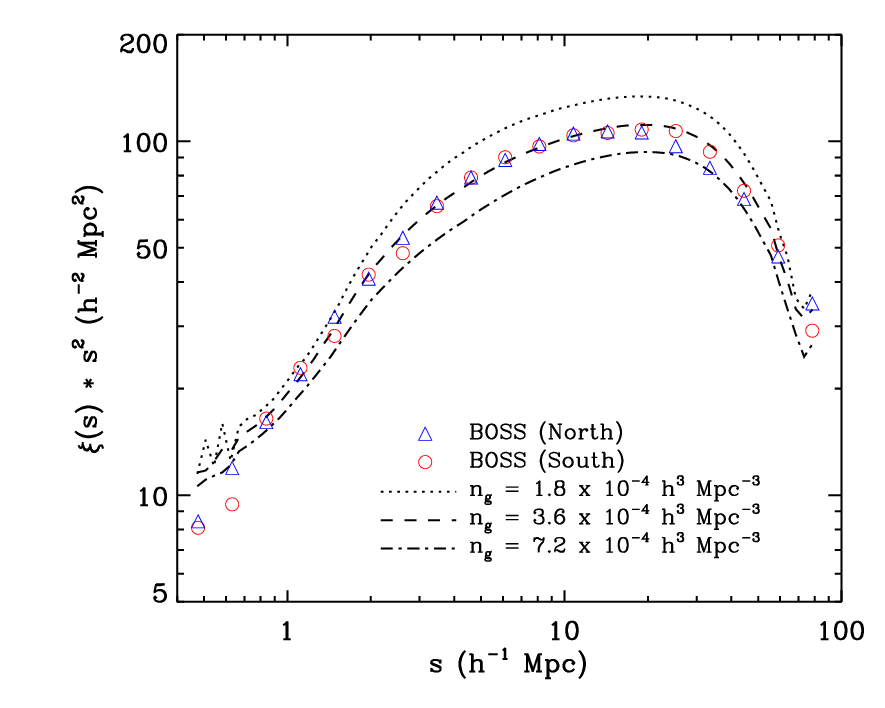}
      \caption{
      Redshift-space correlation function for different number densities 
      of our MultiDark halo catalogues at $z=0.53$ (including scatter) as 
      indicated in the plot (see text). 
      We compare these results with the DR9 BOSS-CMASS north and south 
      galaxy subsamples in the redshift range $0.4<z<0.7$. For clarity the 
      error bars are not shown.}
\label{app:A3}
\end{figure}

\subsection{Halo number density}
\label{app:ac}

To assess the clustering dependence with number density 
we evaluate different redshift-space correlations 
and compare with observations (see Fig.~\ref{app:A3}). 
We compute three different correlation functions using the 
MultiDark halo catalogues at $z=0.53$ assuming the stochasticity model 
presented in Section~\ref{HAM}. We use the following number densities: 
$n_{\rm g}=[1.8,3.6,7.2]\times10^{-4}\,h^3\,\Mpc^{-3}$. 
The dashed line corresponds to our effective number density, i.e. 
$n_* \equiv 3.6\times10^{-4}\,h^3\,\Mpc^{-3}$. As expected, doubling and dividing $n_*$ results 
in a weaker and stronger clustering signal respectively. In these extreme cases, 
the departure from observed BOSS-CMASS number densities at the peak of the redshift 
distribution is typically above observational uncertainties. 
However, at $z=0.53$, typical departures from the effective number density adopted 
in this work (e.g., between north and south subsamples) are smaller than 5\% and 
do not appreciably change our final result.


\section{Tables of correlation functions and covariances}
\label{app:b}

\begin{table*}
\begin{center}
\begin{tabular}{|r|rrrrrrrr|}
\hline\hline
$\sigma$\phantom{00} & $\Xi_{\rm N}$\phantom{00} & $\Xi_{\rm S}$\phantom{00} & $\Xi_{\rm N+S}$\phantom{0} & $\Xi_{\rm MD}$\phantom{0} & $\sigma_{\Xi,{\rm N}}$\phantom{0} & $\sigma_{\Xi,{\rm S}}$\phantom{0} & \phantom{0}$\sigma_{\Xi,{\rm N+S}}$ & $\sigma_{\Xi,{\rm C}}$\phantom{0}\\
\hline
 
  0.553  &  306.519  &  258.526  &  296.364   &  270.370  &   10.593  &   21.021  &  9.409  &  6.636  \\
  0.704  &  233.934  &  205.410  &  227.127   &  199.688  &    7.999  &   14.564  &  6.947  &  4.072  \\
  0.896  &  174.314  &  189.613  &  177.940   &  154.041  &    6.279  &   12.604  &  5.552  &  3.314  \\
  1.142  &  136.240  &  128.103  &  134.317   &  129.333  &    4.887  &    9.055  &  4.262  &  2.935  \\
  1.454  &  110.571  &  122.007  &  113.243   &  105.771  &    3.933  &    7.166  &  3.422  &  1.790  \\
  1.851  &   93.106  &   91.557  &   92.791   &   88.003  &    3.203  &    5.809  &  2.780  &  1.797  \\
  2.358  &   77.150  &   79.987  &   77.836   &   76.210  &    2.846  &    4.852  &  2.403  &  1.364  \\
  3.003  &   67.574  &   68.062  &   67.713   &   65.581  &    2.174  &    3.944  &  1.884  &  1.534  \\
  3.824  &   55.136  &   55.377  &   55.221   &   55.377  &    1.857  &    3.498  &  1.604  &  1.325  \\
  4.870  &   43.813  &   45.420  &   44.212   &   46.636  &    1.567  &    2.951  &  1.356  &  1.168  \\
  6.202  &   37.769  &   40.875  &   38.522   &   39.079  &    1.410  &    2.567  &  1.208  &  1.036  \\
  7.898  &   29.486  &   30.329  &   29.667   &   31.701  &    1.184  &    2.273  &  1.022  &  0.996  \\
 10.058  &   24.699  &   25.506  &   24.906   &   25.899  &    1.092  &    1.998  &  0.931  &  0.962  \\
 12.809  &   18.341  &   19.348  &   18.572   &   19.582  &    1.043  &    1.807  &  0.873  &  0.938  \\
 16.312  &   13.561  &   14.853  &   13.842   &   15.044  &    0.913  &    1.634  &  0.771  &  0.884  \\
 20.773  &    9.455  &   10.960  &    9.779   &   10.777  &    0.809  &    1.437  &  0.677  &  0.831  \\
 26.455  &    6.833  &    7.845  &    7.051   &    7.253  &    0.741  &    1.358  &  0.625  &  0.714  \\
 33.690  &    4.465  &    5.390  &    4.656   &    4.654  &    0.663  &    1.182  &  0.553  &  0.625  \\
 42.904  &    2.789  &    3.289  &    2.888   &    2.925  &    0.599  &    1.058  &  0.491  &  0.502  \\
 54.639  &    1.827  &    2.399  &    1.941   &    1.393  &    0.520  &    0.957  &  0.420  &  0.406  \\
 69.582  &    1.496  &    1.468  &    1.480   &    0.863  &    0.463  &    0.821  &  0.361  &  0.387  \\
 88.614  &    1.087  &    0.975  &    1.059   &    0.835  &    0.429  &    0.745  &  0.324  &  0.317  \\

\hline\hline
\end{tabular}
\end{center}
\caption{Projected correlation function, $\Xi_i$, and standard deviation, $\sigma_{\Xi,i}$, for the BOSS-CMASS north (N), south (S) and 
         combined (N$+$S) samples measured in 22 equally spaced logarithmic. Projected distance $\sigma$ and correlation functions 
         are indicated at the centre of the bin and measured in units of $h^{-1}$ Mpc. We also show the MultiDark (MD) HAM correlation 
         function and {\it Carmen} (C) standard deviation estimates.}
\label{tab:proy_cf}
\end{table*}

\begin{table*}
\begin{center}
  \tabcolsep 7.2pt
\begin{tabular}{|r|rrrrrrrr|}
\hline\hline
$s$\phantom{00} & $\xi_{\rm N}$\phantom{00} & $\xi_{\rm S}$\phantom{00} & $\xi_{\rm N+S}$\phantom{0} & $\xi_{\rm MD}$\phantom{0} & $\sigma_{\xi,{\rm N}}$\phantom{0}  & $\sigma_{\xi,{\rm S}}$\phantom{0} & \phantom{0}$\sigma_{\xi,{\rm N+S}}$ & $\sigma_{\xi,{\rm C}}$\phantom{0}\\
\hline

 0.476  &  37.266  &  35.725  &  37.880  &  55.002  &   5.733  &  10.891  &   5.260  &   1.071  \\
 0.632  &  29.886  &  23.598  &  28.942  &  37.425  &   2.809  &   5.442  &   2.477  &   0.623  \\
 0.839  &  22.851  &  23.385  &  23.270  &  24.742  &   1.399  &   2.595  &   1.243  &   0.510  \\
 1.114  &  17.698  &  18.450  &  18.003  &  17.917  &   0.683  &   1.182  &   0.604  &   0.223  \\
 1.479  &  14.595  &  12.877  &  14.314  &  13.837  &   0.348  &   0.644  &   0.307  &   0.153  \\
 1.965  &  10.598  &  10.881  &  10.725  &  10.508  &   0.196  &   0.374  &   0.174  &   0.088  \\
 2.609  &   7.843  &   7.095  &   7.705  &   7.910  &   0.118  &   0.207  &   0.107  &   0.059  \\
 3.464  &   5.590  &   5.464  &   5.585  &   5.627  &   0.067  &   0.134  &   0.066  &   0.039  \\
 4.600  &   3.736  &   3.733  &   3.744  &   3.789  &   0.039  &   0.075  &   0.038  &   0.029  \\
 6.108  &   2.375  &   2.416  &   2.392  &   2.438  &   0.026  &   0.046  &   0.024  &   0.016  \\
 8.111  &   1.494  &   1.469  &   1.493  &   1.532  &   0.016  &   0.031  &   0.015  &   0.012  \\
10.771  &   0.906  &   0.896  &   0.906  &   0.936  &   0.011  &   0.021  &   0.010  &   0.009  \\
14.303  &   0.521  &   0.516  &   0.521  &   0.559  &   0.008  &   0.015  &   0.007  &   0.007  \\
18.993  &   0.294  &   0.299  &   0.296  &   0.323  &   0.006  &   0.012  &   0.005  &   0.005  \\
25.221  &   0.152  &   0.168  &   0.156  &   0.179  &   0.005  &   0.009  &   0.004  &   0.004  \\
33.491  &   0.075  &   0.083  &   0.077  &   0.093  &   0.004  &   0.007  &   0.003  &   0.003  \\
44.473  &   0.035  &   0.037  &   0.035  &   0.041  &   0.003  &   0.006  &   0.003  &   0.002  \\
59.056  &   0.014  &   0.015  &   0.014  &   0.014  &   0.002  &   0.004  &   0.002  &   0.002  \\
78.421  &   0.006  &   0.005  &   0.006  &   0.005  &   0.002  &   0.003  &   0.002  &   0.002  \\

\hline\hline
\end{tabular}
\end{center}
\caption{Redshift-space correlation function, $\xi_i$, and standard deviation, $\sigma_{\xi,i}$, for the BOSS-CMASS north (N), south (S) 
         and combined (N$+$S) samples measured in 19 equally spaced logarithmic bins. The scale distance $s$ is indicated at the centre 
         of the bin and measured in units of $h^{-1}$ Mpc. We also show the MultiDark (MD) HAM correlation function 
         and {\it Carmen} (C) standard deviation estimates.}
\label{tab:reds_cf}
\end{table*}


\begin{landscape}
\begin{table}
\begin{center}
   \tabcolsep 3.3pt
{\tiny
\begin{tabular}{|c|cccccccccccccccccccccc|}
\hline\hline
 
   $\sigma$  &      0.553 &      0.704 &      0.896 &      1.142 &      1.454 &      1.851 &      2.358 &      3.003 &      3.824 &      4.870 &      6.202 &      7.898 &     10.058 &     12.809 &     16.312 &     20.773 &     26.455 &     33.690 &     42.904 &     54.639 &     69.582 &     88.614\\\hline
      0.553 &      1.000 &  --  &  --  &  --  &  --  &  --  &  --  &  --  &  --  &  --  &  --  &  --  &  --  &  --  &  --  &  --  &  --  &  --  &  --  &  --  &  --  &  -- \\
      0.704 &      0.112 &      1.000 &  --  &  --  &  --  &  --  &  --  &  --  &  --  &  --  &  --  &  --  &  --  &  --  &  --  &  --  &  --  &  --  &  --  &  --  &  --  &  -- \\
      0.896 &      0.072 &      0.105 &      1.000 &  --  &  --  &  --  &  --  &  --  &  --  &  --  &  --  &  --  &  --  &  --  &  --  &  --  &  --  &  --  &  --  &  --  &  --  &  -- \\
      1.142 &      0.129 &      0.070 &      0.151 &      1.000 &  --  &  --  &  --  &  --  &  --  &  --  &  --  &  --  &  --  &  --  &  --  &  --  &  --  &  --  &  --  &  --  &  --  &  -- \\
      1.454 &      0.030 &      0.135 &      0.096 &      0.174 &      1.000 &  --  &  --  &  --  &  --  &  --  &  --  &  --  &  --  &  --  &  --  &  --  &  --  &  --  &  --  &  --  &  --  &  -- \\
      1.851 &      0.034 &      0.149 &      0.147 &      0.127 &      0.152 &      1.000 &  --  &  --  &  --  &  --  &  --  &  --  &  --  &  --  &  --  &  --  &  --  &  --  &  --  &  --  &  --  &  -- \\
      2.358 &      0.111 &      0.211 &      0.134 &      0.167 &      0.215 &      0.255 &      1.000 &  --  &  --  &  --  &  --  &  --  &  --  &  --  &  --  &  --  &  --  &  --  &  --  &  --  &  --  &  -- \\
      3.003 &      0.142 &      0.146 &      0.146 &      0.142 &      0.218 &      0.204 &      0.352 &      1.000 &  --  &  --  &  --  &  --  &  --  &  --  &  --  &  --  &  --  &  --  &  --  &  --  &  --  &  -- \\
      3.824 &      0.069 &      0.117 &      0.112 &      0.185 &      0.162 &      0.311 &      0.370 &      0.408 &      1.000 &  --  &  --  &  --  &  --  &  --  &  --  &  --  &  --  &  --  &  --  &  --  &  --  &  -- \\
      4.870 &      0.102 &      0.086 &      0.153 &      0.170 &      0.216 &      0.208 &      0.374 &      0.332 &      0.402 &      1.000 &  --  &  --  &  --  &  --  &  --  &  --  &  --  &  --  &  --  &  --  &  --  &  -- \\
      6.202 &      0.118 &      0.101 &      0.179 &      0.184 &      0.208 &      0.264 &      0.379 &      0.411 &      0.467 &      0.504 &      1.000 &  --  &  --  &  --  &  --  &  --  &  --  &  --  &  --  &  --  &  --  &  -- \\
      7.898 &      0.131 &      0.113 &      0.156 &      0.197 &      0.241 &      0.291 &      0.453 &      0.407 &      0.461 &      0.522 &      0.648 &      1.000 &  --  &  --  &  --  &  --  &  --  &  --  &  --  &  --  &  --  &  -- \\
     10.058 &      0.114 &      0.104 &      0.155 &      0.229 &      0.277 &      0.290 &      0.442 &      0.401 &      0.488 &      0.518 &      0.656 &      0.710 &      1.000 &  --  &  --  &  --  &  --  &  --  &  --  &  --  &  --  &  -- \\
     12.809 &      0.104 &      0.149 &      0.190 &      0.221 &      0.254 &      0.288 &      0.436 &      0.388 &      0.456 &      0.499 &      0.627 &      0.685 &      0.788 &      1.000 &  --  &  --  &  --  &  --  &  --  &  --  &  --  &  -- \\
     16.312 &      0.119 &      0.163 &      0.179 &      0.242 &      0.228 &      0.288 &      0.400 &      0.367 &      0.456 &      0.481 &      0.583 &      0.633 &      0.771 &      0.812 &      1.000 &  --  &  --  &  --  &  --  &  --  &  --  &  -- \\
     20.773 &      0.115 &      0.152 &      0.157 &      0.192 &      0.192 &      0.248 &      0.364 &      0.363 &      0.419 &      0.427 &      0.553 &      0.602 &      0.720 &      0.761 &      0.837 &      1.000 &  --  &  --  &  --  &  --  &  --  &  -- \\
     26.455 &      0.095 &      0.152 &      0.141 &      0.182 &      0.191 &      0.204 &      0.336 &      0.337 &      0.370 &      0.410 &      0.526 &      0.579 &      0.664 &      0.698 &      0.768 &      0.856 &      1.000 &  --  &  --  &  --  &  --  &  -- \\
     33.690 &      0.086 &      0.093 &      0.101 &      0.178 &      0.165 &      0.179 &      0.281 &      0.292 &      0.325 &      0.364 &      0.472 &      0.495 &      0.592 &      0.597 &      0.677 &      0.737 &      0.838 &      1.000 &  --  &  --  &  --  &  -- \\
     42.904 &      0.084 &      0.077 &      0.086 &      0.208 &      0.142 &      0.147 &      0.260 &      0.254 &      0.282 &      0.326 &      0.408 &      0.438 &      0.513 &      0.509 &      0.592 &      0.646 &      0.735 &      0.857 &      1.000 &  --  &  --  &  -- \\
     54.639 &      0.074 &      0.035 &      0.071 &      0.198 &      0.121 &      0.106 &      0.247 &      0.229 &      0.220 &      0.258 &      0.338 &      0.357 &      0.414 &      0.411 &      0.489 &      0.525 &      0.617 &      0.711 &      0.843 &      1.000 &  --  &  -- \\
     69.582 &      0.013 &      0.002 &      0.078 &      0.144 &      0.111 &      0.030 &      0.170 &      0.178 &      0.175 &      0.196 &      0.261 &      0.320 &      0.323 &      0.318 &      0.376 &      0.397 &      0.491 &      0.567 &      0.671 &      0.820 &      1.000 &  -- \\
     88.614 &      0.028 &     -0.016 &      0.026 &      0.082 &      0.064 &      0.001 &      0.138 &      0.136 &      0.135 &      0.131 &      0.162 &      0.248 &      0.236 &      0.255 &      0.292 &      0.311 &      0.372 &      0.423 &      0.519 &      0.608 &      0.787 &      1.000\\

\hline\hline
\end{tabular}
}
\end{center}
\caption{Correlation matrix for the projected correlation function of the north BOSS-CMASS subsample estimated from an ensemble of mock galaxy catalogues designed to follow the observed geometry and redshift distribution \citep[more details can be found in ][]{Manera2013}. The projected distance $\sigma$ is in units of $h^{-1}\,{\rm Mpc}$.}
\label{tab:proy_cov}
\end{table}
\end{landscape}

\begin{landscape}
\begin{table}
\begin{center}
 \tabcolsep 3.3pt
{\tiny
\begin{tabular}{|c|cccccccccccccccccccccccccccc|}
\hline\hline

   $\sigma$  &      0.553 &      0.704 &      0.896 &      1.142 &      1.454 &      1.851 &      2.358 &      3.003 &      3.824 &      4.870 &      6.202 &      7.898 &     10.058 &     12.809 &     16.312 &     20.773 &     26.455 &     33.690 &     42.904 &     54.639 &     69.582 &     88.614\\\hline
      0.553 &      1.000 &  --  &  --  &  --  &  --  &  --  &  --  &  --  &  --  &  --  &  --  &  --  &  --  &  --  &  --  &  --  &  --  &  --  &  --  &  --  &  --  &  -- \\
      0.704 &      0.127 &      1.000 &  --  &  --  &  --  &  --  &  --  &  --  &  --  &  --  &  --  &  --  &  --  &  --  &  --  &  --  &  --  &  --  &  --  &  --  &  --  &  -- \\
      0.896 &      0.118 &      0.146 &      1.000 &  --  &  --  &  --  &  --  &  --  &  --  &  --  &  --  &  --  &  --  &  --  &  --  &  --  &  --  &  --  &  --  &  --  &  --  &  -- \\
      1.142 &      0.062 &      0.121 &      0.148 &      1.000 &  --  &  --  &  --  &  --  &  --  &  --  &  --  &  --  &  --  &  --  &  --  &  --  &  --  &  --  &  --  &  --  &  --  &  -- \\
      1.454 &      0.032 &      0.140 &      0.124 &      0.147 &      1.000 &  --  &  --  &  --  &  --  &  --  &  --  &  --  &  --  &  --  &  --  &  --  &  --  &  --  &  --  &  --  &  --  &  -- \\
      1.851 &      0.089 &      0.134 &      0.141 &      0.083 &      0.095 &      1.000 &  --  &  --  &  --  &  --  &  --  &  --  &  --  &  --  &  --  &  --  &  --  &  --  &  --  &  --  &  --  &  -- \\
      2.358 &      0.069 &      0.093 &      0.142 &      0.108 &      0.192 &      0.231 &      1.000 &  --  &  --  &  --  &  --  &  --  &  --  &  --  &  --  &  --  &  --  &  --  &  --  &  --  &  --  &  -- \\
      3.003 &      0.096 &      0.184 &      0.154 &      0.090 &      0.192 &      0.296 &      0.317 &      1.000 &  --  &  --  &  --  &  --  &  --  &  --  &  --  &  --  &  --  &  --  &  --  &  --  &  --  &  -- \\
      3.824 &      0.095 &      0.162 &      0.229 &      0.157 &      0.210 &      0.297 &      0.291 &      0.420 &      1.000 &  --  &  --  &  --  &  --  &  --  &  --  &  --  &  --  &  --  &  --  &  --  &  --  &  -- \\
      4.870 &      0.084 &      0.167 &      0.213 &      0.146 &      0.253 &      0.284 &      0.348 &      0.389 &      0.535 &      1.000 &  --  &  --  &  --  &  --  &  --  &  --  &  --  &  --  &  --  &  --  &  --  &  -- \\
      6.202 &      0.158 &      0.178 &      0.226 &      0.183 &      0.255 &      0.284 &      0.359 &      0.455 &      0.554 &      0.542 &      1.000 &  --  &  --  &  --  &  --  &  --  &  --  &  --  &  --  &  --  &  --  &  -- \\
      7.898 &      0.123 &      0.174 &      0.192 &      0.174 &      0.186 &      0.274 &      0.354 &      0.461 &      0.498 &      0.566 &      0.655 &      1.000 &  --  &  --  &  --  &  --  &  --  &  --  &  --  &  --  &  --  &  -- \\
     10.058 &      0.119 &      0.220 &      0.192 &      0.161 &      0.220 &      0.279 &      0.359 &      0.458 &      0.509 &      0.535 &      0.626 &      0.709 &      1.000 &  --  &  --  &  --  &  --  &  --  &  --  &  --  &  --  &  -- \\
     12.809 &      0.125 &      0.186 &      0.197 &      0.148 &      0.215 &      0.294 &      0.320 &      0.490 &      0.485 &      0.522 &      0.629 &      0.667 &      0.764 &      1.000 &  --  &  --  &  --  &  --  &  --  &  --  &  --  &  -- \\
     16.312 &      0.126 &      0.176 &      0.159 &      0.146 &      0.199 &      0.271 &      0.341 &      0.428 &      0.451 &      0.504 &      0.623 &      0.639 &      0.692 &      0.793 &      1.000 &  --  &  --  &  --  &  --  &  --  &  --  &  -- \\
     20.773 &      0.136 &      0.178 &      0.110 &      0.132 &      0.194 &      0.249 &      0.242 &      0.411 &      0.445 &      0.467 &      0.585 &      0.621 &      0.670 &      0.728 &      0.808 &      1.000 &  --  &  --  &  --  &  --  &  --  &  -- \\
     26.455 &      0.083 &      0.163 &      0.098 &      0.115 &      0.165 &      0.224 &      0.227 &      0.346 &      0.359 &      0.409 &      0.484 &      0.553 &      0.623 &      0.650 &      0.721 &      0.838 &      1.000 &  --  &  --  &  --  &  --  &  -- \\
     33.690 &      0.084 &      0.093 &      0.049 &      0.064 &      0.136 &      0.156 &      0.132 &      0.274 &      0.264 &      0.314 &      0.369 &      0.447 &      0.503 &      0.536 &      0.602 &      0.723 &      0.838 &      1.000 &  --  &  --  &  --  &  -- \\
     42.904 &      0.096 &      0.083 &     -0.004 &      0.068 &      0.103 &      0.119 &      0.128 &      0.185 &      0.246 &      0.248 &      0.305 &      0.350 &      0.423 &      0.440 &      0.464 &      0.575 &      0.670 &      0.809 &      1.000 &  --  &  --  &  -- \\
     54.639 &      0.095 &      0.035 &     -0.017 &      0.061 &      0.046 &      0.068 &      0.116 &      0.129 &      0.165 &      0.170 &      0.182 &      0.246 &      0.300 &      0.311 &      0.362 &      0.436 &      0.496 &      0.614 &      0.795 &      1.000 &  --  &  -- \\
     69.582 &      0.080 &      0.049 &      0.049 &      0.072 &      0.018 &      0.055 &      0.094 &      0.115 &      0.150 &      0.147 &      0.148 &      0.210 &      0.235 &      0.241 &      0.285 &      0.325 &      0.348 &      0.446 &      0.562 &      0.774 &      1.000 &  -- \\
     88.614 &      0.053 &      0.071 &      0.025 &      0.058 &      0.018 &      0.066 &      0.108 &      0.076 &      0.117 &      0.156 &      0.171 &      0.188 &      0.179 &      0.193 &      0.235 &      0.259 &      0.247 &      0.291 &      0.342 &      0.498 &      0.730 &      1.000\\

\hline\hline
\end{tabular}
}
\end{center}
\caption{Correlation matrix for the projected correlation function of the south BOSS-CMASS subsample estimated from an ensemble of mock galaxy catalogues designed to follow the observed geometry and redshift distribution \citep[more details can be found in ][]{Manera2013}. The projected distance $\sigma$ is in units of $h^{-1}\,{\rm Mpc}$.}
\label{tab:proy_cov}
\end{table}
\end{landscape}

\begin{landscape}
\begin{table}
\begin{center}
   \tabcolsep 3.3pt
{\tiny
\begin{tabular}{|c|cccccccccccccccccccccc|}
\hline\hline
 
 $\sigma$  &      0.553 &      0.704 &      0.896 &      1.142 &      1.454 &      1.851 &      2.358 &      3.003 &      3.824 &      4.870 &      6.202 &      7.898 &     10.058 &     12.809 &     16.312 &     20.773 &     26.455 &     33.690 &     42.904 &     54.639 &     69.582 &     88.614\\\hline
      0.553 &      1.000 &  --  &  --  &  --  &  --  &  --  &  --  &  --  &  --  &  --  &  --  &  --  &  --  &  --  &  --  &  --  &  --  &  --  &  --  &  --  &  --  &  -- \\
      0.704 &      0.114 &      1.000 &  --  &  --  &  --  &  --  &  --  &  --  &  --  &  --  &  --  &  --  &  --  &  --  &  --  &  --  &  --  &  --  &  --  &  --  &  --  &  -- \\
      0.896 &      0.079 &      0.105 &      1.000 &  --  &  --  &  --  &  --  &  --  &  --  &  --  &  --  &  --  &  --  &  --  &  --  &  --  &  --  &  --  &  --  &  --  &  --  &  -- \\
      1.142 &      0.109 &      0.079 &      0.144 &      1.000 &  --  &  --  &  --  &  --  &  --  &  --  &  --  &  --  &  --  &  --  &  --  &  --  &  --  &  --  &  --  &  --  &  --  &  -- \\
      1.454 &      0.030 &      0.135 &      0.098 &      0.163 &      1.000 &  --  &  --  &  --  &  --  &  --  &  --  &  --  &  --  &  --  &  --  &  --  &  --  &  --  &  --  &  --  &  --  &  -- \\
      1.851 &      0.042 &      0.139 &      0.141 &      0.112 &      0.141 &      1.000 &  --  &  --  &  --  &  --  &  --  &  --  &  --  &  --  &  --  &  --  &  --  &  --  &  --  &  --  &  --  &  -- \\
      2.358 &      0.092 &      0.178 &      0.124 &      0.139 &      0.204 &      0.242 &      1.000 &  --  &  --  &  --  &  --  &  --  &  --  &  --  &  --  &  --  &  --  &  --  &  --  &  --  &  --  &  -- \\
      3.003 &      0.127 &      0.152 &      0.139 &      0.119 &      0.209 &      0.226 &      0.339 &      1.000 &  --  &  --  &  --  &  --  &  --  &  --  &  --  &  --  &  --  &  --  &  --  &  --  &  --  &  -- \\
      3.824 &      0.064 &      0.114 &      0.127 &      0.168 &      0.168 &      0.300 &      0.337 &      0.407 &      1.000 &  --  &  --  &  --  &  --  &  --  &  --  &  --  &  --  &  --  &  --  &  --  &  --  &  -- \\
      4.870 &      0.088 &      0.098 &      0.150 &      0.151 &      0.221 &      0.221 &      0.356 &      0.337 &      0.421 &      1.000 &  --  &  --  &  --  &  --  &  --  &  --  &  --  &  --  &  --  &  --  &  --  &  -- \\
      6.202 &      0.120 &      0.114 &      0.176 &      0.172 &      0.212 &      0.266 &      0.360 &      0.418 &      0.476 &      0.495 &      1.000 &  --  &  --  &  --  &  --  &  --  &  --  &  --  &  --  &  --  &  --  &  -- \\
      7.898 &      0.123 &      0.119 &      0.149 &      0.176 &      0.220 &      0.286 &      0.414 &      0.410 &      0.455 &      0.518 &      0.639 &      1.000 &  --  &  --  &  --  &  --  &  --  &  --  &  --  &  --  &  --  &  -- \\
     10.058 &      0.107 &      0.123 &      0.146 &      0.196 &      0.260 &      0.285 &      0.409 &      0.406 &      0.480 &      0.506 &      0.641 &      0.698 &      1.000 &  --  &  --  &  --  &  --  &  --  &  --  &  --  &  --  &  -- \\
     12.809 &      0.098 &      0.152 &      0.177 &      0.191 &      0.239 &      0.288 &      0.387 &      0.404 &      0.449 &      0.488 &      0.617 &      0.666 &      0.774 &      1.000 &  --  &  --  &  --  &  --  &  --  &  --  &  --  &  -- \\
     16.312 &      0.111 &      0.165 &      0.158 &      0.203 &      0.215 &      0.281 &      0.363 &      0.370 &      0.440 &      0.471 &      0.580 &      0.617 &      0.745 &      0.799 &      1.000 &  --  &  --  &  --  &  --  &  --  &  --  &  -- \\
     20.773 &      0.109 &      0.156 &      0.132 &      0.159 &      0.182 &      0.250 &      0.312 &      0.363 &      0.410 &      0.419 &      0.550 &      0.585 &      0.697 &      0.740 &      0.822 &      1.000 &  --  &  --  &  --  &  --  &  --  &  -- \\
     26.455 &      0.084 &      0.155 &      0.116 &      0.148 &      0.177 &      0.209 &      0.286 &      0.325 &      0.351 &      0.392 &      0.504 &      0.550 &      0.639 &      0.670 &      0.746 &      0.842 &      1.000 &  --  &  --  &  --  &  --  &  -- \\
     33.690 &      0.075 &      0.094 &      0.074 &      0.131 &      0.148 &      0.175 &      0.219 &      0.271 &      0.292 &      0.333 &      0.433 &      0.455 &      0.551 &      0.559 &      0.642 &      0.715 &      0.826 &      1.000 &  --  &  --  &  --  &  -- \\
     42.904 &      0.071 &      0.081 &      0.049 &      0.157 &      0.121 &      0.144 &      0.197 &      0.220 &      0.253 &      0.288 &      0.369 &      0.386 &      0.465 &      0.461 &      0.539 &      0.603 &      0.700 &      0.835 &      1.000 &  --  &  --  &  -- \\
     54.639 &      0.059 &      0.033 &      0.029 &      0.147 &      0.089 &      0.098 &      0.176 &      0.185 &      0.178 &      0.205 &      0.275 &      0.286 &      0.348 &      0.339 &      0.421 &      0.460 &      0.556 &      0.662 &      0.812 &      1.000 &  --  &  -- \\
     69.582 &     -0.001 &      0.007 &      0.045 &      0.100 &      0.071 &      0.030 &      0.098 &      0.133 &      0.131 &      0.140 &      0.192 &      0.239 &      0.248 &      0.231 &      0.297 &      0.313 &      0.405 &      0.492 &      0.600 &      0.775 &      1.000 &  -- \\
     88.614 &      0.004 &     -0.007 &     -0.016 &      0.033 &      0.025 &      0.004 &      0.063 &      0.079 &      0.081 &      0.076 &      0.103 &      0.161 &      0.149 &      0.154 &      0.198 &      0.209 &      0.263 &      0.319 &      0.402 &      0.501 &      0.718 &      1.000\\

\hline\hline
\end{tabular}
}
\end{center}
\caption{Correlation matrix for the projected correlation function of the combined (north+south) BOSS-CMASS subsample estimated from an ensemble of mock galaxy catalogues designed to follow the observed geometry and redshift distribution \citep[more details can be found in ][]{Manera2013}. The projected distance $\sigma$ is in units of $h^{-1}\,{\rm Mpc}$.}
\label{tab:proy_cov}
\end{table}
\end{landscape}


\begin{landscape}
\begin{table}
\begin{center}
 \tabcolsep 3.3pt
{\tiny
\begin{tabular}{|c|ccccccccccccccccccc|}
\hline\hline

     $s$  &      0.476 &      0.632 &      0.839 &      1.114 &      1.479 &      1.965 &      2.609 &      3.464 &      4.600 &      6.108 &      8.111 &     10.771 &     14.303 &     18.993 &     25.221 &     33.491 &     44.473 &     59.056 &     78.421\\\hline
      0.476 &      1.000 &  --  &  --  &  --  &  --  &  --  &  --  &  --  &  --  &  --  &  --  &  --  &  --  &  --  &  --  &  --  &  --  &  --  &  -- \\
      0.632 &      0.121 &      1.000 &  --  &  --  &  --  &  --  &  --  &  --  &  --  &  --  &  --  &  --  &  --  &  --  &  --  &  --  &  --  &  --  &  -- \\
      0.839 &      0.085 &      0.103 &      1.000 &  --  &  --  &  --  &  --  &  --  &  --  &  --  &  --  &  --  &  --  &  --  &  --  &  --  &  --  &  --  &  -- \\
      1.114 &      0.061 &      0.103 &      0.115 &      1.000 &  --  &  --  &  --  &  --  &  --  &  --  &  --  &  --  &  --  &  --  &  --  &  --  &  --  &  --  &  -- \\
      1.479 &      0.128 &      0.114 &      0.160 &      0.122 &      1.000 &  --  &  --  &  --  &  --  &  --  &  --  &  --  &  --  &  --  &  --  &  --  &  --  &  --  &  -- \\
      1.965 &      0.083 &      0.053 &      0.066 &      0.056 &      0.138 &      1.000 &  --  &  --  &  --  &  --  &  --  &  --  &  --  &  --  &  --  &  --  &  --  &  --  &  -- \\
      2.609 &      0.040 &     -0.010 &      0.022 &      0.063 &      0.112 &      0.084 &      1.000 &  --  &  --  &  --  &  --  &  --  &  --  &  --  &  --  &  --  &  --  &  --  &  -- \\
      3.464 &      0.057 &      0.078 &      0.015 &      0.032 &      0.100 &      0.150 &      0.214 &      1.000 &  --  &  --  &  --  &  --  &  --  &  --  &  --  &  --  &  --  &  --  &  -- \\
      4.600 &      0.078 &      0.067 &      0.146 &      0.112 &      0.220 &      0.175 &      0.123 &      0.288 &      1.000 &  --  &  --  &  --  &  --  &  --  &  --  &  --  &  --  &  --  &  -- \\
      6.108 &      0.045 &      0.048 &      0.118 &      0.125 &      0.185 &      0.176 &      0.184 &      0.237 &      0.385 &      1.000 &  --  &  --  &  --  &  --  &  --  &  --  &  --  &  --  &  -- \\
      8.111 &      0.084 &      0.047 &      0.140 &      0.132 &      0.208 &      0.197 &      0.149 &      0.282 &      0.407 &      0.488 &      1.000 &  --  &  --  &  --  &  --  &  --  &  --  &  --  &  -- \\
     10.771 &      0.024 &      0.045 &      0.162 &      0.120 &      0.212 &      0.184 &      0.140 &      0.234 &      0.364 &      0.465 &      0.599 &      1.000 &  --  &  --  &  --  &  --  &  --  &  --  &  -- \\
     14.303 &      0.053 &      0.032 &      0.146 &      0.136 &      0.218 &      0.187 &      0.146 &      0.176 &      0.368 &      0.471 &      0.551 &      0.704 &      1.000 &  --  &  --  &  --  &  --  &  --  &  -- \\
     18.993 &      0.050 &      0.013 &      0.174 &      0.138 &      0.201 &      0.187 &      0.081 &      0.195 &      0.319 &      0.406 &      0.501 &      0.651 &      0.762 &      1.000 &  --  &  --  &  --  &  --  &  -- \\
     25.221 &      0.029 &     -0.005 &      0.120 &      0.086 &      0.167 &      0.150 &      0.096 &      0.122 &      0.243 &      0.345 &      0.428 &      0.575 &      0.678 &      0.791 &      1.000 &  --  &  --  &  --  &  -- \\
     33.491 &     -0.004 &     -0.021 &      0.137 &      0.054 &      0.138 &      0.085 &      0.051 &      0.081 &      0.214 &      0.263 &      0.351 &      0.497 &      0.575 &      0.690 &      0.819 &      1.000 &  --  &  --  &  -- \\
     44.473 &      0.007 &     -0.002 &      0.123 &      0.098 &      0.113 &      0.068 &      0.045 &      0.095 &      0.220 &      0.241 &      0.313 &      0.423 &      0.482 &      0.586 &      0.678 &      0.827 &      1.000 &  --  &  -- \\
     59.056 &     -0.010 &      0.010 &      0.084 &      0.064 &      0.059 &      0.017 &      0.055 &      0.118 &      0.133 &      0.174 &      0.220 &      0.280 &      0.336 &      0.423 &      0.487 &      0.609 &      0.817 &      1.000 &  -- \\
     78.421 &     -0.021 &      0.012 &      0.038 &      0.015 &      0.022 &      0.000 &      0.061 &      0.088 &      0.109 &      0.124 &      0.143 &      0.194 &      0.230 &      0.285 &      0.321 &      0.408 &      0.571 &      0.764 &      1.000\\

\hline\hline
\end{tabular}
}
\end{center}
\caption{Correlation matrix for the redshift-space correlation function of the north BOSS-CMASS subsample estimated from an ensemble of mock galaxy catalogues designed to follow the observed geometry and redshift distribution \citep[more details can be found in][]{Manera2013}. The distance scale $s$ is in units of $h^{-1}$\,Mpc.}
\label{tab:reds_cov}
\end{table}
\end{landscape}

\begin{landscape}
\begin{table}
\begin{center}
 \tabcolsep 3.3pt
{\tiny
\begin{tabular}{|c|ccccccccccccccccccc|}
\hline\hline

    $s$  &      0.476 &      0.632 &      0.839 &      1.114 &      1.479 &      1.965 &      2.609 &      3.464 &      4.600 &      6.108 &      8.111 &     10.771 &     14.303 &     18.993 &     25.221 &     33.491 &     44.473 &     59.056 &     78.421\\\hline
      0.476 &      1.000 &  --  &  --  &  --  &  --  &  --  &  --  &  --  &  --  &  --  &  --  &  --  &  --  &  --  &  --  &  --  &  --  &  --  &  -- \\
      0.632 &      0.053 &      1.000 &  --  &  --  &  --  &  --  &  --  &  --  &  --  &  --  &  --  &  --  &  --  &  --  &  --  &  --  &  --  &  --  &  -- \\
      0.839 &      0.080 &      0.102 &      1.000 &  --  &  --  &  --  &  --  &  --  &  --  &  --  &  --  &  --  &  --  &  --  &  --  &  --  &  --  &  --  &  -- \\
      1.114 &      0.123 &      0.081 &      0.143 &      1.000 &  --  &  --  &  --  &  --  &  --  &  --  &  --  &  --  &  --  &  --  &  --  &  --  &  --  &  --  &  -- \\
      1.479 &      0.039 &      0.034 &      0.040 &      0.014 &      1.000 &  --  &  --  &  --  &  --  &  --  &  --  &  --  &  --  &  --  &  --  &  --  &  --  &  --  &  -- \\
      1.965 &      0.006 &      0.031 &      0.081 &      0.129 &      0.072 &      1.000 &  --  &  --  &  --  &  --  &  --  &  --  &  --  &  --  &  --  &  --  &  --  &  --  &  -- \\
      2.609 &      0.019 &     -0.041 &      0.097 &      0.016 &      0.126 &      0.107 &      1.000 &  --  &  --  &  --  &  --  &  --  &  --  &  --  &  --  &  --  &  --  &  --  &  -- \\
      3.464 &      0.006 &      0.100 &      0.144 &      0.137 &      0.054 &      0.141 &      0.216 &      1.000 &  --  &  --  &  --  &  --  &  --  &  --  &  --  &  --  &  --  &  --  &  -- \\
      4.600 &      0.031 &      0.126 &      0.123 &      0.114 &      0.079 &      0.125 &      0.126 &      0.276 &      1.000 &  --  &  --  &  --  &  --  &  --  &  --  &  --  &  --  &  --  &  -- \\
      6.108 &      0.080 &      0.095 &      0.186 &      0.123 &      0.120 &      0.153 &      0.198 &      0.284 &      0.344 &      1.000 &  --  &  --  &  --  &  --  &  --  &  --  &  --  &  --  &  -- \\
      8.111 &      0.108 &      0.098 &      0.098 &      0.155 &      0.130 &      0.119 &      0.179 &      0.295 &      0.321 &      0.487 &      1.000 &  --  &  --  &  --  &  --  &  --  &  --  &  --  &  -- \\
     10.771 &      0.057 &      0.146 &      0.127 &      0.159 &      0.116 &      0.110 &      0.155 &      0.278 &      0.309 &      0.438 &      0.586 &      1.000 &  --  &  --  &  --  &  --  &  --  &  --  &  -- \\
     14.303 &      0.044 &      0.039 &      0.136 &      0.146 &      0.150 &      0.084 &      0.093 &      0.202 &      0.304 &      0.397 &      0.532 &      0.680 &      1.000 &  --  &  --  &  --  &  --  &  --  &  -- \\
     18.993 &     -0.012 &      0.052 &      0.103 &      0.135 &      0.105 &      0.109 &      0.072 &      0.193 &      0.297 &      0.365 &      0.480 &      0.626 &      0.779 &      1.000 &  --  &  --  &  --  &  --  &  -- \\
     25.221 &      0.004 &      0.048 &      0.118 &      0.108 &      0.106 &      0.131 &      0.067 &      0.167 &      0.235 &      0.308 &      0.422 &      0.538 &      0.663 &      0.794 &      1.000 &  --  &  --  &  --  &  -- \\
     33.491 &     -0.009 &     -0.015 &      0.026 &      0.098 &      0.063 &      0.113 &      0.069 &      0.114 &      0.159 &      0.185 &      0.321 &      0.405 &      0.509 &      0.647 &      0.804 &      1.000 &  --  &  --  &  -- \\
     44.473 &     -0.023 &     -0.021 &      0.018 &      0.115 &      0.034 &      0.084 &      0.075 &      0.081 &      0.112 &      0.110 &      0.218 &      0.268 &      0.373 &      0.486 &      0.609 &      0.792 &      1.000 &  --  &  -- \\
     59.056 &      0.011 &     -0.047 &      0.023 &      0.086 &      0.018 &      0.046 &      0.042 &      0.069 &      0.049 &      0.057 &      0.140 &      0.195 &      0.241 &      0.337 &      0.414 &      0.570 &      0.762 &      1.000 &  -- \\
     78.421 &     -0.016 &     -0.097 &      0.033 &      0.015 &      0.015 &     -0.009 &     -0.006 &     -0.019 &     -0.001 &     -0.024 &      0.039 &      0.056 &      0.096 &      0.174 &      0.229 &      0.322 &      0.454 &      0.686 &      1.000\\

\hline\hline
\end{tabular}
}
\end{center}
\caption{Correlation matrix for the redshift-space correlation function of the south BOSS-CMASS subsample estimated from an ensemble of mock galaxy catalogues designed to follow the observed geometry and redshift distribution \citep[more details can be found in][]{Manera2013}. The distance scale $s$ is in units of $h^{-1}$\,Mpc.}
\label{tab:reds_cov}
\end{table}
\end{landscape}

\begin{landscape}
\begin{table}
\begin{center}
 \tabcolsep 3.3pt
{\tiny
\begin{tabular}{|c|ccccccccccccccccccc|}
\hline\hline

    $s$  &      0.476 &      0.632 &      0.839 &      1.114 &      1.479 &      1.965 &      2.609 &      3.464 &      4.600 &      6.108 &      8.111 &     10.771 &     14.303 &     18.993 &     25.221 &     33.491 &     44.473 &     59.056 &     78.421\\\hline
      0.476 &      1.000 &  --  &  --  &  --  &  --  &  --  &  --  &  --  &  --  &  --  &  --  &  --  &  --  &  --  &  --  &  --  &  --  &  --  &  -- \\
      0.632 &      0.087 &      1.000 &  --  &  --  &  --  &  --  &  --  &  --  &  --  &  --  &  --  &  --  &  --  &  --  &  --  &  --  &  --  &  --  &  -- \\
      0.839 &      0.125 &      0.107 &      1.000 &  --  &  --  &  --  &  --  &  --  &  --  &  --  &  --  &  --  &  --  &  --  &  --  &  --  &  --  &  --  &  -- \\
      1.114 &      0.095 &      0.127 &      0.153 &      1.000 &  --  &  --  &  --  &  --  &  --  &  --  &  --  &  --  &  --  &  --  &  --  &  --  &  --  &  --  &  -- \\
      1.479 &      0.125 &      0.166 &      0.145 &      0.134 &      1.000 &  --  &  --  &  --  &  --  &  --  &  --  &  --  &  --  &  --  &  --  &  --  &  --  &  --  &  -- \\
      1.965 &      0.099 &      0.021 &      0.107 &      0.132 &      0.175 &      1.000 &  --  &  --  &  --  &  --  &  --  &  --  &  --  &  --  &  --  &  --  &  --  &  --  &  -- \\
      2.609 &      0.077 &      0.038 &      0.083 &      0.137 &      0.163 &      0.148 &      1.000 &  --  &  --  &  --  &  --  &  --  &  --  &  --  &  --  &  --  &  --  &  --  &  -- \\
      3.464 &      0.085 &      0.097 &      0.107 &      0.130 &      0.141 &      0.192 &      0.292 &      1.000 &  --  &  --  &  --  &  --  &  --  &  --  &  --  &  --  &  --  &  --  &  -- \\
      4.600 &      0.122 &      0.137 &      0.175 &      0.168 &      0.234 &      0.206 &      0.226 &      0.382 &      1.000 &  --  &  --  &  --  &  --  &  --  &  --  &  --  &  --  &  --  &  -- \\
      6.108 &      0.116 &      0.116 &      0.188 &      0.171 &      0.213 &      0.240 &      0.288 &      0.344 &      0.452 &      1.000 &  --  &  --  &  --  &  --  &  --  &  --  &  --  &  --  &  -- \\
      8.111 &      0.162 &      0.094 &      0.185 &      0.178 &      0.236 &      0.260 &      0.234 &      0.358 &      0.492 &      0.531 &      1.000 &  --  &  --  &  --  &  --  &  --  &  --  &  --  &  -- \\
     10.771 &      0.082 &      0.104 &      0.196 &      0.141 &      0.206 &      0.216 &      0.233 &      0.339 &      0.430 &      0.504 &      0.646 &      1.000 &  --  &  --  &  --  &  --  &  --  &  --  &  -- \\
     14.303 &      0.104 &      0.053 &      0.196 &      0.134 &      0.217 &      0.236 &      0.182 &      0.245 &      0.408 &      0.476 &      0.583 &      0.718 &      1.000 &  --  &  --  &  --  &  --  &  --  &  -- \\
     18.993 &      0.073 &      0.023 &      0.186 &      0.129 &      0.196 &      0.215 &      0.100 &      0.239 &      0.354 &      0.399 &      0.503 &      0.640 &      0.754 &      1.000 &  --  &  --  &  --  &  --  &  -- \\
     25.221 &      0.064 &      0.023 &      0.134 &      0.074 &      0.154 &      0.188 &      0.138 &      0.172 &      0.282 &      0.326 &      0.444 &      0.554 &      0.648 &      0.783 &      1.000 &  --  &  --  &  --  &  -- \\
     33.491 &      0.005 &     -0.025 &      0.121 &      0.039 &      0.129 &      0.104 &      0.078 &      0.095 &      0.230 &      0.218 &      0.337 &      0.463 &      0.540 &      0.675 &      0.811 &      1.000 &  --  &  --  &  -- \\
     44.473 &     -0.001 &     -0.025 &      0.101 &      0.092 &      0.096 &      0.082 &      0.057 &      0.092 &      0.203 &      0.180 &      0.285 &      0.367 &      0.448 &      0.558 &      0.671 &      0.828 &      1.000 &  --  &  -- \\
     59.056 &      0.002 &     -0.029 &      0.057 &      0.068 &      0.040 &      0.046 &      0.041 &      0.106 &      0.148 &      0.148 &      0.209 &      0.262 &      0.318 &      0.411 &      0.488 &      0.627 &      0.817 &      1.000 &  -- \\
     78.421 &     -0.003 &     -0.037 &      0.040 &      0.047 &      0.009 &      0.031 &      0.043 &      0.076 &      0.140 &      0.120 &      0.168 &      0.198 &      0.217 &      0.275 &      0.334 &      0.437 &      0.581 &      0.758 &      1.000\\

\hline\hline
\end{tabular}
}
\end{center}
\caption{Correlation matrix for the redshift-space correlation function of the combined (north+south) BOSS-CMASS subsample estimated from an ensemble of mock galaxy catalogues designed to follow the observed geometry and redshift distribution \citep[more details can be found in][]{Manera2013}. The distance scale $s$ is in units of $h^{-1}$\,Mpc.}
\label{tab:reds_cov}
\end{table}
\end{landscape}

\end{document}